\documentclass[iop,apj,numberedappendix]{emulateapj}

\pdfoutput=1

\usepackage{enumerate}
\usepackage{natbib,ifthen}
\usepackage{graphicx}
\usepackage{fancyhdr}
\usepackage{amssymb,amsmath}
\usepackage{epstopdf}
\usepackage{chngcntr}
\usepackage{IEEEtrantools}

\shorttitle{HH 111/HH 121 Protostellar System}
\shortauthors{Sewi{\l}o et al.}

\begin{document}


\title{Very Large Array Ammonia Observations of the HH 111/HH 121 Protostellar System: a Detection of a New Source With a Peculiar Chemistry }


\setcounter{affil}{1}

\newcounter{nasa}\setcounter{nasa}{\value{affil}}\stepcounter{affil}
\altaffiltext{\arabic{nasa}}{NASA Goddard Space Flight Center, 8800 Greenbelt Rd, Greenbelt, MD 20771, USA, marta.m.sewilo@nasa.gov}
\setcounter{affil}{2}
\newcounter{uva}\setcounter{uva}{\value{affil}}\stepcounter{affil}
\altaffiltext{\arabic{uva}}{Department of Astronomy, University of Virginia, PO Box 400325, Charlottesville, VA 22904, USA}
\setcounter{affil}{3}
\newcounter{nrao}\setcounter{nrao}{\value{affil}}\stepcounter{affil}
\altaffiltext{\arabic{nrao}}{National Radio Astronomy Observatory, 520 Edgemont Rd, Charlottesville, VA 22903, USA}
\setcounter{affil}{4}
\newcounter{mp}\setcounter{mp}{\value{affil}}\stepcounter{affil}
\altaffiltext{\arabic{mp}}{The Center for Astrochemical Studies of the Max Planck Intitute for Extraterrestial Physics, Giessenbachstrasse 1, 85748 Garching, Germany}
\setcounter{affil}{5}
\newcounter{yunnan}\setcounter{yunnan}{\value{affil}}\stepcounter{affil}
\altaffiltext{\arabic{yunnan}}{Department of Astronomy, Yunnan University, and Key Laboratory of Astroparticle Physics of Yunnan Province, Kunming, 650091, China}

\author{Marta Sewi{\l}o\altaffilmark{\arabic{nasa}}, Jennifer Wiseman\altaffilmark{\arabic{nasa}}, Remy Indebetouw\altaffilmark{\arabic{uva},\arabic{nrao}}, Steven B. Charnley\altaffilmark{\arabic{nasa}}, Jaime E. Pineda\altaffilmark{\arabic{mp}}, Johan E. Lindberg\altaffilmark{\arabic{nasa}}, Sheng-Li Qin\altaffilmark{\arabic{yunnan}}}


\begin{abstract}
\indent  We present the results of Very Large Array NH$_{3}$ $(J,K)=(1,1)$ and $(2,2)$ observations of the HH\,111/HH\,121 protostellar system. HH\,111, with a spectacular collimated optical jet, is one of the most well-known Herbig-Haro objects. We report the detection of a new source (NH$_{3}-$S) in the vicinity of HH\,111/HH\,121 ($\sim$0.03 pc from the HH\,111 jet source) in two epochs of the ammonia observations.  This constitutes the first detection of this source, in a region which has been thoroughly covered previously by both continuum and spectral line interferometric observations.  We study the kinematic and physical properties of HH\,111 and the newly discovered NH$_{3}-$S.  We also use HCO$^{+}$ and HCN $(J=4-3)$ data obtained with the James Clerk Maxwell Telescope and archival Atacama Large Millimeter/submillimeter Array  $^{13}$CO, $^{12}$CO, and C$^{18}$O $(J=2-1)$, N$_2$D$^{+}$ $(J=3-2)$, and $^{13}$CS $(J=5-4)$ data to gain insight into the nature of NH$_{3}-$S.  The chemical structure of NH$_3-$S shows evidence for ``selective freeze-out'', an inherent characteristic of dense cold cores.  The inner part of NH$_3-$S shows subsonic non-thermal velocity dispersions indicating a ``coherent core'', while they increase in the direction of the jets.  Archival near- to far-infrared data show no indication of any embedded source in NH$_3-$S.  The properties of NH$_3-$S and its location in the infrared dark cloud suggest that it is a starless core located in a turbulent medium with turbulence induced by Herbig-Haro jets and associated outflows.  More data is needed to fully understand the physical and chemical properties of NH$_3-$S and if/how its evolution is affected by nearby jets. 
\end{abstract}

\section{Introduction}
\label{s:intro}

\begin{figure*}[ht!]
\centering
\includegraphics[width=0.8\textwidth]{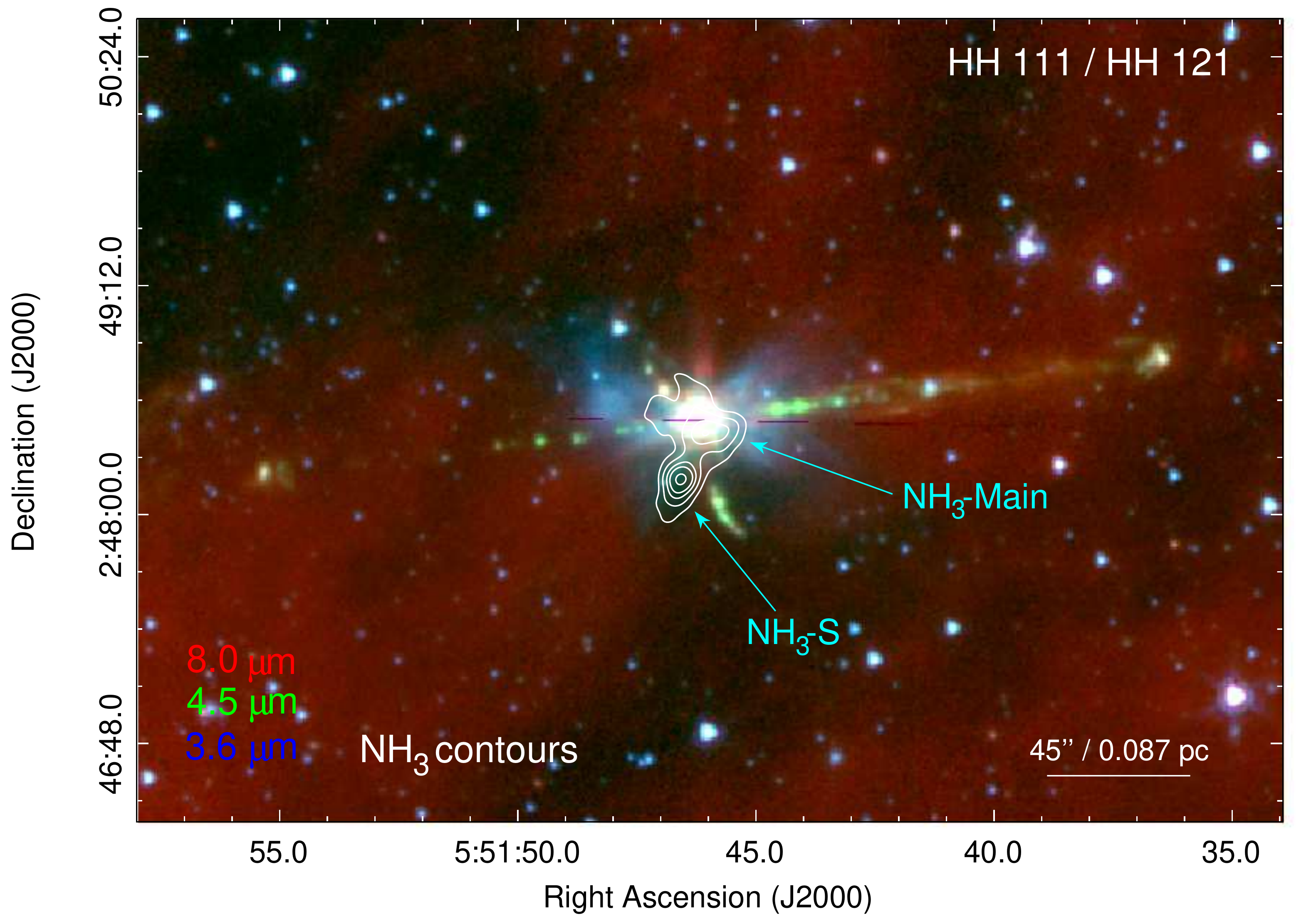}
\caption{A three-color composite image of HH\,111/HH\,121, combining the {\it Spitzer} IRAC 8 $\mu$m ({\it red}), 4.5 $\mu$m ({\it green}), and 3.6 $\mu$m ({\it blue}) images. The HH\,111 ($\sim$E-W) and HH\,121 ($\sim$N-S) jets are clearly seen in the 4.5 $\mu$m emission. Two ammonia sources are indicated with arrows and labeled. The contours represent the NH$_{3}$ (1,\,1) emission;  the contour levels are (20, 40, 60, 75, 95)\% of the NH$_{3}$ (1,\,1) integrated intensity peak of  86.9 mJy beam$^{-1}$ km s$^{-1}$ (see Figure~\ref{f:mom0}).  The {\it Spitzer}/IRAC resolution is $\sim$2$''$. The VLA synthesized beam is $\sim$7$\rlap.{''}$9 $\times$ 6$\rlap.{''}$5.  The linear scale shown in the lower right corresponds to a distance of 400 pc. North is up and east to the left.
 \label{f:spitzer}}
\end{figure*}

The Herbig-Haro object HH\,111 is one of the prototypical examples of highly-collimated optical jet sources \citep{reipurth1989}. It is located in the L1617 dark cloud of the Orion B molecular cloud at a distance of 400 pc (e.g., \citealt{sandstrom2007}). The infrared source IRAS 05491+0247 (or VLA-1 in \citealt{reipurth1999}) is the driving source of the jet. This Class I protostar with an infalling, flattened envelope and circumstellar disk has the luminosity of $\sim$25 L$_{\odot}$ \citep{reipurth1992} and is deeply embedded in a 30~M$_{\odot}$ molecular cloud core \citep{reipurth1991,stapelfeldt1993}. The total extent of the HH\,111 jet complex is 367$''$ (or $\sim$0.7 pc; \citealt{reipurth1999}); it consists of a blueshifted, highly-collimated and bright optical jet, a redshifted faint counterjet, and several bow shocks.  The HH\,111 jet originates in the high-extinction region and its base is associated with a reflection nebula illuminated by the protostar VLA--1.  The proper motions along the jet are large ($\sim$300--600 km~s$^{-1}$) and it moves at an inclination angle of 10$^{\circ}$ to the plane of the sky. The dynamical age of the complex is 800 years \citep{reipurth1992}. A second pair of bipolar jets (HH\,121) was discovered in the near-infrared \citep{gredel1993}; it intersects HH\,111 near the position of the central source at an angle of 61$^{\circ}$, suggesting that the driving source of the HH\,111 jet may be a binary.  \citet{reipurth1999} argue that the quadrupolar morphology of VLA-1 in the 3.6 cm images suggests that it is a close binary with a projected separation of $<$0$\rlap.{''}$1 ($\sim$40 AU at 400 pc). 

The HH\,111 jet is associated with a large well-collimated molecular outflow (e.g., \citealt{reipurth1991}; \citealt{cernicharo1996}; \citealt{nagar1997}).  Based on the CO kinematic data, \citet{cernicharo1996} concluded that the CO flow surrounds the Herbig-Haro jet. A second well-defined bipolar molecular flow in the region coincides with the HH\,121 infrared jet.

HH\,111 has been a target of multiple interferometric observations (e.g., the Submillimeter Array, Owens Valley Radio Observatory, Nobeyama Millimeter Array; see Section~\ref{s:nh3})  with resolutions ranging from less than one to a few arcsec. The morphology, chemistry, and kinematics of the envelope and the disk of the source exciting HH\,111 jet have been studied in detail. 

In this work, we present the results of the Very Large Array NH$_{3}$ (1,\,1) and (2,\,2) observations of the HH\,111/HH\,121 protostellar system and its surroundings. Ammonia offers a valuable probe of both gas density and temperature.  We discuss the distribution, kinematics, and physical properties of the gas.  We report the discovery of an NH$_{3}$ source in the vicinity of HH\,111/HH\,121 ($\sim$15$''$ or $\sim$6000 AU) and explore its nature using the ancillary mid- to far-infrared, and (sub)millimeter continuum and molecular line data.  These multiwavelength observations combined with the ammonia data allow us to determine the chemical structure of the newly discovered NH$_{3}$ source, its physical parameters (including the temperature, density, and mass), the velocity structure, and non-thermal motions, and to assess the stellar content.  The location of the source close to two Herbig-Haro objects suggests that the environment may be an important factor in its formation and evolution.  Theoretical models show that although the outflow-driven turbulence (or ``protostellar turbulence'') can suppress/delay global star formation, they can induce star formation on small scales by dynamical compression of pre-existing dense cores (e.g., \citealt{nakamura2007}).  

The observations, data reduction, and the ancillary data are described in Section~\ref{s:data}.  In Section~\ref{s:results}, we present a detailed analysis of the NH$_{3}$ data and discuss the results in the context of the physical and chemical characteristics of the region.  A discussion on the nature of the newly discovered  NH$_{3}$ source is provided in Section~\ref{s:nh3s}.  The summary and conclusions are given in Section~\ref{s:summary}. 

\begin{deluxetable*}{p{7cm}cc}
\centering
\tablecaption{Instrumental Parameters for the VLA NH$_{3}$ Observations of HH\,111 \label{t:data}}
\tablewidth{0pt}
\tablehead{
\multicolumn{1}{c}{Parameter} &
\colhead{Epoch 1} &
\colhead{Epoch 2}}
\startdata
Program ID \dotfill & AW512 & AW543 \\
Observation dates \dotfill & 1999 June 1--2 &  2000  August 24\\
Total observing time (hr) \dotfill & $\sim$3.8  &  $\sim$2.2 \\
Configuration \dotfill & D\tablenotemark{a,b} & D\tablenotemark{b} \\
Number of antennas  \dotfill & 27\tablenotemark{a} & 27 \\
Rest frequency of the NH$_{3}$ lines (MHz):  &  &   \\
\hspace{3mm} (1,\,1) \dotfill & 23,694.506 & 23,694.506\\
\hspace{3mm} (2,\,2) \dotfill & 23,722.634 & 23,722.634\\
Correlator mode \dotfill & 2AD & 2AD \\
Bandwidth (kHz) \dotfill & 1562.5001 & 1550.2931 \\
Number of channels \dotfill & 127 & 127 \\
Channel separation (kHz) \dotfill & 12.207 & 12.207\\
Velocity resolution (km s$^{-1}$) \dotfill & 0.154 & 0.154 \\
FWHM of the primary beam ($'$) \dotfill & 2.1 &  2.1 \\
FWHM of the synthesized beam ($'' \times ''$), PA ($^{\circ}$) \dotfill &  & \\
\hspace{3mm} (1,\,1) \dotfill & 8.38 $\times$ 5.81, 42.3 & 7.86 $\times$ 6.54, -10.0\\
\hspace{3mm} (2,\,2) \dotfill & \nodata & 8.02 $\times$ 6.46, -4.6 \\
Flux density calibrator (Jy): & & \\
\hspace{3mm} 0542+498 / 3C147 (IF 1) \dotfill & 1.73 & 1.81\\
\hspace{3mm} 0542+498 / 3C147 (IF 2) \dotfill & 1.73 & 1.85\\
Phase calibrator (Jy):  & & \\
\hspace{3mm} 0532+075 (IF 1) \dotfill &  1.67 $\pm$ 0.06 & 1.05 $\pm$ 0.01\\ 
\hspace{3mm} 0532+075 (IF 2) \dotfill & 1.77 $\pm$ 0.07 & 1.04 $\pm$ 0.01\\ 
Bandpass calibrator (Jy): & & \\
\hspace{3mm} 0319+415 / 3C84 (IF 1) \dotfill & 11.5 $\pm$ 0.4 & 11.3 $\pm$ 0.2\\
\hspace{3mm} 0319+415 / 3C84 (IF 2) \dotfill & 12.3 $\pm$ 0.5 & 11.3 $\pm$ 0.2 
\enddata
\tablenotetext{a}{The array was in transition to the A configuration. As a result, the data from four antennas that were already moved had to be discarded.}
\tablenotetext{b}{The largest angular scale structure that can be imaged in full 12 hour synthesis observations in the D-array at 22 GHz is 66$''$. \label{t:obs}}
\end{deluxetable*}

\section{The Data}
\label{s:data}

In this section, we describe the Very Large Array (VLA) NH$_{3}$ (1,\,1) and (2,\,2) and the James Clerk Maxwell Telescope (JCMT) HCO$^{+}$ and HCN observations and data reduction. We also describe the analysis of the archival Atacama Large Millimeter/submm Array (ALMA) Band 6 data and provide information on the ancillary {\it Spitzer} Space Telescope mid-infrared and the {\it Herschel} Space Observatory far-infrared/submm data.

\subsection{VLA}

The ammonia data were obtained with the `historical' Very Large Array (VLA) of the National Radio Astronomy Observatory\footnote{The National Radio Astronomy Observatory is a facility of the National Science Foundation operated under cooperative agreement by Associated Universities, Inc.} in June 1999 (AW\,512) and August 2000 (AW\,543) in the D configuration.  The ammonia ($J$, $K$) = (1,\,1) and (2,\,2) inversion transitions with rest frequencies of 23,694.506 MHz and 23,722.634 MHz, respectively, were observed simultaneously.  The instrumental parameters, as well as flux densities of the flux, bandpass and phase calibrators, are summarized in Table~\ref{t:obs}.  The data were calibrated using the Astronomical Image Processing System (AIPS) software package.

The calibrated VLA data were further analyzed using the Common Astronomy Software Applications (CASA) package \citep{mcmullin2007}. The data were imaged and deconvolved interactively using the CASA task  \texttt{clean}. A natural weighting was used and a 20 k$\lambda$ taper was applied to the UV data. The resulting synthesized beams are listed in  Table~\ref{t:obs}. The data cubes were corrected for the primary beam attenuation using the CASA task \texttt{impbcor}.  Both NH$_{3}$ (1,\,1) and (2,\,2) line emission were detected in Epoch 2 and only the (1,\,1) line in Epoch 1.  Due to some technical difficulties and bad weather conditions during the observations, the overall quality of the Epoch 1 data is significantly lower than that of Epoch 2.  As a consequence, only the Epoch 2 data will be used in further analysis. However, within uncertainties the NH$_3$ (1,\,1) integrated flux density and (2,\,2) upper limit from Epoch 1 agree with the detections in Epoch 2. The noise levels in the Epoch 2 NH$_{3}$ (1,\,1) and (2,\,2) data cubes determined from the line-free channels are 7.9 mJy beam$^{-1}$ and 7.7 mJy beam$^{-1}$, respectively.

\subsection{JCMT}
\label{s:jcmtdata}

HCO$^{+}$ (4--3) and HCN (4--3) emission from the HH\,111/HH\,121 protostellar system was observed using the Heterodyne Array Receiver Program (HARP) and the Auto-Correlation Spectral  Imaging System  (ACSIS; \citealt{buckle2009})  at  the James Clerk Maxwell Telescope (JCMT) on Mauna Kea, Hawaii, on October 16 and 17, 2016 (Project ID: M16BP057).  The on-source integration time was 53 min. and 49 min. for the HCO$^{+}$ and HCN observations, respectively. 

HARP has 16 detectors (receptors) arranged in a 4$\times$4 configuration with an on-sky projected beam separation of 30$''$; 14 out of 16 detectors were functional during the observations.  The half power beam width of each receptor is approximately 14$''$.  The observations at 356.734 GHz (HCO$^{+}$ 4--3) and 354.505 GHz (HCN 4--3)  were carried out with the HARP/ACSIS jiggle beam-switching mode using the map-centered HARP4 jiggle pattern.   The resulting 2$'$$\times$2$'$ HCO$^{+}$ and HCN images are centered on (RA, Dec.; J2000) = (5$^{\rm h}$51$^{\rm m}$46$\rlap.^{\rm s}$325, $+$2$^{\circ}$48$'$24$\rlap.{''}$47).  At 345~GHz the main beam efficiency for HARP is 0.64.  The weather conditions were dry with $\tau_{225GHz}$$\sim$0.045 for both observing runs. ACSIS was in the single sub-band mode with a bandwidth of 250 MHz separated into 8193 30.5 kHz channels. The resulting velocity resolution is 0.026 km~s$^{-1}$.   
 
The  data  were  reduced  using  the  ORAC  Data  Reduction pipeline (\texttt{ORAC-DR}) described in  \citet{jenness2015}.

\subsection{ALMA}
\label{s:almadata}

\begin{deluxetable*}{p{3cm}cccc}
\centering
\tablecaption{Summary of the ALMA Archival Molecular Line Data \label{t:moldata}}
\tablewidth{0pt}
\tablehead{
\multicolumn{1}{c}{Molecule} &
\multicolumn{1}{c}{Transition} &
\colhead{Frequency} &
\colhead{$\Delta v$\tablenotemark{a}} &
\colhead{Synth. Beam: ($\Theta_B$, PA)} \\
\colhead{} &
\colhead{} &
\colhead{(GHz)} &
\colhead{(km s$^{-1}$)} &
\colhead{($'' \times ''$, $^{\circ}$)} 
}
\startdata
$^{13}$CO  \dotfill & (2--1) & 220.39868 & 0.2 & 0.84 $\times$ 0.67, -79.3 \\
C$^{18}$O \dotfill & (2--1)& 219.56035 & 0.2 &  0.84 $\times$ 0.72, -81.2 \\
N$_{2}$D$^{+}$  \dotfill &  (3--2)& 231.32183 & 0.7 &  0.81 $\times$ 0.67, -87.4\\
$^{13}$CS  \dotfill & (5--4) & 231.22069 & 0.7 & 0.81 $\times$ 0.67, -87.4 
\enddata
\tablenotetext{a}{The final velocity resolution.}
\end{deluxetable*}

HH\,111 was observed with ALMA in Band 6 as part of project 2012.1.00013.S, using both the 12m and 7m arrays. The {\it J}=2--1 transition of $^{12}$CO, $^{13}$CO, and C$^{18}$O were observed simultaneously with 230~GHz continuum.  The correlator settings for $^{13}$CO and C$^{18}$O used 30.518~kHz channels, and a spectral resolution (with online Hanning smoothing) of 0.083~km s$^{-1}$.  HH\,111 (corresponding to the ammonia source we refer to as NH$_{3}$--Main in this paper) was the target of these observations; however, the images are large enough to cover the newly discovered ammonia source, which we dubbed NH$_{3}$--S.  

The project was executed three times in May 2014 using the 12m array with baselines from 20 to 558~m, for a total time on source of 144 minutes.  Absolute flux calibration of the three executions was performed using Ganymede, J0510+180 (1.27~Jy at 220~GHz), and Callisto.  Bandpass and phase calibration used J0607$-$0834 (1.5~Jy) and J0532$+$0732 (630~mJy), respectively. For the 7m array, data from 18 executions between December 15 2013 and December 14 2014 were of good quality, incorporating baselines from 9 to 49~m and a total of 414 minutes on source. Absolute flux calibration used Ganymede, Callisto, Pallas, J0510$+$180, and J0423$-$013 (one execution only, 840~mJy at 220~GHz). J0750$+$1231 (1~Jy at 220~GHz) was used for bandpass calibration, and phase calibration used either J0532$+$0732 (500-1400~mJy at 220~GHz during the course of these observations) or J0607$-$0834 (1.3~Jy at 220~GHz). The complete list of ALMA Science Data Model (ASDM) UIDs used is provided in the Appendix.

The calibration in the archive was performed with several different CASA versions as the data were taken, so to ensure correct data weighting, we retrieved the raw visibility data and calibrated them using the ALMA calibration pipeline included in the CASA 4.5.3 package.  

Continuum was subtracted in the uv domain from each line spectral window, and the 7m and 12m data were simultaneously imaged and deconvolved interactively. Synthesized beams for $^{13}$CO and C$^{18}$O are 0$\rlap.{''}$84$\times$0$\rlap.{''}$67 and 0$\rlap.{''}$84$\times$0$\rlap.{''}$72, respectively.  The noise levels in the images away from line center are 4.5 and 3.5~mJy beam$^{-1}$, respectively, but $^{13}$CO shows evidence of significant resolved-out large-scale flux near the line center, and an effective image fidelity floor of $\sim$15~mJy beam$^{-1}$ at those velocities. The velocity resolution of the $^{13}$CO and C$^{18}$O data cubes is 0.2 km~s$^{-1}$.

The ALMA observations also covered other spectral lines with lower spectral resolution ($\sim$0.7 km~s$^{-1}$).  These include N$_{2}$D$^{+}$ (3-2) and $^{13}$CS (5-4) molecular lines (see Table~\ref{t:moldata}). We imaged these lines as described above; the synthesized beam for both N$_{2}$D$^{+}$ and $^{13}$CS is 0$\rlap.{''}$81$\times$0$\rlap.{''}$67. The data cubes were corrected for the primary beam attenuation using the CASA task \texttt{impbcor}.  The noise levels in the final N$_{2}$D$^{+}$ and $^{13}$CS  images are 1.8 and 1.1~mJy beam$^{-1}$, respectively. 

The ALMA C$^{18}$O data for HH\,111 were presented in \citet{lee2016}.  Here we use the C$^{18}$O data, as well as previously unpublished $^{13}$CO, $^{12}$CO, and  $^{13}$CS data, qualitatively to investigate the nature of the newly discovered source NH$_{3}-$S (see Section~\ref{s:nh3s}).  We discuss the ALMA N$_{2}$D$^{+}$ observations in more detail (see Section~\ref{s:n2dp}).

\subsection{Ancillary Archival Data: Spitzer and Herschel}

We use archival data from the \emph{Spitzer} Space Telescope and the \emph{Herschel} Space Observatory.

The \emph{Spitzer} data have been downloaded from the \emph{Spitzer} Heritage Archive (SHA) and include two epochs of observations.  The first data set was obtained in 2005  (GO 3315, PI: A. Noriega-Crespo)  and includes the observations with both the ``Infrared Array Camera''   (IRAC, \citealt{fazio2004}; 3.6, 4.5, 5.8, and 8.0 $\mu$m)  and ``Multiband Imaging Photometer for Spitzer'' (MIPS, \citealt{rieke2004};  24, 70, and 160 $\mu$m).  The spatial resolution of the IRAC observations is $\sim$2$''$, the resolutions of the MIPS  24, 70, and 160 $\mu$m data are 6$''$, 18$''$, and 40$''$, respectively. The 2012 data  (GO 80109, PI: J. Kirkpatrick) were taken during the warm \emph{Spitzer} mission and thus only 3.6 and 4.5 $\mu$m data are available. The 2005 IRAC post basic calibrated data (Post-BCD) have been presented in \citet{noriega2011}. Here we use the \emph{Spitzer} Enhanced Imaging Products (SEIP),  ``Super Mosaics'' and ``Source List'', provided in the SHA for the cryogenic mission. We use the Post-BCD Level 2 mosaics from the 2012 observations. 

The  \emph{Herschel} Photodetector  Array  Camera  and  Spectrometer (PACS; \citealt{poglitsch2010}) 70 and 160 $\mu$m and the Spectral and Photometric Imaging Receiver (SPIRE; \citealt{griffin2010}) 250, 350, and 500 $\mu$m mosaics have been downloaded from the \emph{Herschel} Science Archive (Proposal ID: OT1\_tbell\_1).  The data were obtained in 2003 and are Level 3 (SPIRE) and Level 2 (PACS) processed (the pipeline used the Standard Product Generation software v13.0.0); no further data processing has been done for our analysis.  The nominal spatial resolution of the \emph{Herschel} data ranges from $\sim$5$''$ at 70 $\mu$m to $\sim$35$''$ at 500 $\mu$m.

\section{Observational Results}
\label{s:results}

\begin{figure*}[ht!]
\centering
\includegraphics[width=0.48\textwidth]{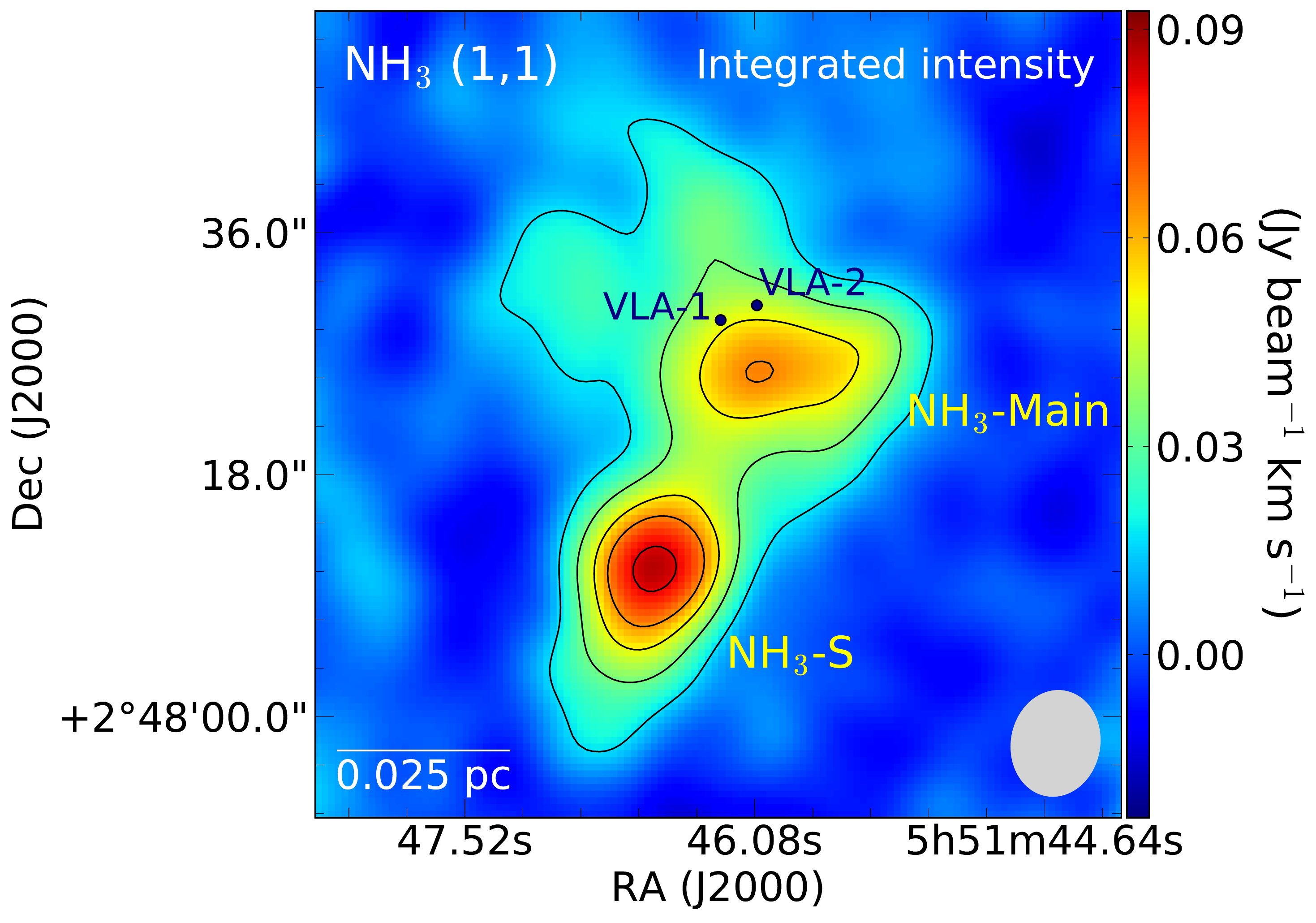}
\hfill
\includegraphics[width=0.485\textwidth]{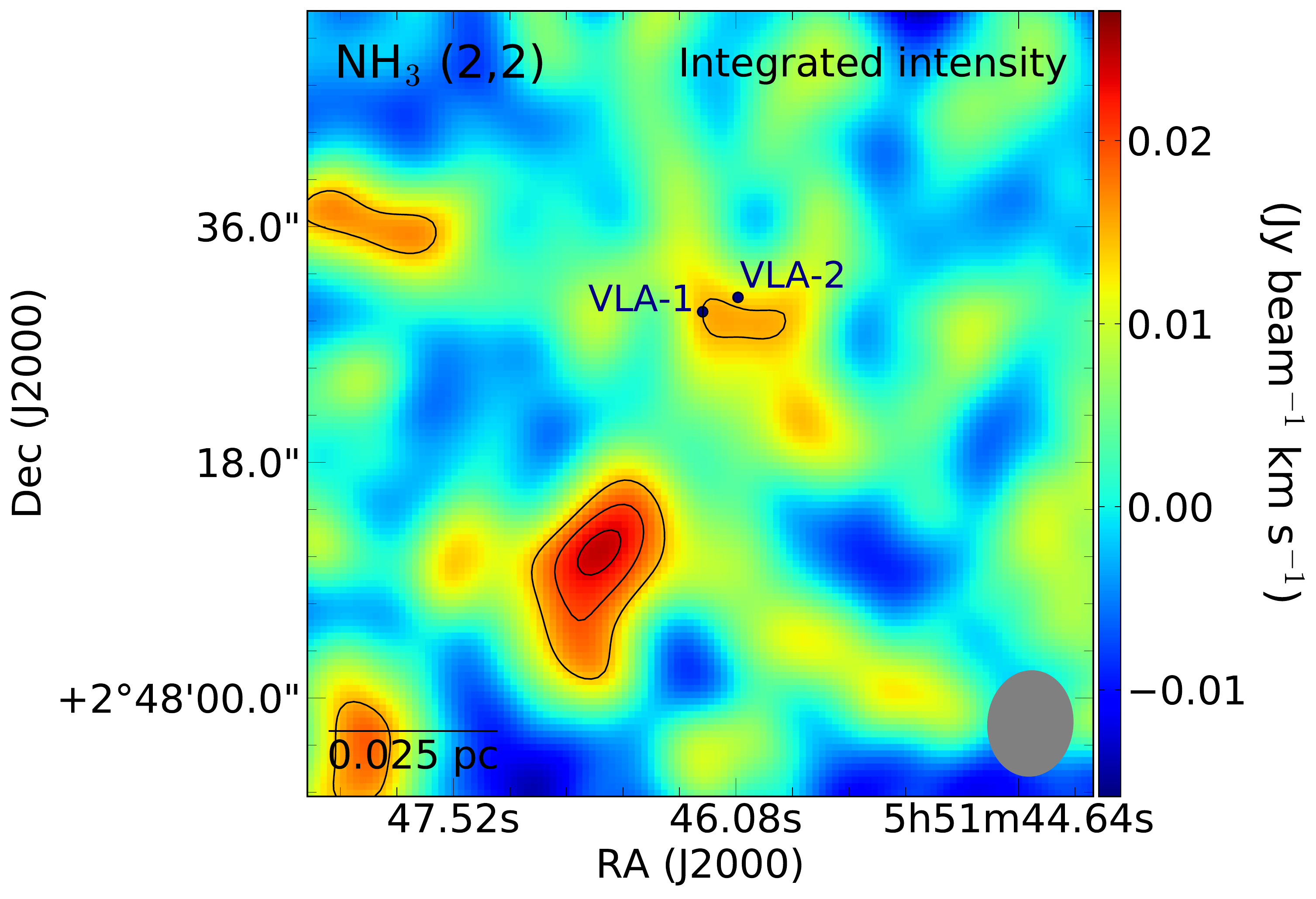}
\caption{The Epoch 2 integrated intensity images of the main NH$_{3}$  (1,\,1) and (2,\,2) line components ({\it left} and {\it right}, respectively).  Two radio continuum sources associated with HH\,111 are indicated with filled circles and labeled (see also Fig.~\ref{f:fir2mm}).  The NH$_{3}$ (1,\,1) contour levels are (20, 40, 60, 75, 95)\% of the integrated intensity peak of  86.9 mJy beam$^{-1}$ km s$^{-1}$.  The NH$_{3}$ (2,\,2) contour levels are (60, 80, 95)\% of the integrated intensity peak of  24.7 mJy beam$^{-1}$ km s$^{-1}$. The VLA synthesized beam is shown in the lower right corner in each image. \label{f:mom0}}
\end{figure*}

\begin{figure}[ht!]
\centering
\includegraphics[width=0.48\textwidth]{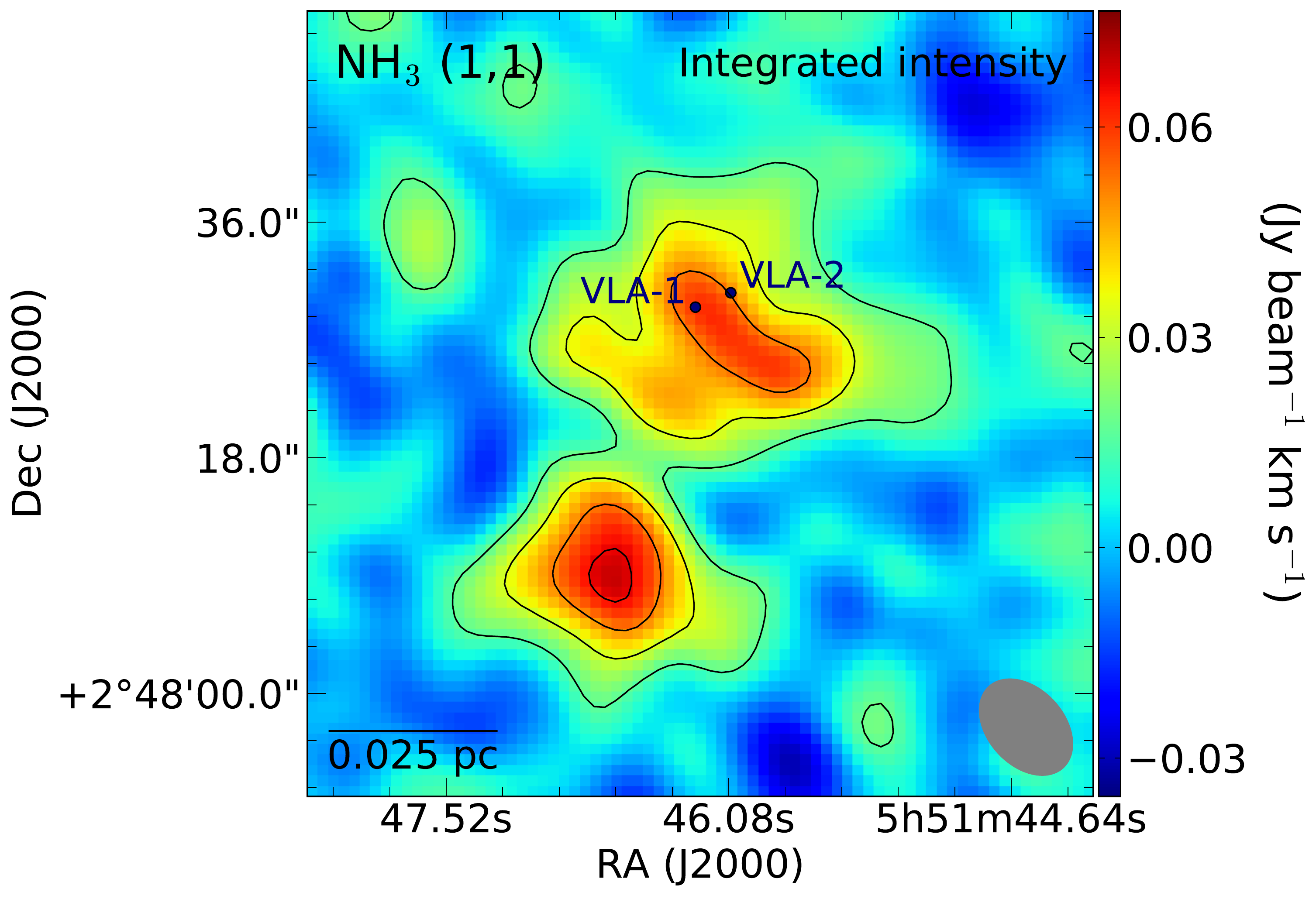} 
\caption{The integrated intensity image for the Epoch 1 observations (AW512; see Table~\ref{t:obs}). Both sources, NH$_{3}$--Main and NH$_{3}$--S, are detected. The contour levels are (25, 50, 75, 95)\% of the intensity peak of 0.07 Jy beam$^{-1}$ km s$^{-1}$. The size of the synthesized beam shown in the lower right corner of the image is 8$\rlap.{''}$38$\times$5$\rlap.{''}$81.  \label{f:epoch1}}
\end{figure}

\subsection{NH$_{3}$ Emission}
\label{s:nh3}

The NH$_{3}$ (1,\,1) and (2,\,2) images shown in Fig.~\ref{f:mom0} reveal two sources. One of the ammonia sources coincides with HH\,111, tracing an envelope of  the source associated with the jet.  The second source located $\sim$15$''$ 
($\sim$0.029 pc or 6000 AU at 400 pc) to the south-east is a new detection, even more prominent; we dubbed this source NH$_{3}-$S to distinguish it from the source associated with HH\,111 that we refer to as NH$_{3}-$Main throughout the paper.   The NH$_3$ emission in NH$_3-$Main does not peak directly on the protostar, but is offset by $\sim$5$''$ ($\sim$0.01 pc or 2000 AU) toward southwest.   This offset may indicate the drop in the NH$_3$ abundance at the location of the protostar caused by depletion or destruction of the molecules (e.g., \citealt{belloche2002};  \citealt{tobin2011}).  Both NH$_{3}-$S and NH$_{3}-$Main were also detected in the Epoch 1 data (see Fig.~\ref{f:epoch1}).

\begin{figure*}[ht!]
\centering
\includegraphics[width=0.9\textwidth]{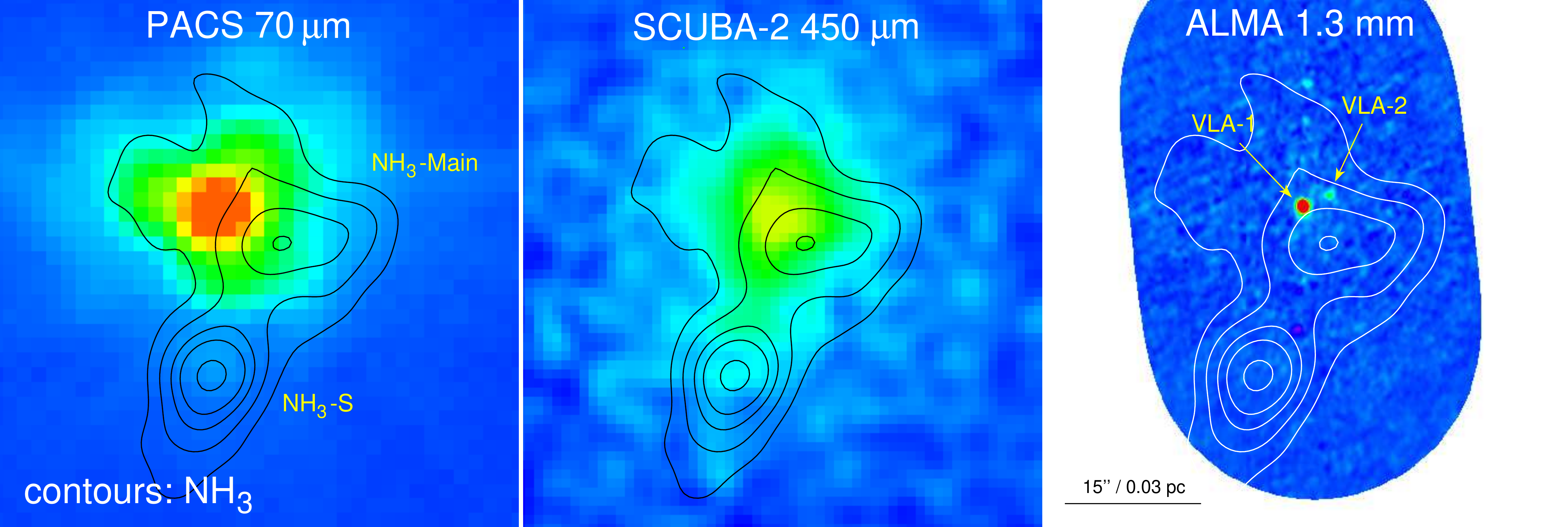}
\caption{The archival {\it Herschel} PACS 70 $\mu$m ({\it left}; HPBW$\sim$6$''$), JCMT SCUBA-2 450 $\mu$m ({\it center}; $\sim$14$''$), and ALMA 1.3 mm continuum ({\it right}; $\sim$0$\rlap.{''}$7) images of HH\,111/HH\,121.  The contours represent the NH$_{3}$ (1,\,1) emission; the contour levels as in Fig.~\ref{f:mom0}.   NH$_3-$S is not detected in these bands. Two VLA sources associated with HH\,111, VLA-1 and VLA-2, are indicated with arrows in the 1.3 mm image.   \label{f:fir2mm}}
\end{figure*}

The HH\,111/HH\,121 protostellar system and its surroundings have been thoroughly covered by observations over the broad wavelength range (from the optical to cm wavelengths), yet NH$_{3}-$S remained undetected until our NH$_{3}$ observations.  NH$_{3}-$S has not been reported as detected in any of the single dish and interferometric molecular line observations (e.g., CO, $^{13}$CO, C$^{18}$O, SO, CS;  see Section~\ref{s:intro}) or the cm- and mm-wave  continuum observations with the VLA (e.g., \citealt{reipurth1999} at 3.6 cm; a resolution of $\sim$0$\rlap.{''}$4) and SMA (e.g., \citealt{lee2010} at 1.3 mm; $\sim$1$''$). The molecular lines observed interferometrically include: 
CO {\it J}=1--0 (\citealt{lefloch2007};  $\sim$3$''$),   
CO 2--1 (\citealt{lee2011}; $\sim$0$\rlap.{''}$6),
$^{13}$CO 1--0 (\citealt{stapelfeldt1993}; $\sim$7$''$), 
$^{13}$CO 2--1 (\citealt{lee2009}; $\sim$3$''$),
C$^{18}$O 2--1 (\citealt{lee2009,lee2010,lee2011}; $\sim$0$\rlap.{''}$3--3$''$), 
SO 5$_{6}$--4$_{5}$ (\citealt{lee2009,lee2010,lee2011}; $\sim$0$\rlap.{''}$3--3$''$).
We have also checked the 2MASS ({\it JHK$_{S}$}),  {\it Spitzer} (3.6--160 $\mu$m), {\it WISE} (3.5--22 $\mu$m),  and {\it Herschel} (70--500 $\mu$m) archival data and found that no source was detected at the position of NH$_{3}-$S with these facilities.  Example images of the region showing a non-detection of NH$_{3}-$S are shown in Fig.~\ref{f:fir2mm}.

\subsubsection{Ammonia Line Profile Fitting}
\label{s:physpar}

\begin{figure*}[ht!]
\centering
\includegraphics[width=0.35\textwidth]{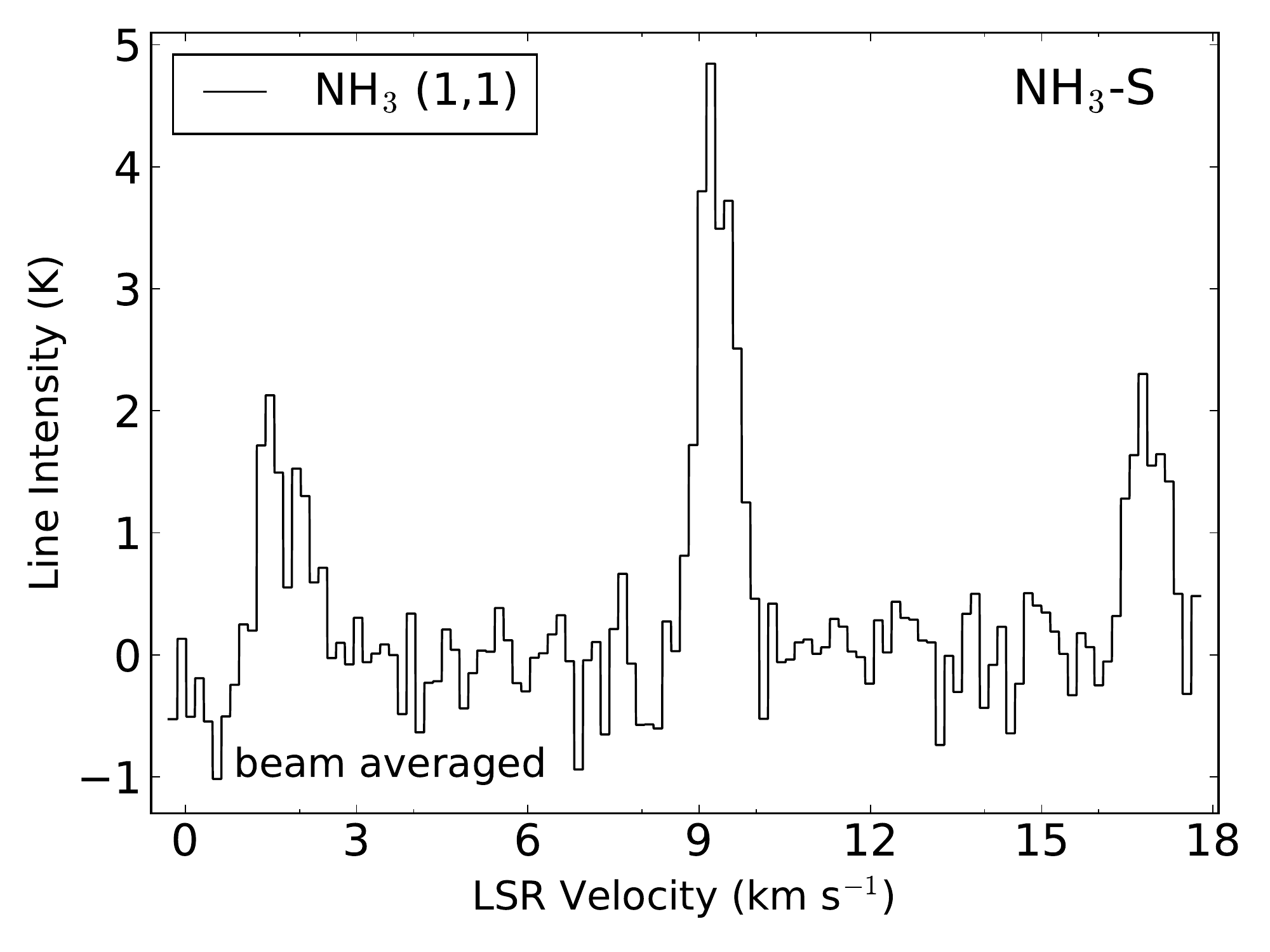}
\includegraphics[width=0.35\textwidth]{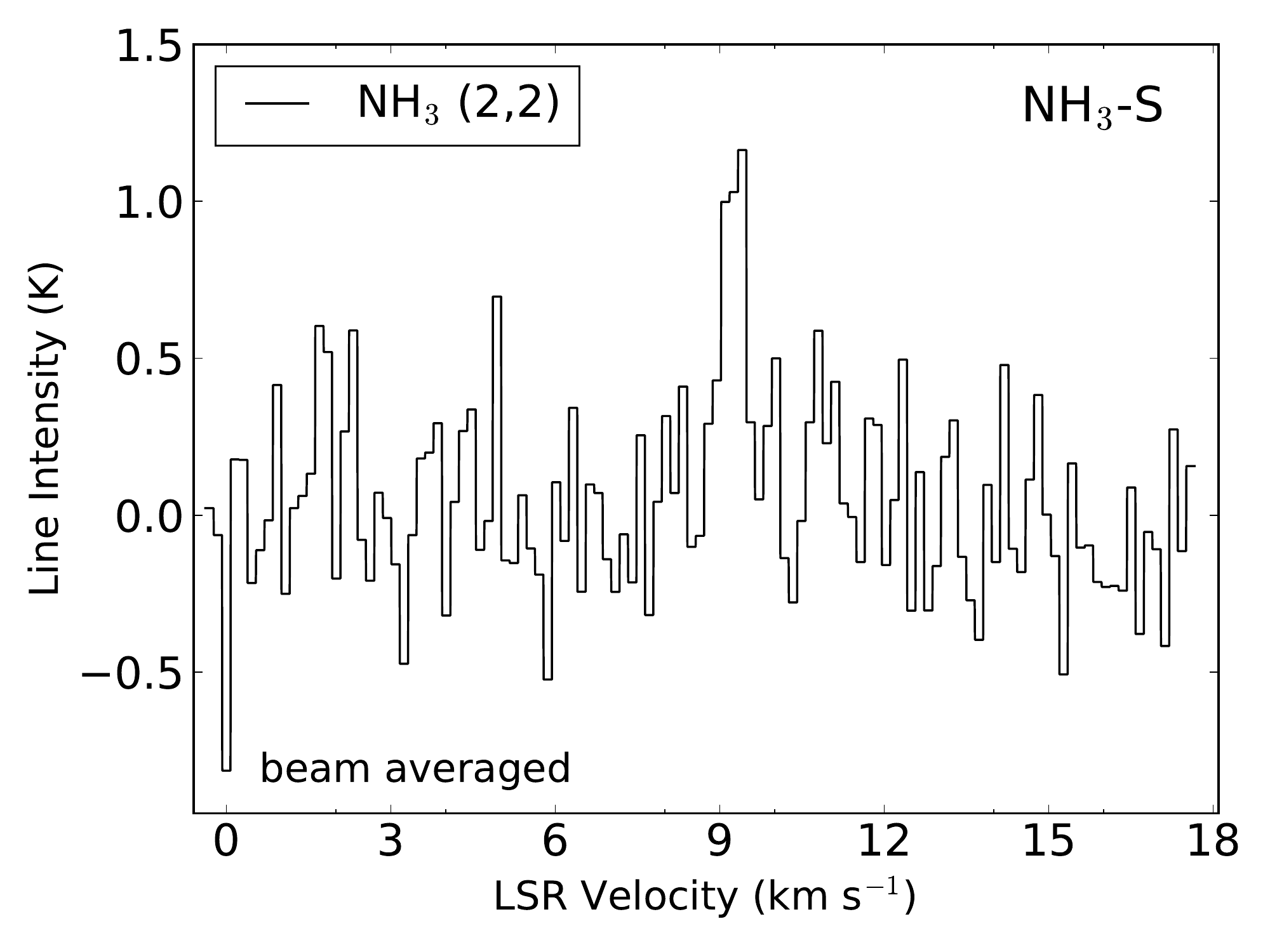} \\
\includegraphics[width=0.35\textwidth]{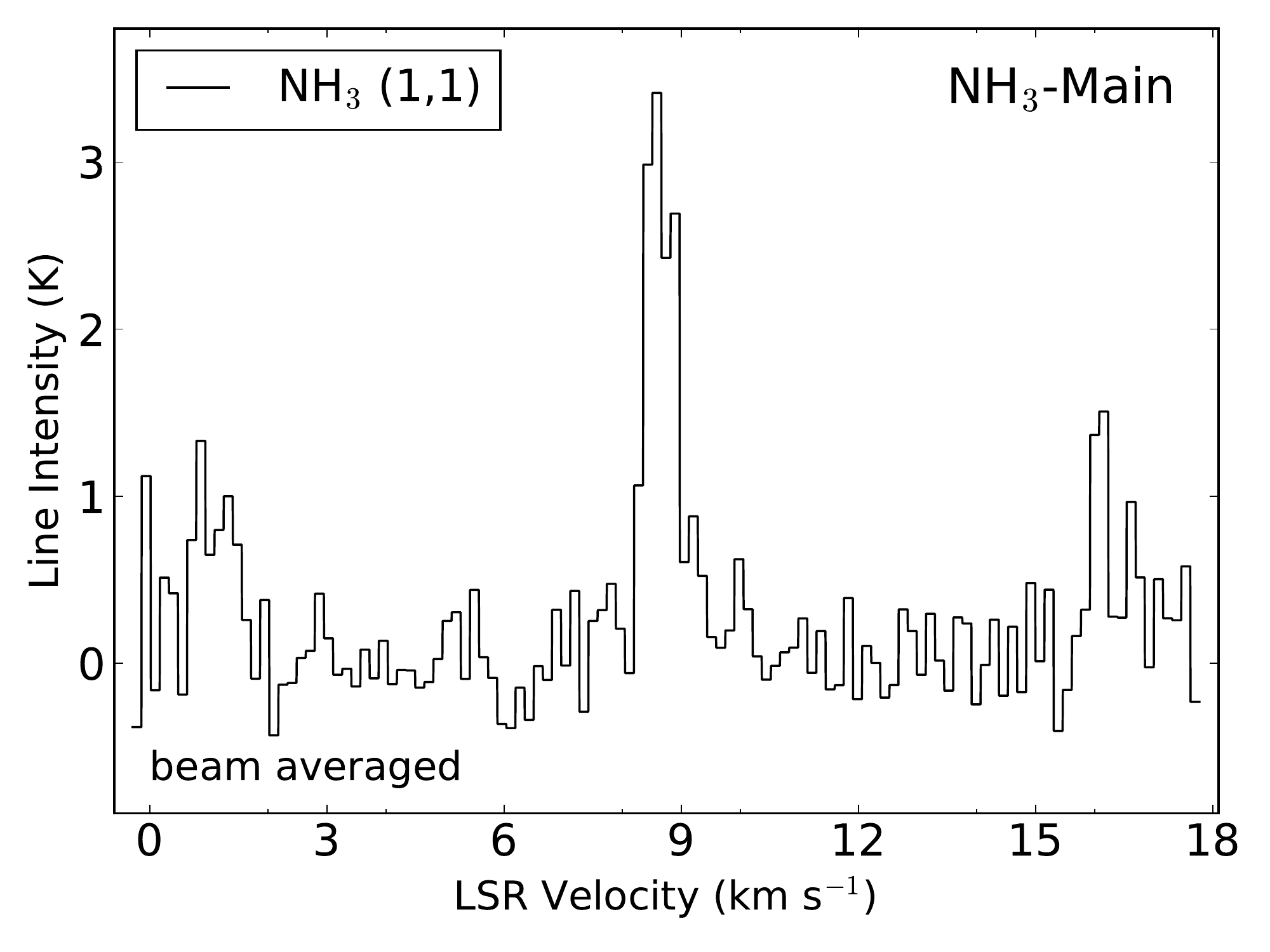}
\includegraphics[width=0.35\textwidth]{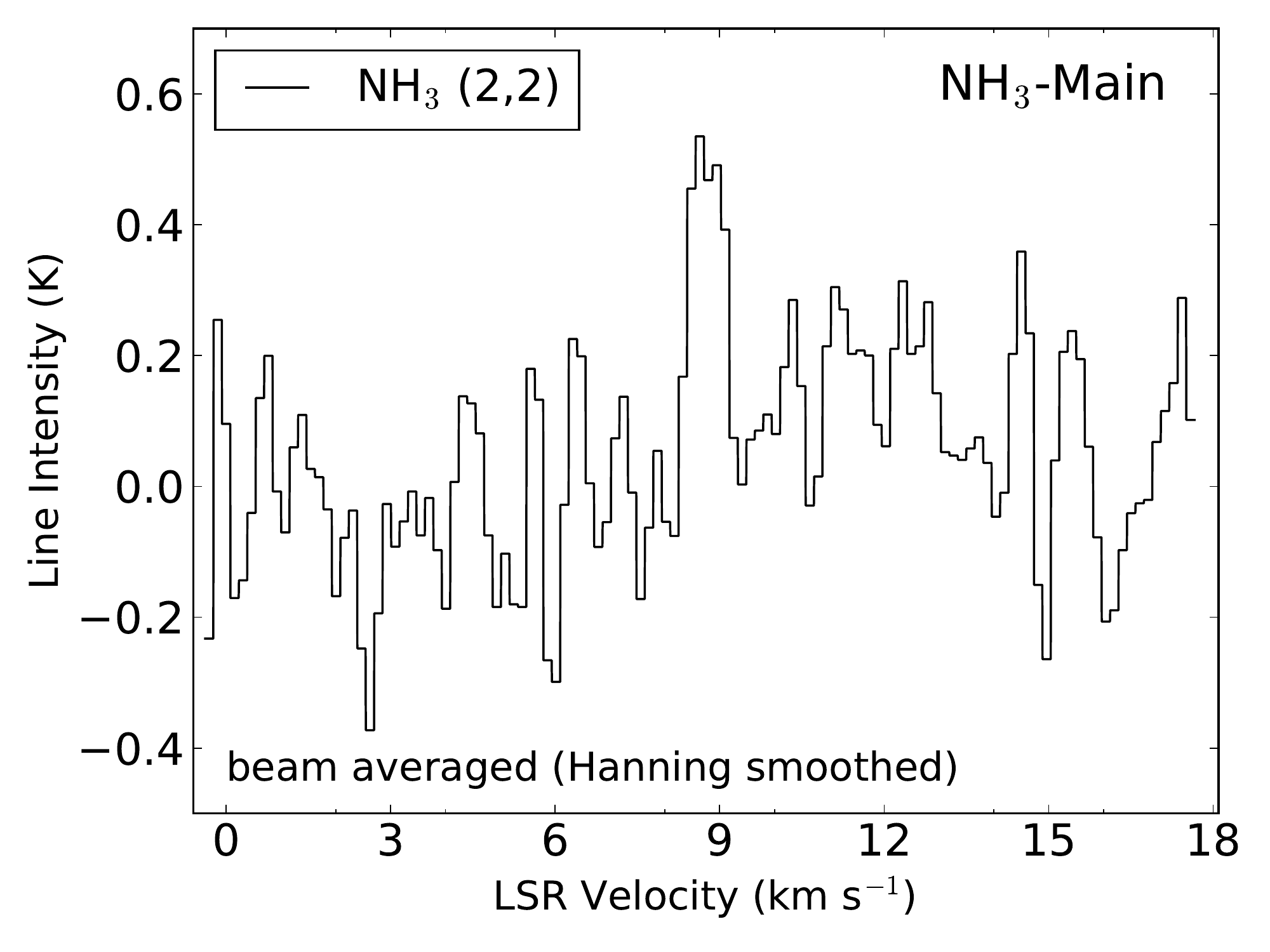}
\caption{The example NH$_{3}$ (1,\,1) and (2,\,2) spectra for NH$_{3}-$S ({\it top panel}) and NH$_{3}-$Main ({\it bottom panel}). Each spectrum was extracted as a mean over the area corresponding to one synthesized beam, centered on the NH$_{3}$ (1,\,1) emission peak. \label{f:exspec} }
\end{figure*}

Ammonia constitutes an ideal probe of physical conditions in the ambient molecular material (e.g., \citealt{ho1979}; \citealt{ho1983};  \citealt{harju1993}; \citealt{busquet2009}).  Ammonia has the distinctive spectrum with a main line and a pair of satellite lines on each side of it; in total, these lines are composed of 18 distinct hyperfine components. The optical depth of the transition can be calculated directly from the brightness temperature ratio of the satellite to main lines of the NH$_3$ spectrum, facilitating the calculation of the excitation temperature.  When two or more NH$_{3}$ transitions are observed, it is possible to calculate a rotational temperature describing the relative populations of different energy states,  and column densities. The rotational temperatures can be converted to kinetic temperatures based on the models of collisional excitation of NH$_{3}$ (e.g., \citealt{danby1988}; \citealt{tafalla2004}).

Our VLA ammonia observations cover the main line and the inner satellite components of the NH$_{3}$ spectrum.  We detected both the main line and satellite components for the (1,\,1), and the main line only for the (2,\,2) transition (see Fig.~\ref{f:exspec}). We use the NH$_{3}$ (1,\,1) and (2,\,2) lines to determine physical properties of NH$_{3}-$Main and NH$_{3}-$S. 

Here we fit simultaneously all observed hyperfine components of the NH$_3$ (1,\,1) and (2,\,2) lines using a forward model presented by Friesen \& Pineda et al. (submitted). This method describes the emission at every position with a  centroid velocity ($v_{LSR}$), velocity dispersion ($\sigma_{v}$),  kinetic temperature ($T_{kin}$), excitation temperature ($T_{ex}$), and the total NH$_3$ column density ($N(\rm{NH_3})$).  We adopt the local thermodynamic equilibrium (LTE) value of 1 for the ortho- to para-NH$_3$ ratio.  The kinetic temperature is derived from the rotational temperature assuming that mostly the (1,\,1) and (2,\,2) levels are populated. The model is implemented in the Python analysis toolkit  \texttt{pyspeckit} as `cold-ammonia' \citep{pyspeckit}.  We fit the NH$_3$ line profiles on a pixel-by-pixel basis to construct the maps of physical parameters.  The fitting was done for pixels with a signal-to-noise ratio for the NH$_3$ (1,\,1) line larger than 5.

Figures~\ref{f:mom1}  and \ref{f:mom2} show the maps of, respectively, the NH$_3$ (1,\,1) line velocity ($v_{LSR}$) and the full width at half maximum  ($\Delta v$ = $\sqrt{8\,{\rm ln}2}\,\sigma_{v}$, where $\sigma_{v}$ is the velocity dispersion, directly related to gas temperature, see Section~\ref{s:nonth}).  Figure~\ref{f:physpar} shows maps of  $T_{kin}$,  $N(\rm{NH_3})$,  and $T_{ex}$. The maps of physical parameters estimated based on both the NH$_{3}$ (1,\,1) and (2,\,2) lines,  $T_{kin}$  and $N(\rm{NH_3})$, cover the relatively small areas  where the NH$_{3}$ (2,\,2) line was detected (see Fig.~\ref{f:physpar}).  The maps of $T_{ex}$  in Fig.~\ref{f:physpar} are shown for the same area as for $T_{kin}$ and $N(\rm{NH_3})$.

\subsubsection{Velocity Structure}
\label{s:kinematics}

\begin{figure*}[ht!]
\centering
\includegraphics[width=0.33\textwidth]{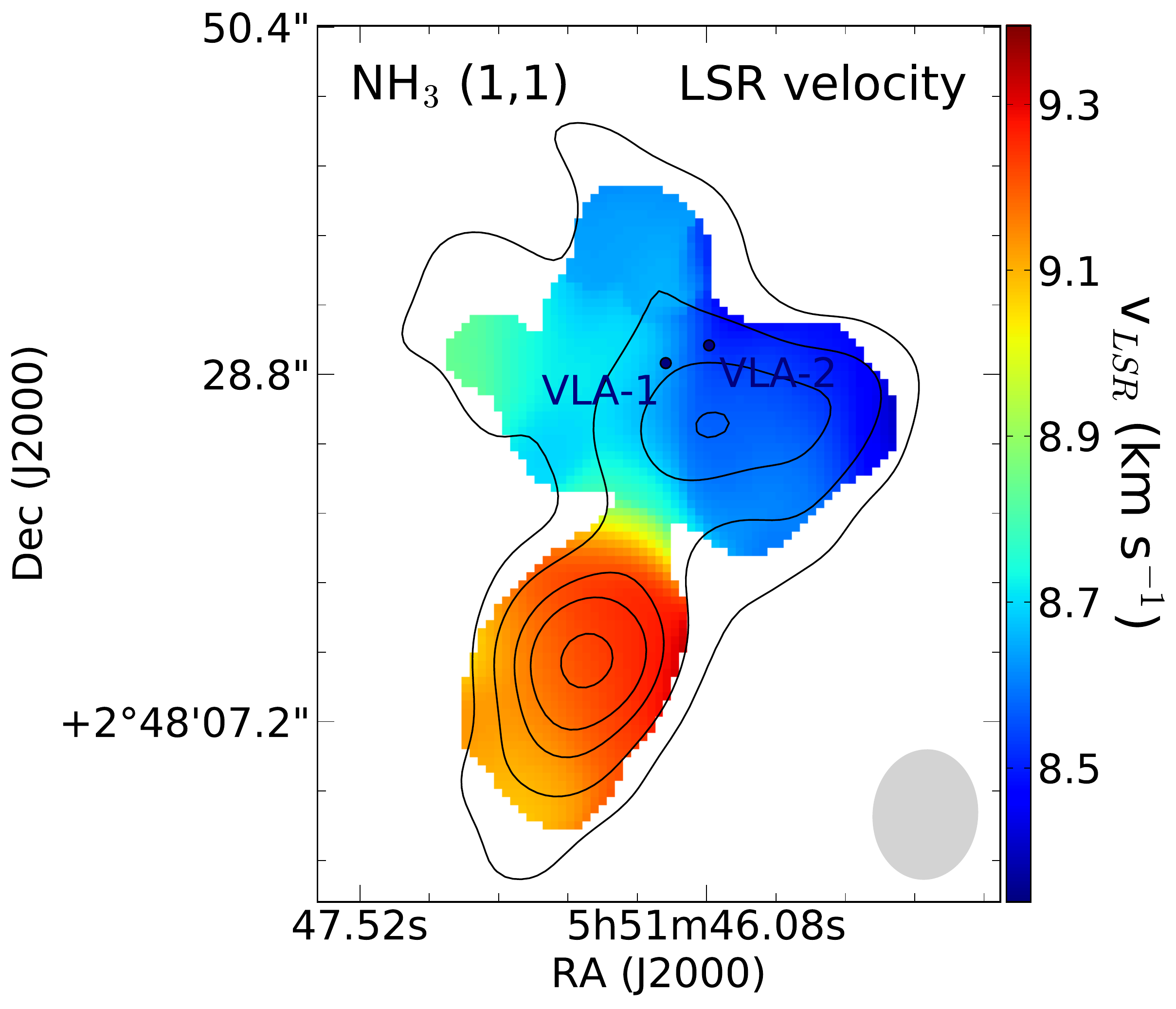}
\includegraphics[width=0.33\textwidth]{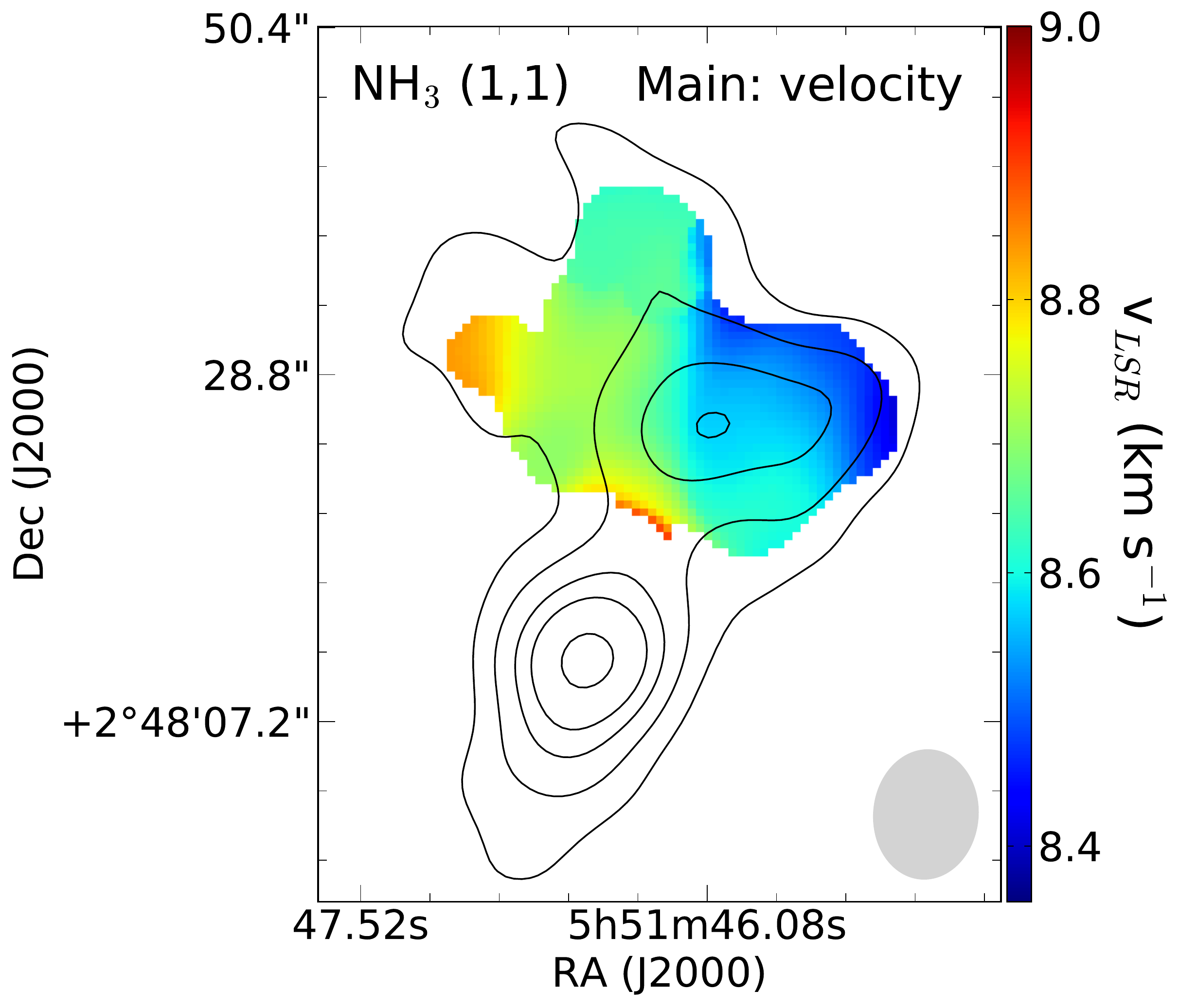}
\includegraphics[width=0.33\textwidth]{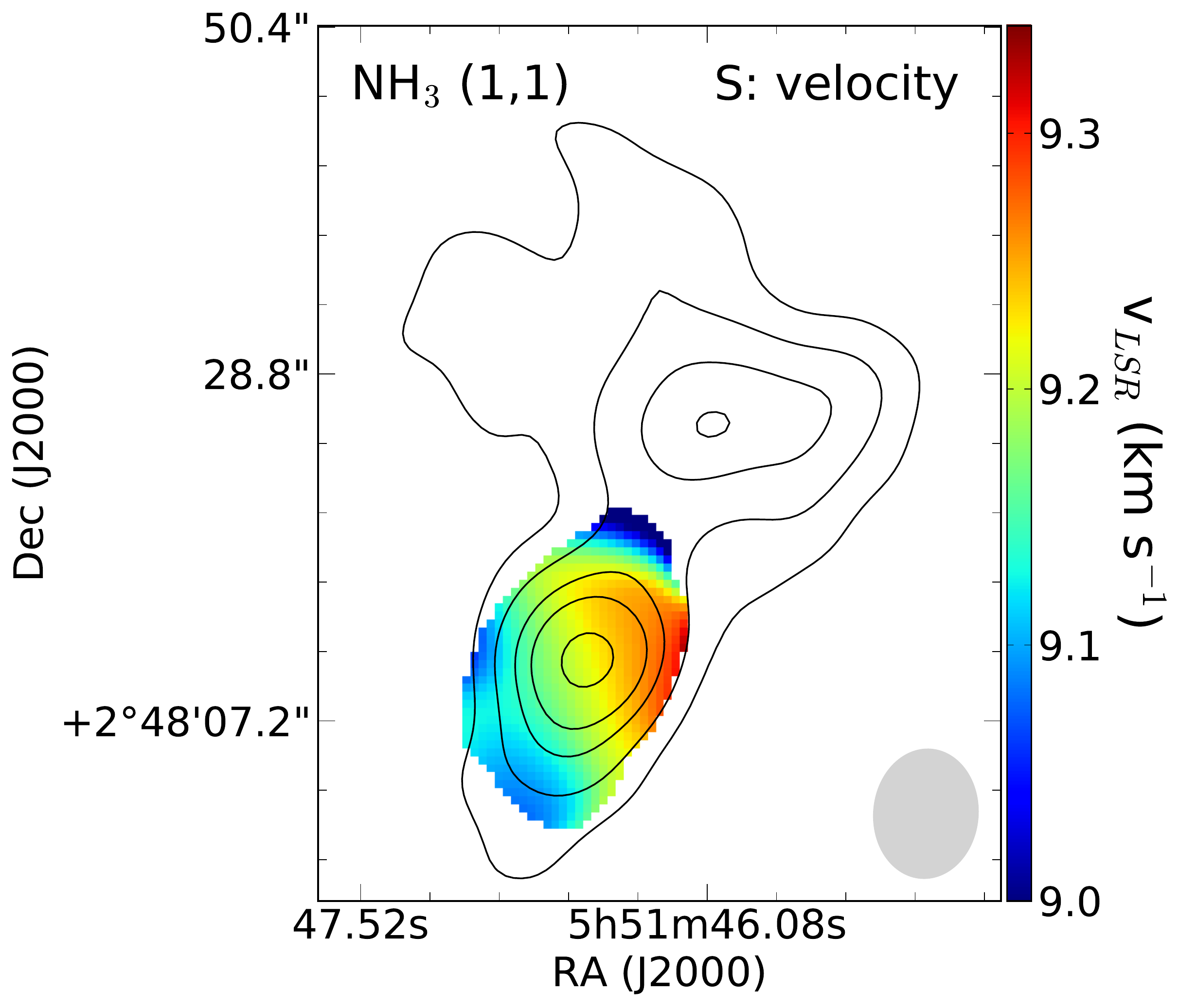}
\caption{The {\it left} panel shows  the NH$_{3}$ (1,\,1) LSR velocity map of the HH\,111/HH\,121 protostellar system, which clearly shows that NH$_{3}-$Main and NH$_{3}-$S are spatially and kinematically distinct sources.  The velocities were determined for pixels with the signal-to-noise ratio larger than 5 (see Section~\ref{s:physpar}).  The  {\it middle} and {\it right} panels show the velocity maps of individual sources: NH$_{3}-$Main and NH$_{3}-$S, respectively; these images provide a more detailed look at the velocity distribution for each source.  The {\it middle} panel shows regions with velocities lower than 8.9 km s$^{-1}$, while the {\it right} panel those with velocities larger or equal to 8.9 km s$^{-1}$. The NH$_{3}$ (1,\,1) contours as in Fig.~\ref{f:mom0}. The VLA synthesized beam is shown in the lower right corner in each image.  \label{f:mom1}}
\end{figure*}

The $v_{LSR}$ map shows that NH$_3-$S and NH$_3-$Main appear to be kinematically distinct sources with a velocity difference of about 1~km~s$^{-1}$ (Fig.~\ref{f:mom1}).   The mean $v_{LSR}$ of NH$_3-$S and NH$_3-$Main is 9.2~km~s$^{-1}$ and 8.6~km~s$^{-1}$, respectively (see Table~\ref{t:physpar}).  A clear velocity gradient roughly from south-east to north-west is detected in NH$_3-$S.  A velocity gradient of 7.3~km~s$^{-1}$~pc$^{-1}$ was measured along the line with the position angle of 104$\rlap.^{\circ}$6  and length of 0.027 pc intersecting the peak of the NH$_3$ (1,\,1) emission. 

Although velocity gradients can be identified in the $v_{LSR}$ map for NH$_3-$Main, they are less organized than in NH$_3-$S.  In general, the western part of the source has lower velocities than the eastern part. There is no evidence for a velocity gradient in the direction perpendicular to the HH\,111 jet that would trace the rotation of the protostellar envelope of  VLA-1;  such a gradient has been detected with C$^{18}$O (see \citealt{lee2010} and Fig.~\ref{f:C18Omom}).  The kinematics of this region are very complex due to the presence of two outflows, as well as the infall and rotation in the protostellar envelope. 

\begin{figure}[ht!]
\centering
\includegraphics[width=0.4\textwidth]{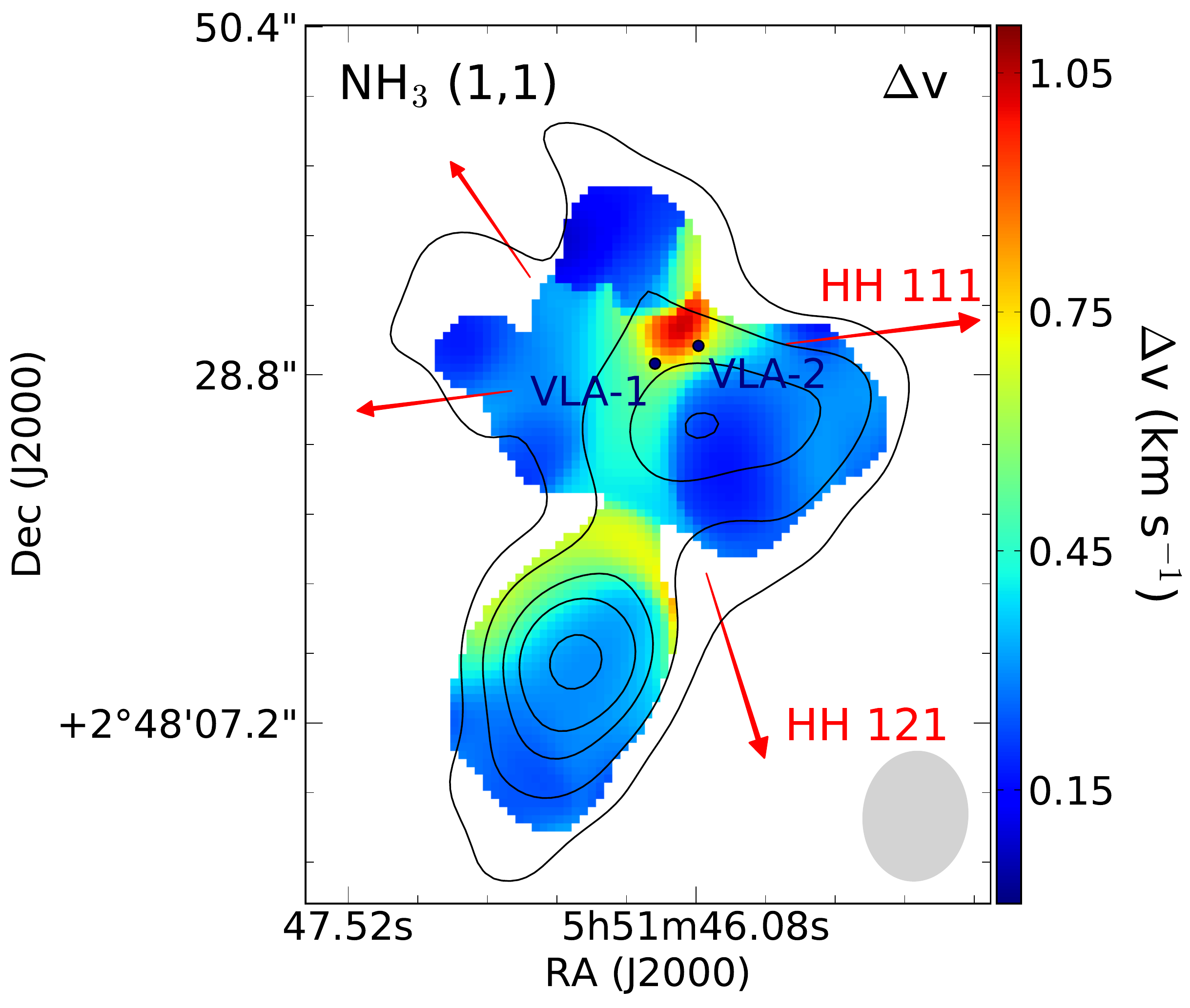}
\caption{The NH$_{3}$ (1,\,1)  line width ($\Delta$v, a full width at half maximum corrected for instrumental broadening) map of HH\,111/HH\,121.  The line widths were determined for pixels with the signal-to-noise ratio larger than 5 (see Section~\ref{s:physpar}).  The red arrows show the approximate directions of the HH\,111 and HH\,121 jets.  The NH$_{3}$ (1,\,1) contours as in Fig.~\ref{f:mom0}. The VLA synthesized beam is shown in the lower right corner. \label{f:mom2}}
\end{figure}

A sharp transition in velocity between NH$_3-$Main and NH$_3-$S (Fig.~\ref{f:mom1}) and increased linewidths at this location (Fig.~\ref{f:mom2}) are likely artifacts, the result of the hyperfine fitting of the line profiles formed by significant blending of two velocity components;  the individual lines are not clearly distinguishable.    

\citet{tobin2011} found two distinct velocity components in the N$_2$H$^{+}$ images of four protostars:  L673, HH\,211, HH\,108, and RNO\,43.  They also observed regions with artificially broadened lines where these velocity components overlap. The second velocity component in L673, HH\,211, HH\,108, and RNO\,43 is located at a distance of $\sim$0.05 pc from the protostar, comparable to the distance between VLA-1 and NH$_3-$S in HH\,111/HH\,121 ($\sim$0.04 pc).   \citet{tobin2011} suggest that the reason for two velocity components in a single region could be related to the initial conditions in the clouds.   It is in agreement with the the theory of turbulent star formation in which cloud cores are initially created and confined by the ram pressure from convergent large-scale flows (e.g, \citealt{padoan2001}; \citealt{maclow2004}; \citealt{klessen2005}; \citealt{pineda2015}).  These cores are transient, dynamically evolving density fluctuations that will either collapse and transform into stars (if they accumulate enough mass), re-expand and dissolve into the surrounding environment, or be destroyed by  shock fronts \citep{klessen2005}.  The velocity difference of $\lesssim$1~km~s$^{-1}$ between NH$_3-$Main and NH$_3-$S can be explained by the turbulent star formation model.

\subsubsection{Physical Parameters}
\label{s:physpar}

\begin{figure*}
\includegraphics[width=0.33\textwidth]{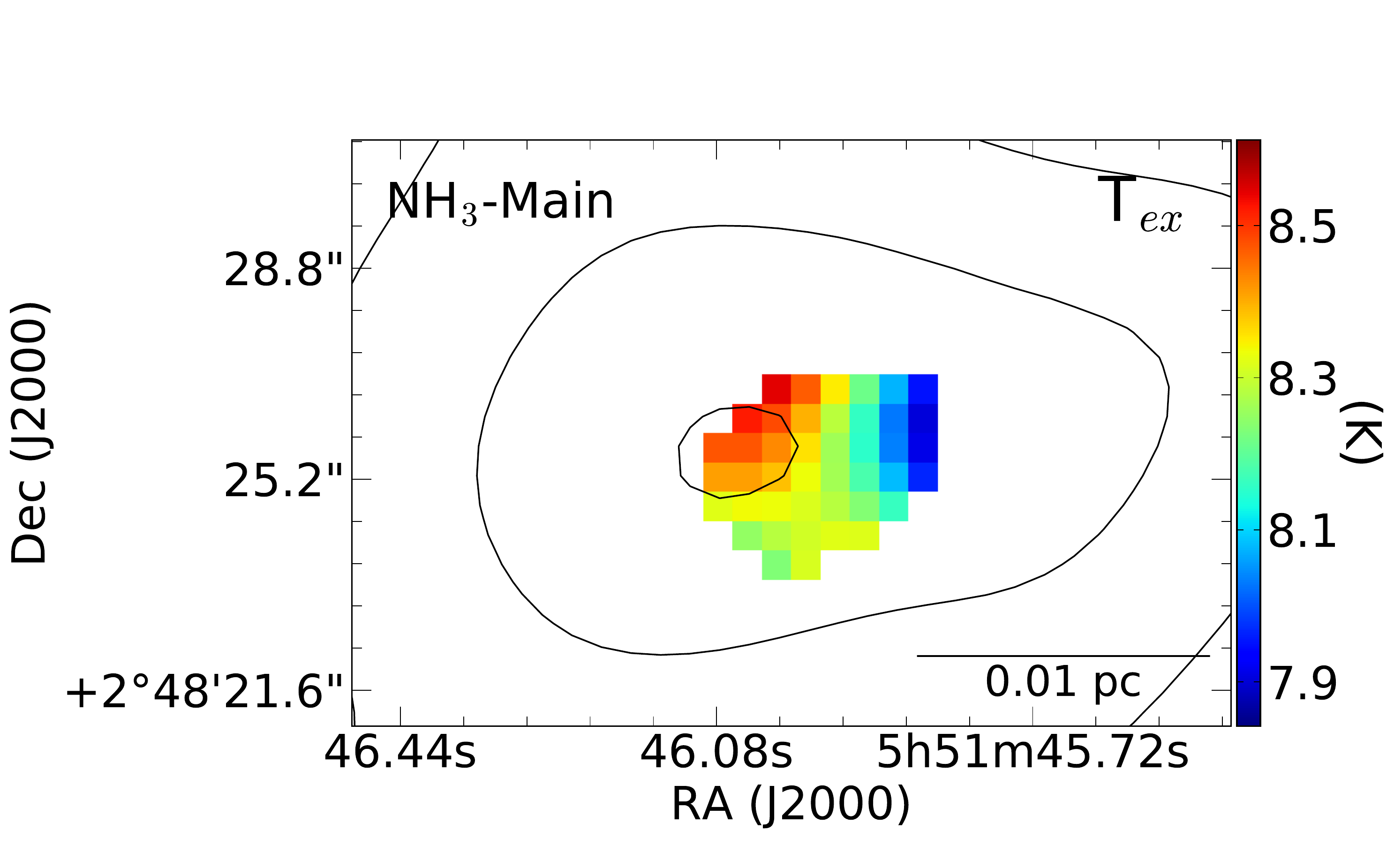}
\includegraphics[width=0.33\textwidth]{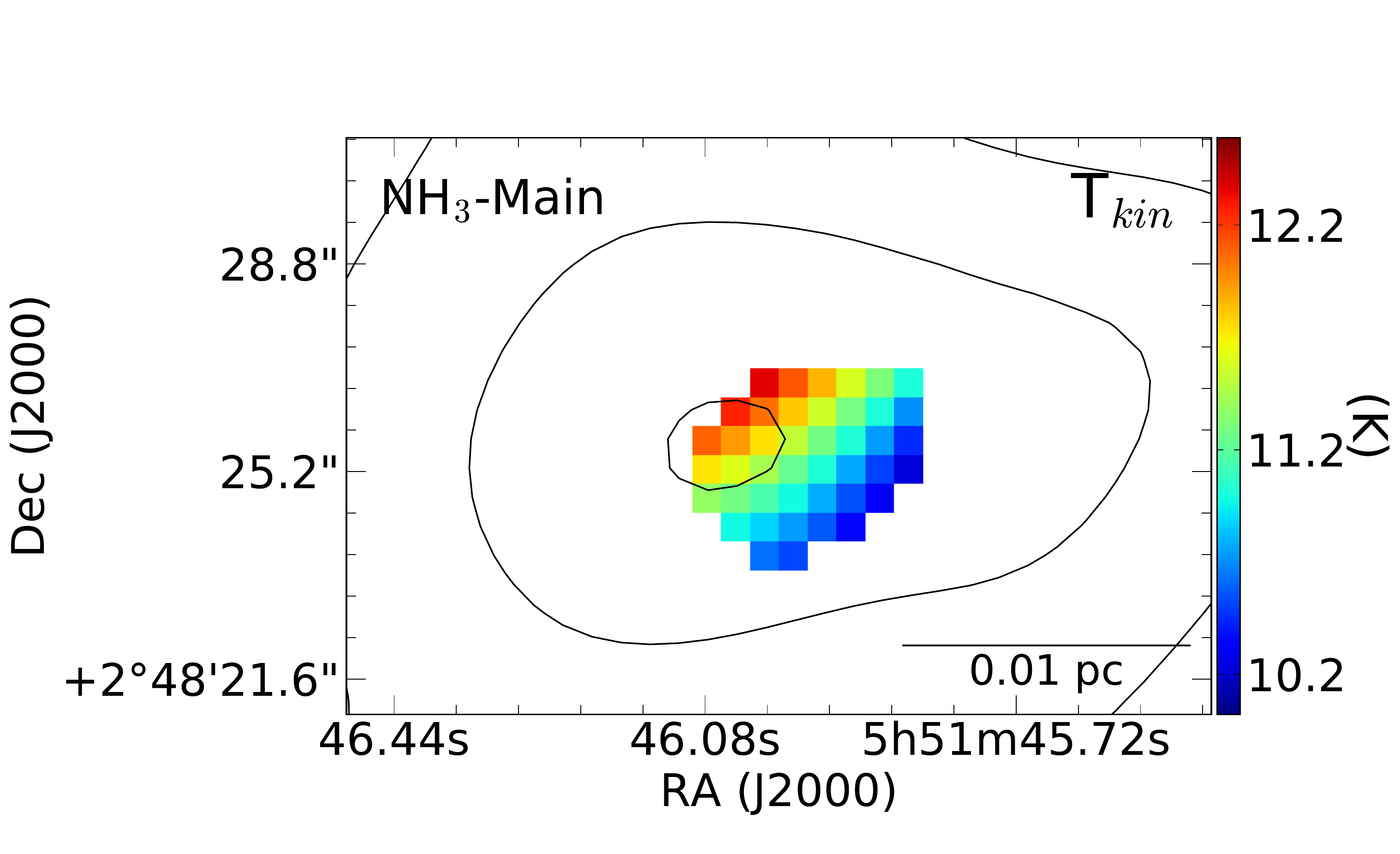}
\includegraphics[width=0.33\textwidth]{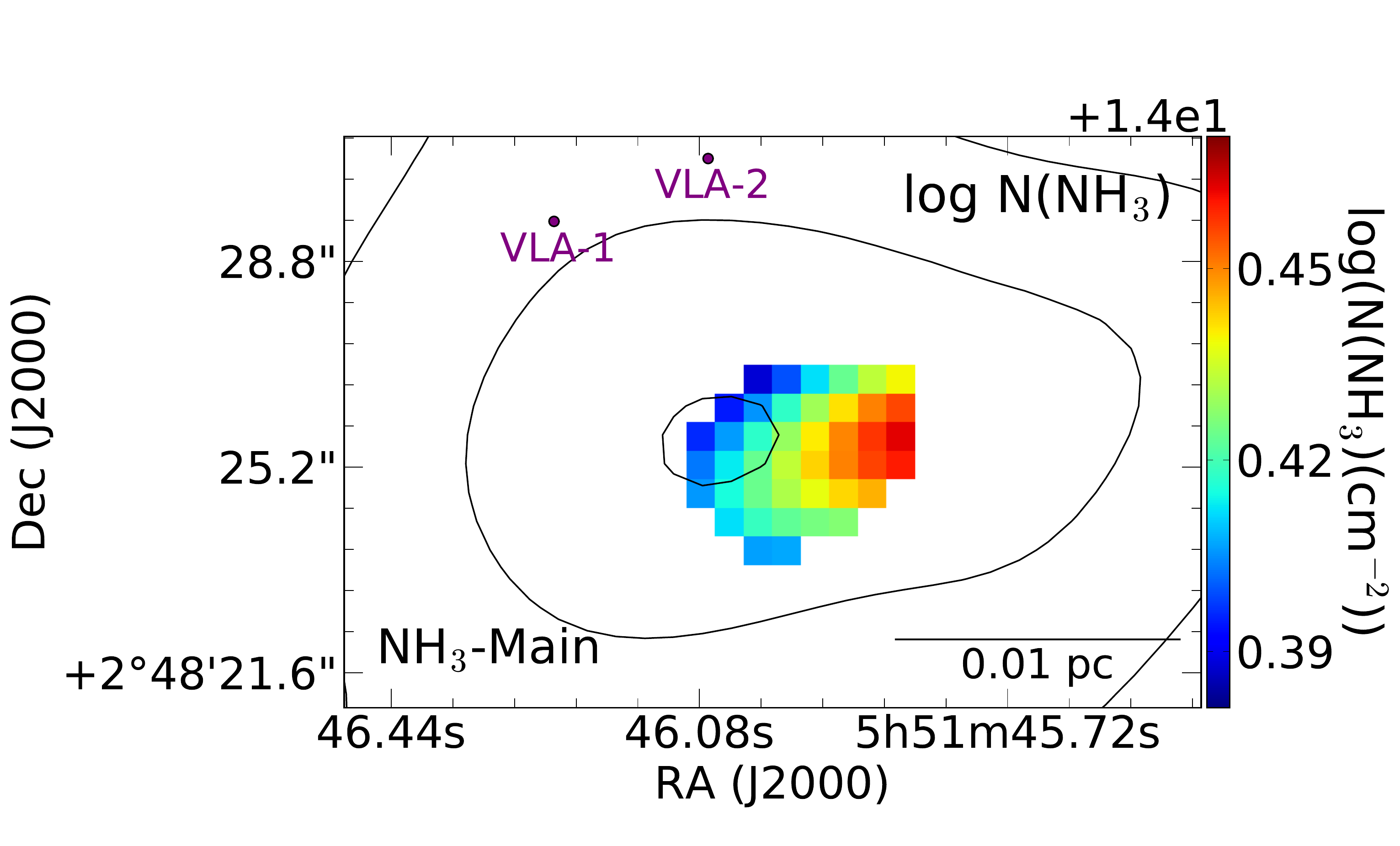}
\includegraphics[width=0.33\textwidth]{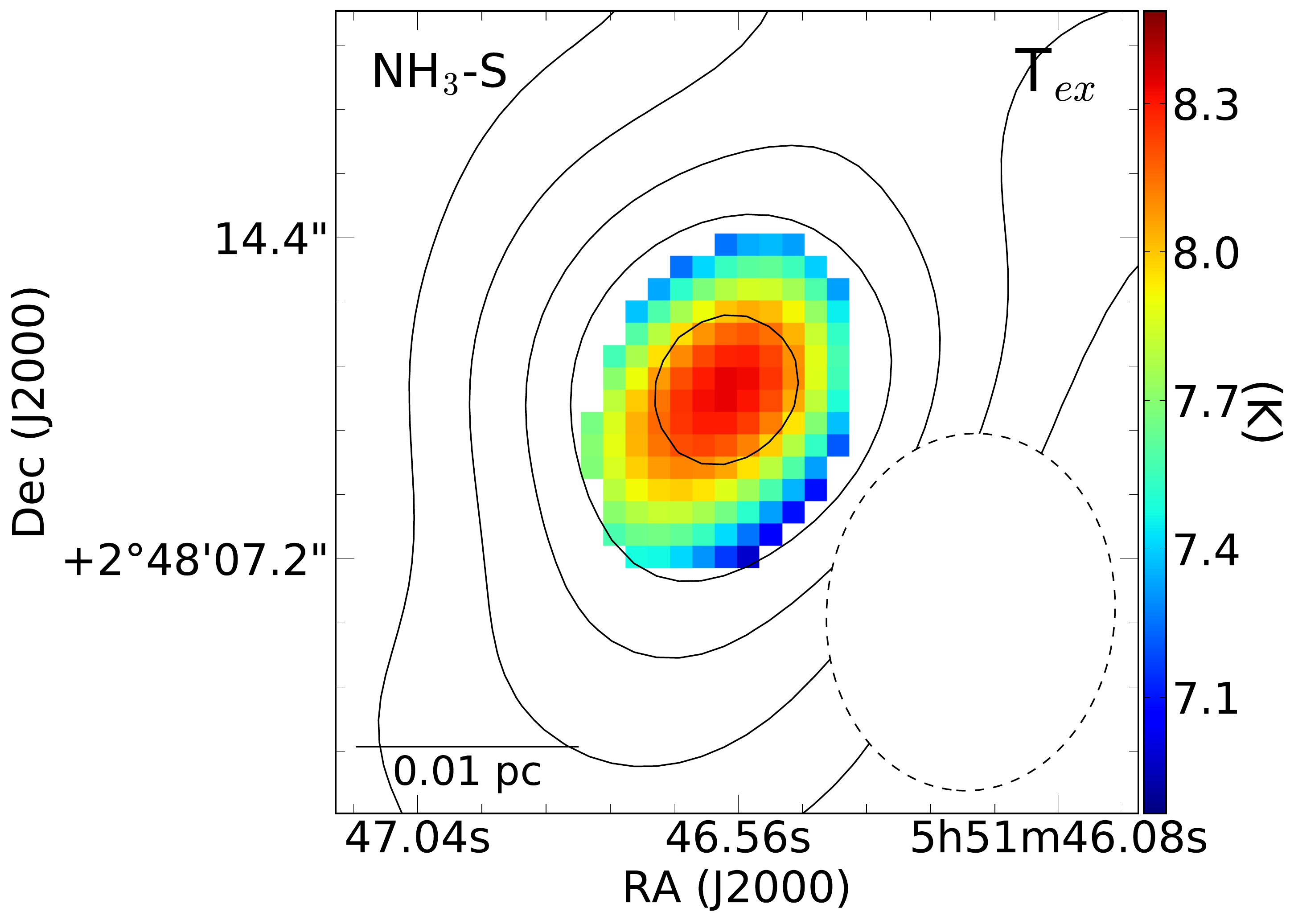}
\includegraphics[width=0.33\textwidth]{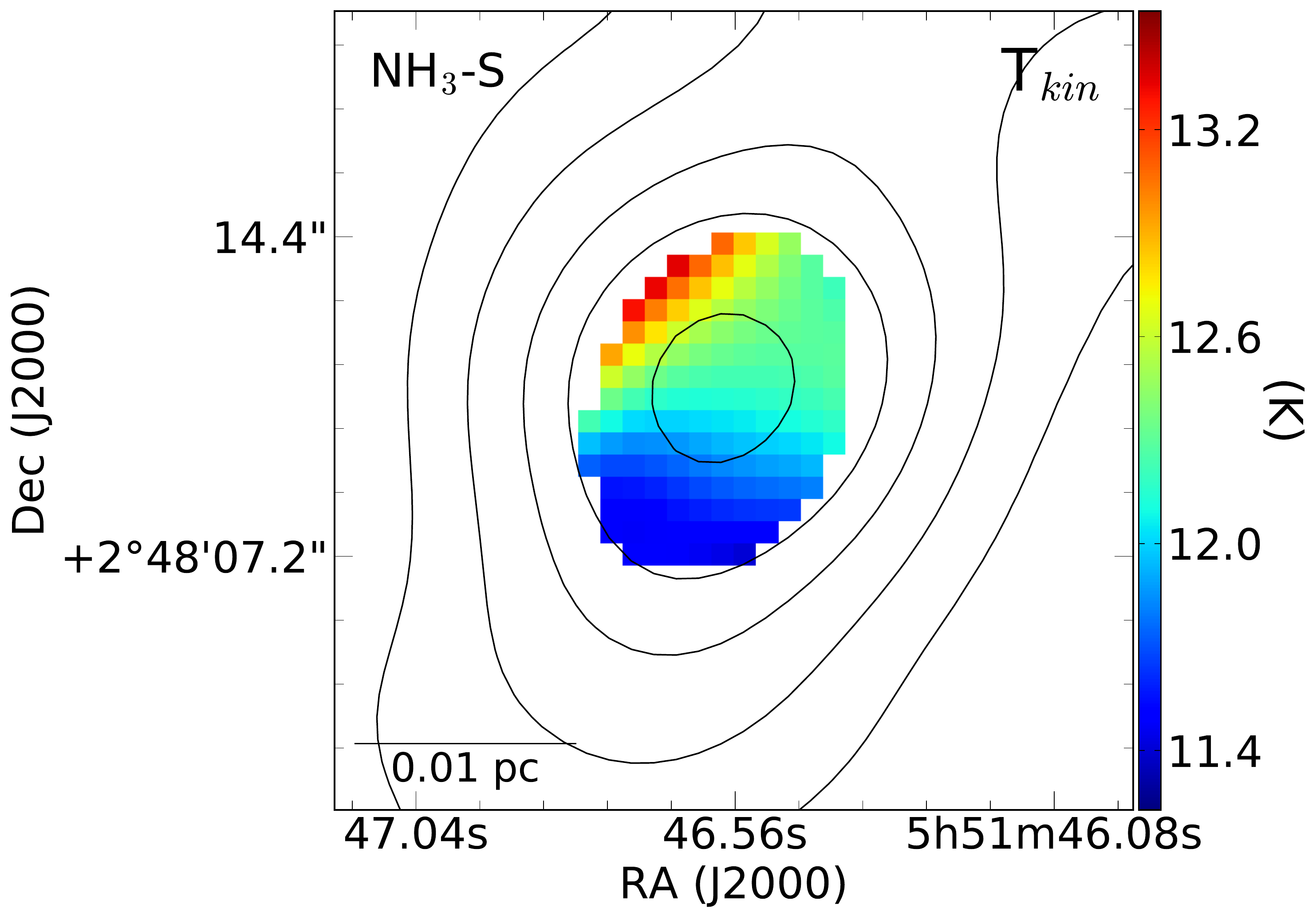}
\includegraphics[width=0.33\textwidth]{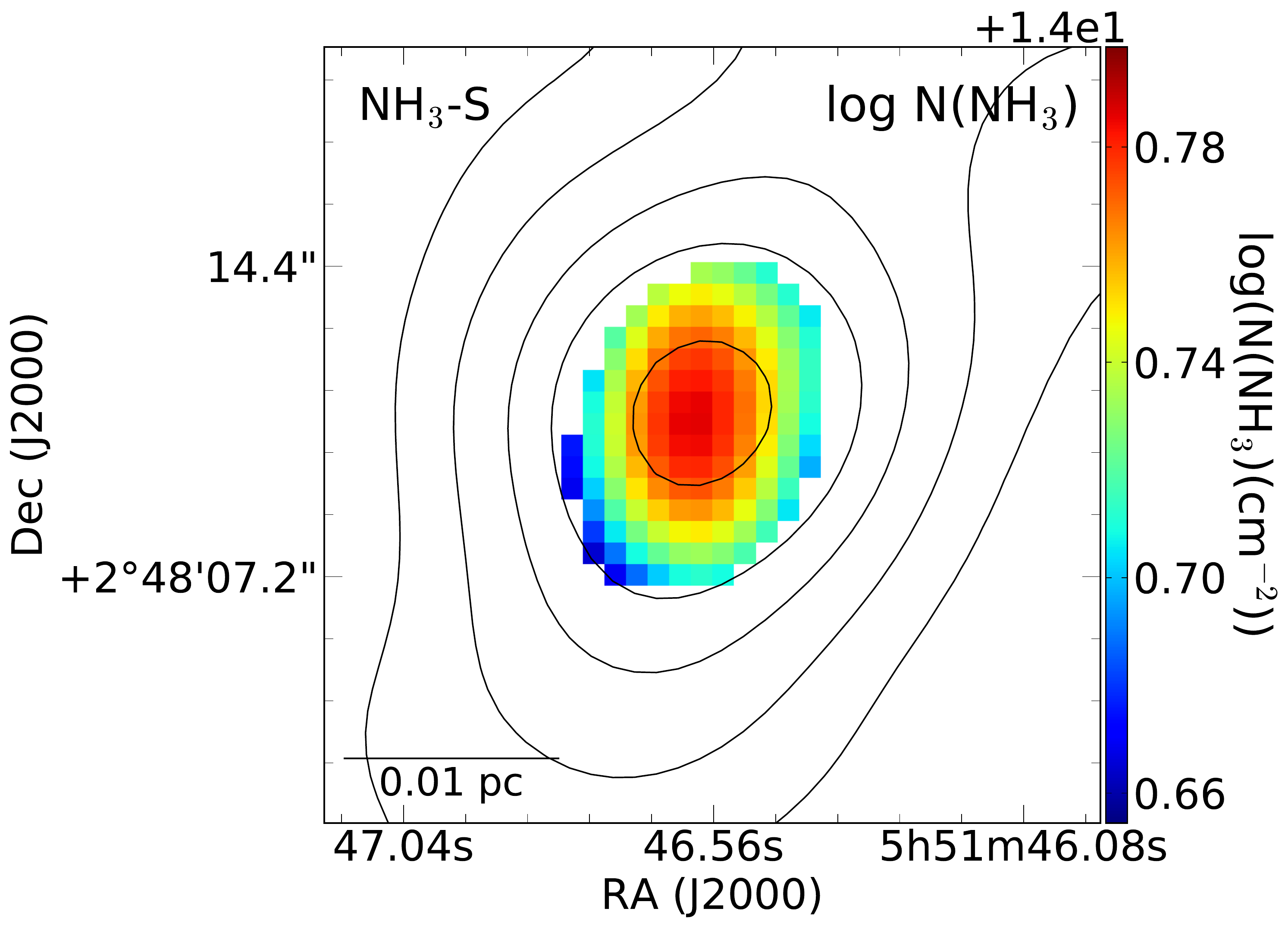}
\caption{The maps of the excitation temperature ($T_{ex}$; {\it left}), the kinetic temperature ($T_{kin}$; {\it center}), and the NH$_3$ column density ($N(NH_3)$; {\it right}) for NH$_3-$Main ({\it top panel}) and NH$_3-$S ({\it bottom panel}).  The VLA synthesized beam is shown in the lower right corner of the leftmost image in the bottom panel.  The color scales are not the same for NH$_3-$Main and NH$_3-$S.  The NH$_3$ (1,\,1) contour levels as in Fig.~\ref{f:mom0}. \label{f:physpar}}
\end{figure*}

The physical parameters averaged over the areas shown in Fig.~\ref{f:physpar} together with standard deviations are listed in Table~\ref{t:physpar} for NH$_3-$Main and NH$_3-$S. 

For NH$_3-$S,  physical parameters are determined for the central part of the core, roughly corresponding to the area of the synthesized beam (see Fig.~\ref{f:physpar}).  The distribution of both $T_{ex}$ and $N({\rm NH_3})$ is centrally peaked and well-correlated with the peak of the NH$_3$ emission.  The distribution of $T_{kin}$, however, shows a gradient across the core from $\sim$11.4 K in the south to $\sim$13.4 K in the north-east. 

Due to the much fainter NH$_3$ (2,\,2) line emission from NH$_3-$Main, it was possible to determine $T_{kin}$ and $N({\rm NH_3})$ for only a small fraction of the source area, corresponding to $\sim$20\% of the beam.  The  $T_{kin}$, $N({\rm NH_3})$, and $T_{ex}$  maps show gradients of these quantities over this small area (see Fig.~\ref{f:physpar}), with the maximum $T_{kin}$ and $T_{ex}$ associated with the peak of the NH$_3$ (1,\,1) emission.  As a consequence of the small source coverage, reliable trends in these physical parameters for NH$_3-$Main cannot be determined. 

For both NH$_3-$Main and NH$_3-$S, $T_{ex}$ is lower than $T_{kin}$, indicating that the ammonia inversion lines are mostly sub-thermally excited; the density is too low for the level populations to go to LTE (e.g., \citealt{evans1989}; \citealt{shirley2015}). This is consistent with the results found by \citet{friesen2017}.

The CASA task \texttt{imfit} was used to fit a two-dimensional Gaussian component to the NH$_3$ (1,\,1) integrated intensity emission from NH$_3-$S  to estimate the source's angular diameter. The deconvolved major and minor axes $FWHM$s are 11$\rlap.{''}$6 and 5$\rlap.{''}$9 (0.023 $\times$ 0.011 pc), respectively, with an estimated uncertainty of $\sim$20\%. We did not obtain a satisfactory fit for NH$_3-$Main, possibly due to its more complex geometry. 

To estimate the sizes of both NH$_3-$S and NH$_3-$Main, we drew a polygon around the contour at the half-maximum level for each source and derived the area within the contour ($A$), which we used to estimate the ``effective'' angular diameter of the source (or $FWHM_{eff}$) using the equation $FWHM_{eff} = 2 \sqrt{A/\pi}$ (see e.g., \citealt{sanchez2013}; \citealt{kauffmann2013}). Assuming the sources are Gaussian, we calculated the deconvolved sizes $\theta$ from $\theta = \sqrt{FWHM_{eff}^2 - HPBW^2}$, where $HPBW$ (the half-power beam width) is the geometric mean of the minor and major axes of the synthesized beam (see Table~\ref{t:data}).  The estimated sizes of NH$_3-$S and NH$_3-$Main are 9$\rlap.{''}$5 and 14$\rlap.{''}$4 or 0.018 pc and 0.028 pc, respectively.

The molecular mass of the core can be calculated from the equation: 
\begin{equation}
\label{e:mass}
M_{N({\rm NH_3})} = \frac{N({\rm NH_3})}{X}  \mu m_H A,
\end{equation}
where $N({\rm NH_3})$ is the total ammonia column density, $X$ is the [NH$_3$/H$_2$] abundance ratio, $\mu$ is a mean molecular weight per hydrogen molecule ($\mu$=2.8),  $m_{\rm{H}}$ is the mass of the hydrogen atom, and $A$ is the area of the source.   We adopt $X$ = 10$^{-8}$; we note, however, that the observed fractional abundance values range from a few times 10$^{-9}$ to a few times 10$^{-8}$  for dense, cold regions (e.g., \citealt{harju1993}; \citealt{larsson2003};  \citealt{foster2009};  \citealt{friesen2009}).  Using the \texttt{imfit} results to determine the source area, we estimate the molecular mass of the NH$_3-$S core of 0.25 M$_{\odot}$.  If we use the areas within the contours at the half-maximum level (see above), we obtain the NH$_3-$S and NH$_3-$Main  masses of 0.33 M$_{\odot}$ and 0.37 M$_{\odot}$, respectively. 

\begin{deluxetable*}{p{2.5cm}cccccccc}
\centering
\tablecaption{The Peak Intensity Positions, Line-Center Velocities, and Average Physical Parameters of the Ammonia Sources \label{t:physpar}}
\tablewidth{0pt}
\tablehead{
\colhead{Source} &
\colhead{R.A. (J2000)} &
\colhead{Decl. (J2000)} &
\colhead{$v_{{\rm LSR}}$} &
\colhead{$Size$} &
\colhead{$T_{kin}$} &
\colhead{$T_{ex}$} &
\colhead{$N(\rm{NH_3})$} & 
\colhead{$M_{core}$\tablenotemark{a}} \\
\colhead{} &
\colhead{($^{\rm h}~^{\rm m}~^{\rm s}$)} &
\colhead{($^{\circ}~'~''$)} &
\colhead{(km s$^{-1}$)} &
\colhead{(pc)} &
\colhead{(K)} &
\colhead{(K)} &
\colhead{(10$^{14}$ cm$^{-2}$)} &
\colhead{(M$_{\odot}$)}
}
\startdata
NH$_3-$Main \dotfill &  5:51:46.045 & $+$2:48:25.72 & 8.6 (0.1) & 0.028 & 11.2 (0.6) & 8.3 (0.2) & 2.7 (0.1) & 0.37 \\
NH$_3-$S   \dotfill    & 5:51:46.579 & $+$2:48:11.23 & 9.2 (0.1)  & 0.018 & 12.1 (0.5) & 7.8 (0.3) & 5.5 (0.4) & 0.33
\enddata
\tablenotetext{a}{Masses of the ammonia cores estimated using Eq.~\ref{e:mass} with $A$ being equal to the areas within the NH$_3$ (1,\,1) contour at the half-maximum level for a corresponding source.  The linear sizes listed in the `$Size$' column were estimated from `$A$' (see text for details).}
\end{deluxetable*}

Masses can also be estimated based on the ALMA 1.3 mm continuum data.  In high density regions it is expected that $T_{kin}$ is approximately equal to the dust temperature ($T_{d}$) due to the good coupling between the gas and dust. Using the dust temperature we can estimate the upper limit for the cloud mass for NH$_3-$S from the ALMA 1.3 mm continuum data. Assuming optically thin dust continuum emission, the dust mass can be estimated from the equation:   
\begin{equation}
M_{dust} = \frac{S_{\nu}\,D^2}{\kappa_\nu\,B_{\nu}(T_d)},
\end{equation}
where  $S_{\nu}$ is the integrated flux density, $D$ is the distance to the source, $B_{\nu}(T_d)$ is the Planck function, and $\kappa_{\nu}$ is the dust opacity per unit mass (e.g., \citealt{hildebrand1983}; \citealt{shirley2000}).  The clump mass ($M_{clump}$) can de derived by multiplying the dust mass by the gas-to-dust ratio $R_{gd}$:  $M_{clump} = M_{dust}\,R_{gd}$.   We used the \citet{ossenkopf1994} MRN distribution \citep{mathis1977} with thin ice mantles after 10$^{5}$ years of coagulation at a gas density of 10$^{6}$ cm$^{-3}$ model for dust opacity.  For 1.3 mm, $\kappa_{1.3{\rm mm}}$ equals to 0.899 cm$^2$ g$^{-1}$ (or 0.009 cm$^2$ g$^{-1}$ for $R_{gd}$=100) for protostellar cores.  Assuming $R_{gd}$ of 100, the clump mass can be expressed by the formula: 
\begin{eqnarray}
M_{clump} [M_{\odot}] = 0.12\, (e^{14.39(\lambda/{\rm mm})^{-1}(T/{\rm K})^{-1}} - 1) \\
\times \left(\frac{\kappa_{\nu}}{0.01\,{\rm cm^2\,g^{-1}}}\right)^{-1}\,\left(\frac{S_\nu}{{\rm Jy}}\right)\,\left(\frac{D}{{\rm 100\,pc}}\right)^2\,\left(\frac{\lambda}{{\rm mm}}\right)^3 \nonumber
\end{eqnarray}
Since NH$_3-$S has not been detected at 1.3 mm with ALMA, we adopt 3$\times$ the image rms for $S_{1.3{\rm mm}}$ to calculate the mass upper limit. Adopting a distance of 400 pc, the temperature of 12.1 K, and the flux density of 1.8 mJy beam$^{-1}$,  the upper limit for the NH$_3-$S clump mass is 0.013 M$_{\odot}$ \emph{per beam}, which corresponds to $\sim$1.7~M$_{\odot}$ if we adopt a source size determined from the ammonia data ($\sim$127.5 ALMA beams at 1.3 mm).  As the observations show,  in general there is a good correspondence between the distribution of the ammonia and dust emission (see e.g., \citealt{friesen2009}).   
For pre-stellar dense clumps and cores $\kappa_{1.3{\rm mm}}$ = 0.5 cm$^2$ g$^{-1}$ (or 0.005 cm$^2$ g$^{-1}$ if $R_{gd}$ is taken into account) is assumed in literature (e.g.,  \citealt{preibisch1993}; \citealt{andre1996}; \citealt{motte1998}).  If we adopt this value of $\kappa_{1.3{\rm mm}}$, the  estimate of the mass upper limit increases to 0.023 M$_{\odot}$ per beam or $\sim$2.9 M$_{\odot}$.  


The value of $\kappa_{\nu}$  is uncertain as it depends sensitively on the properties of the dust grains  (see e.g., \citealt{henning1995}), e.g. the size, shape, chemical composition, the physical structure of the grains, as well as the dust temperature.  \citet{ossenkopf1994} argue that $\kappa_{\nu}$ can deviate from their tabulated values by a factor of  $\lesssim$2 in environments with different physical conditions. Taking into account the uncertainties in the distance, dust temperature, flux density, as well as the assumed gas-to-dust ratio, we estimate that there is a factor of 3--4 uncertainty in the gas mass estimate.  



\subsubsection{Non-thermal Linewidths}
\label{s:nonth}

Using the kinetic temperature ($T_{kin}$), we can determine the thermal component of the line profile from a source in the LTE. The thermal velocity dispersion ($\sigma_{th}$) can be estimated using the relation: $\sigma_{th}=\sqrt{k_{B}\,T_{kin} / (\mu_{NH_3}\,m_H)}$, where $k_{B}$ is the Boltzmann constant, $T_{kin}$ is the kinetic temperature, $\mu_{NH_3}$ is the molecular weight of the NH$_{3}$ molecule in atomic units ($\mu_{NH_3}$=17.03), and $m_H$ is the mass of the hydrogen atom.  The full-width at half-maximum (FWHM) line width  can be derived by multiplying the velocity dispersion by $\sqrt{8\,{\rm ln}2}$.  The thermal velocity dispersion is $\sim$0.08 km~s$^{-1}$ and $\sim$0.07 km~s$^{-1}$ for NH$_{3}-$S and NH$_{3}-$Main, assuming $T_{kin}$ = 12.1 K and  $T_{kin}$ = 11.2 K, respectively.   The non-thermal velocity dispersion ($\sigma_{nth}$) can be derived using the equation:  $\sigma_{nth}=\sqrt{\sigma_{obs}^2 - \sigma_{th}^2 }$, where $\sigma_{obs}$  and $\sigma_{th}$ are the observed (corrected for instrumental broadening) and thermal velocity dispersions, respectively.   The mean values of $\sigma_{nth}$ are 0.16 km~s$^{-1}$ and 0.14 km~s$^{-1}$ for NH$_{3}-$S and NH$_{3}-$Main, respectively.  The respective $\sigma_{nth}$ standard deviations are 0.06 km~s$^{-1}$ and 0.07 km~s$^{-1}$.  

\begin{figure*}[ht!]
\centering
\includegraphics[width=0.49\textwidth]{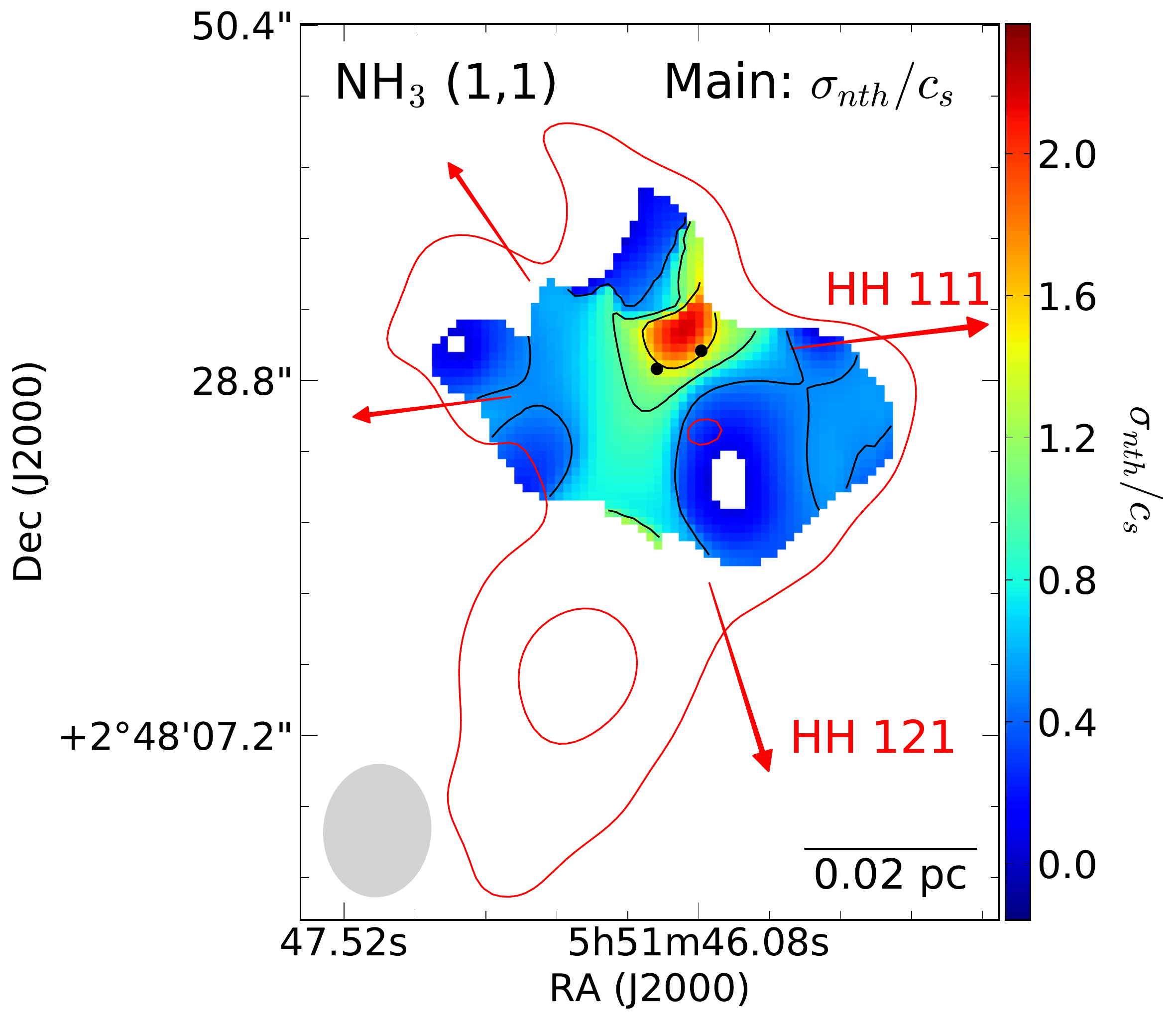}
\includegraphics[width=0.49\textwidth]{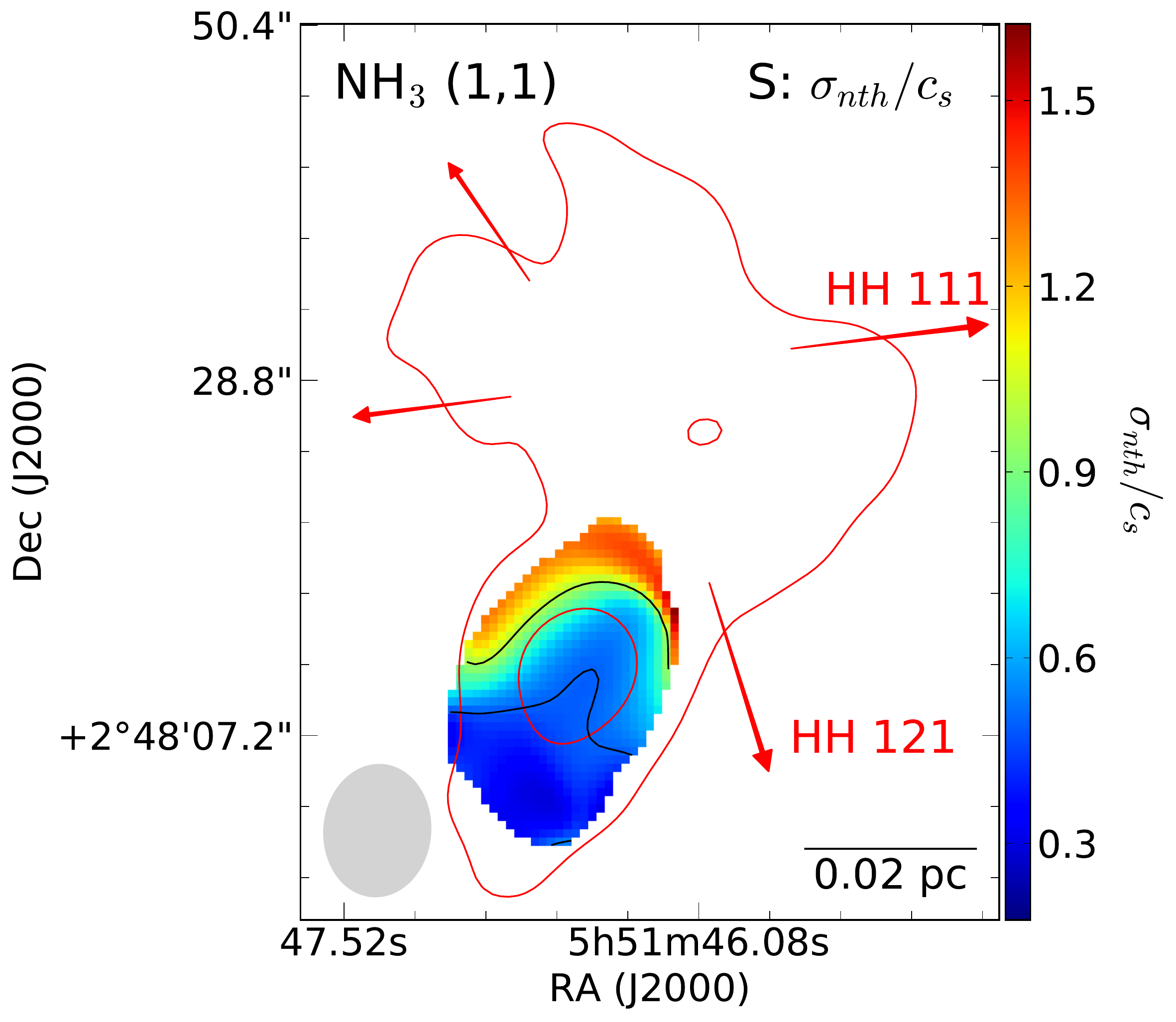}
\caption{The maps of the non-thermal to thermal velocity dispersion ($\sigma_{nth}/c_s$) for NH$_{3}-$Main ({\it left}) and  NH$_{3}-$S ({\it right}).  The red contours correspond to the NH$_{3}$ (1,\,1) integrated intensity with the contour levels of 20\% and 75\% of the peak (see Fig.~\ref{f:mom0}).  The positions of sources VLA-1 and VLA-2 are indicated with black filled circles in the left panel.  The black contours correspond to $\sigma_{nth}/c_s$; the contour levels are  (0.5, 1.0) for  NH$_{3}-$S  and (0.5, 1.0, 1.5) for NH$_{3}-$Main. The red arrows show the approximate directions of the HH\,111 and HH\,121 jets; $\sigma_{nth}/c_s$ peaks toward the outflows/jets. The VLA synthesized beam is shown at the lower left. \label{f:nonth}}
\end{figure*}

We compare  $\sigma_{nth}$ to a thermal sound speed, $c_{s}=\sqrt{k_{B}\,T_{kin} / (\mu m_H)}$, where $\mu$ is a molecular weight of a mean particle, $\mu$=2.33.   Figure~\ref{f:nonth} shows maps of $\sigma_{nth}/c_{s}$ for NH$_{3}-$S and NH$_{3}-$Main.  The figure shows that for NH$_{3}-$S, non-thermal line widths are smaller or equal to the thermal line width over most of the core (the core is ``quiescent''), except its northern rim where the core starts having ``transonic'' non-thermal line-of-sight velocity dispersions ($1 < \sigma_{nth}/c_{s} \leq 2$; e.g., \citealt{klessen2005}).  In NH$_{3}-$Main, the turbulent velocity dispersion increases from the subsonic values on the eastern and western side of the source to the transonic values in the central strip and is supersonic in a small area in the north in the vicinity of sources VLA-1 and VLA-2. 

\begin{figure*}[ht!]
\centering
\includegraphics[width=0.49\textwidth]{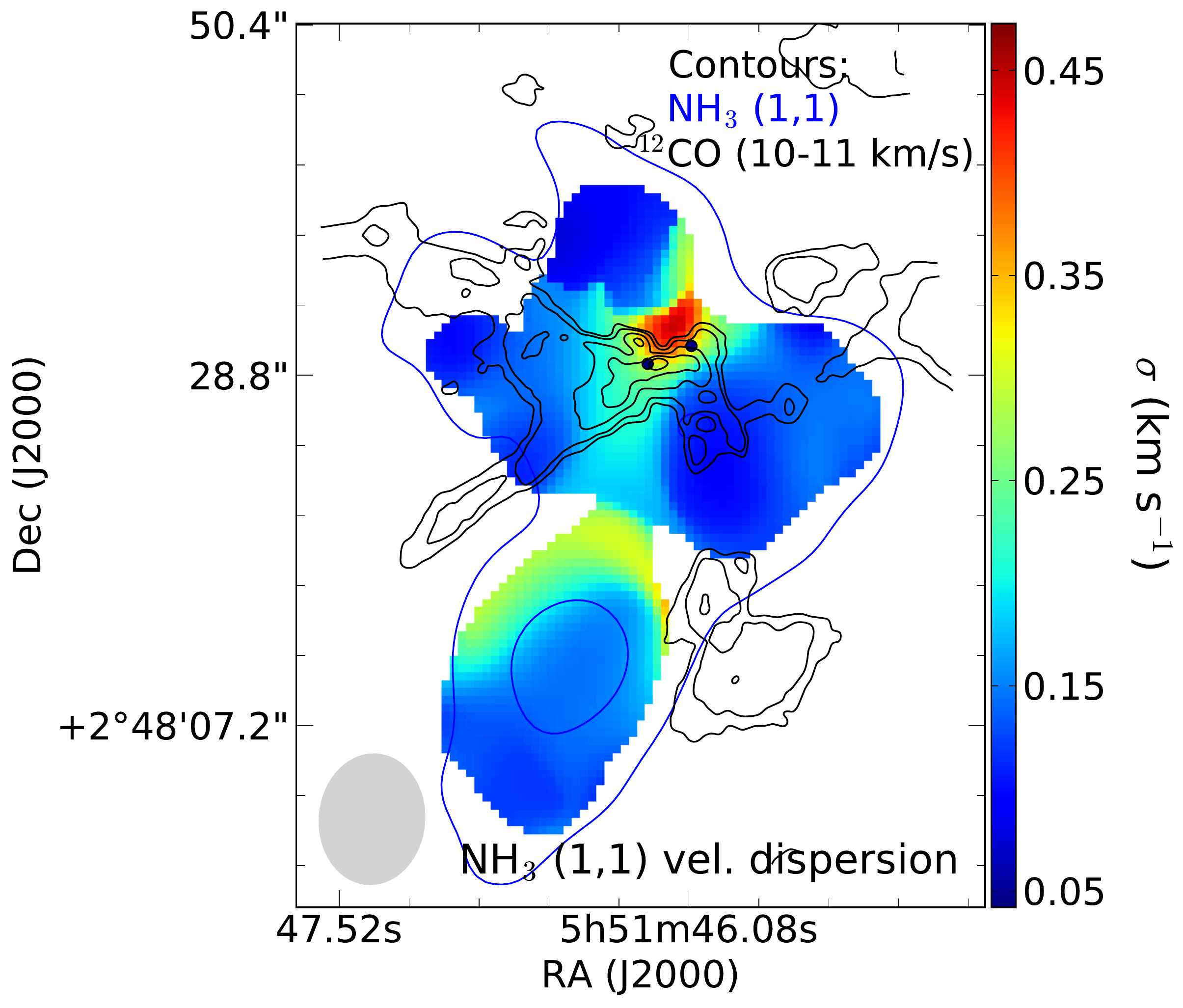}
\includegraphics[width=0.49\textwidth]{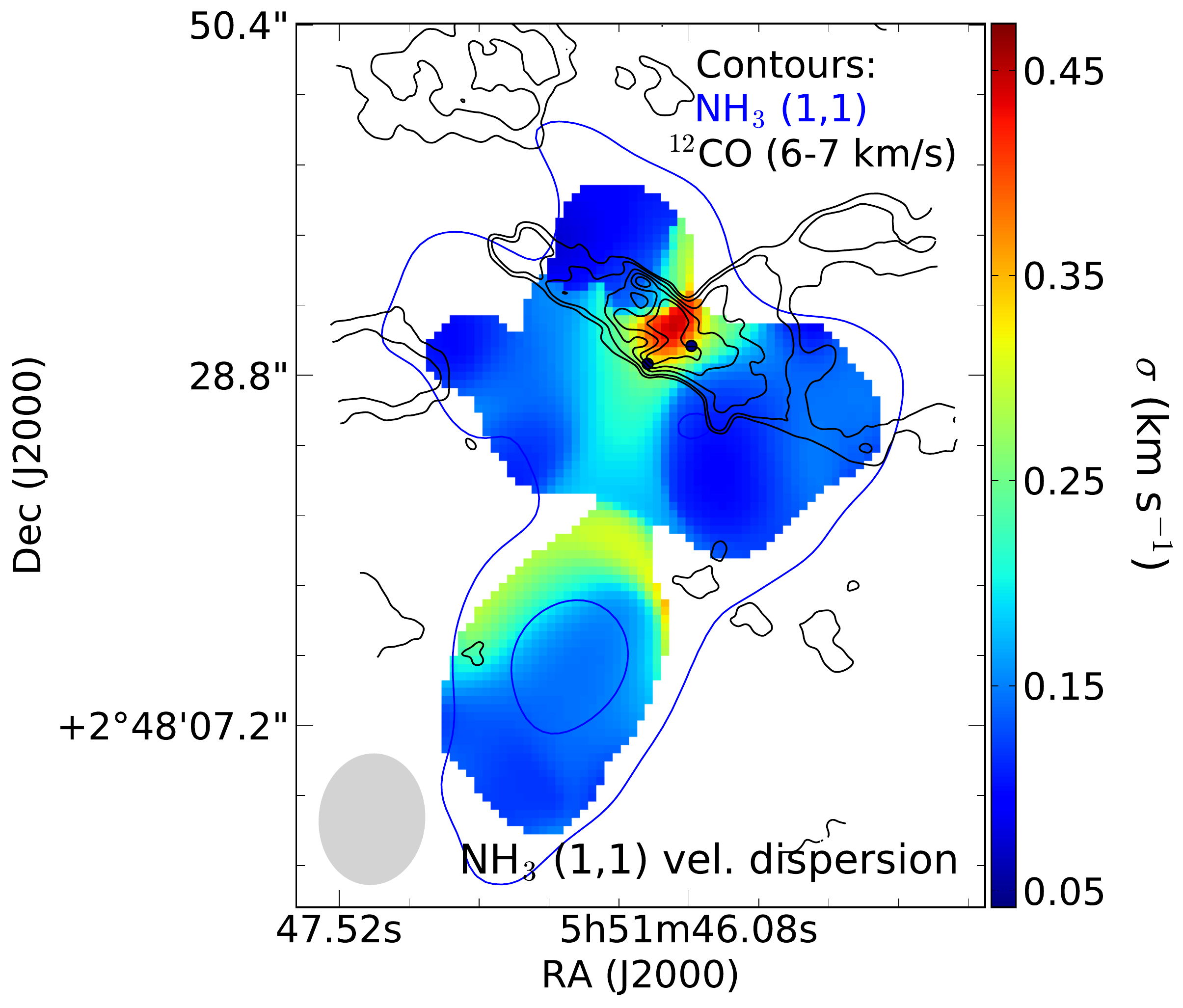}
\caption{The NH$_3$ (1,\,1) velocity dispersion map with the $^{12}$CO (2-1) contours overlaid. The $^{12}$CO contours in the {\it left} and {\it right} panels correspond to the $^{12}$CO integrated intensity for velocity ranges 10.0 - 11.0 km~s$^{-1}$  and 6.0 - 7.0 km~s$^{-1}$ , respectively.  The $^{12}$CO contour levels are (10, 20, 40, 60, 80)\% $\times$ the $^{12}$CO integrated intensity peak in each velocity range: 1.296 Jy beam$^{-1}$ km s$^{-1}$ and 1.355 Jy beam$^{-1}$ km s$^{-1}$ for the  {\it left} and {\it right}, respectively. The blue contours correspond to the NH$_{3}$ (1,\,1) integrated intensity with the contour levels of 20\% and 75\% of the peak (see Fig.~\ref{f:mom0}).  The $^{12}$CO emission traces molecular outflows associated with the HH\,111 and HH\,121 jets. No $^{12}$CO emission was detected toward NH$_{3}-$S. The VLA synthesized beam is shown at the lower left.  \label{f:12co}}
\end{figure*}

The largest turbulent velocity dispersions in NH$_{3}-$Main and NH$_{3}-$S occur in regions where the impact of the Herbig-Haro jets and molecular outflows on the environment is expected to be large.  This is illustrated in Fig.~\ref{f:12co} where we compare the distribution of  $\sigma_{nth}/c_s$ to the $^{12}$CO emission in two representative velocity ranges that trace the outflows associated with both HH\,111 and HH\,121 jets. In NH$_3-$Main, the region of enhanced turbulent velocity dispersion coincides with the base of the HH\,111 jet. The increased linewidths in the inner envelope have been observed toward other young objects, including HH\,211, L\,1157, and L1451-mm  (\citealt{tanner2011}; \citealt{tobin2011}; \citealt{pineda2011}). 

 The distribution of the $^{12}$CO emission suggests a possibility that VLA-2 is the source of the HH\,121 jet;  this conclusion, however, needs to be supported by a detailed analysis of the $^{12}$CO data, which is out of scope of this paper.  NH$_3-$S is located between the HH\,111 jet in the north and the HH\,121 jet in the north-west and west; the location of the region of the enhanced turbulent velocity dispersions along the northern rim of the source and on the north-east and north-west indicates that they may be the result of the turbulence induced by the jets. This scenario will be discussed in more detail in Section~\ref{s:nh3s}.

\subsection{Carbon-bearing Molecules}
\label{s:almaco}

\begin{figure*}[ht!]
\centering
\includegraphics[width=0.45\textwidth]{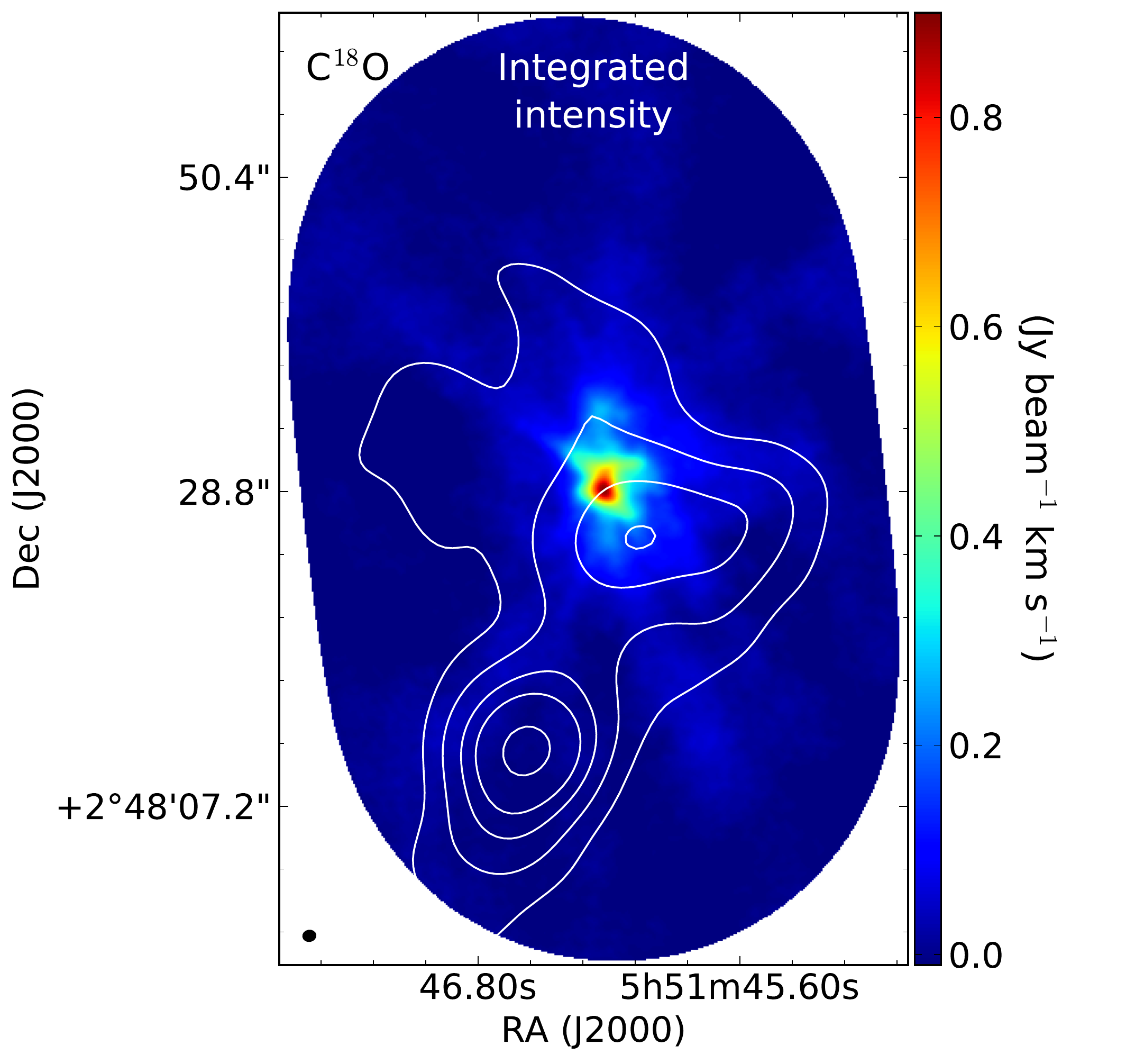}
\includegraphics[width=0.45\textwidth]{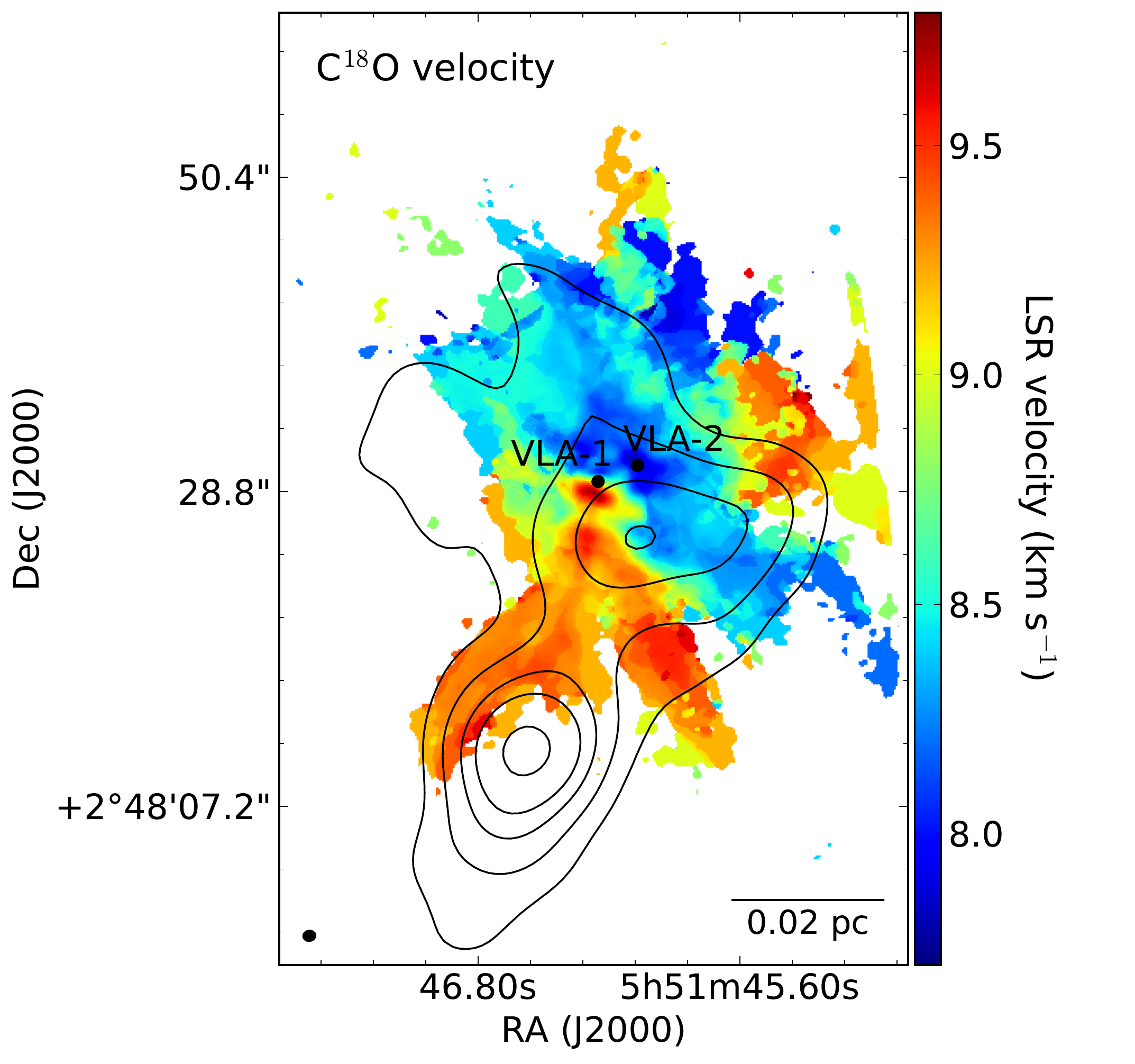}
\caption{The ALMA C$^{18}$O  integrated intensity (moment 0; {\it left}) and the LSR velocity (moment 1; {\it right}) maps. The C$^{18}$O emission in the velocity range from 7.0 to 11.0 km~s$^{-1}$ was used for moment 0 and moment 1 calculations.  Only pixels with a signal-to-noise ratio larger than 10 were included in the moment 1 calculations. The white contours represent the NH$_{3}$ (1,\,1) emission; the contour levels as in Fig.~\ref{f:mom0}.  No C$^{18}$O emission was detected toward the center of NH$_{3}-$S.  In both images, the ALMA beam is shown in the lower left corner. \label{f:C18Omom}}
\end{figure*}

\begin{figure*}[ht!]
\centering
\includegraphics[width=0.45\textwidth]{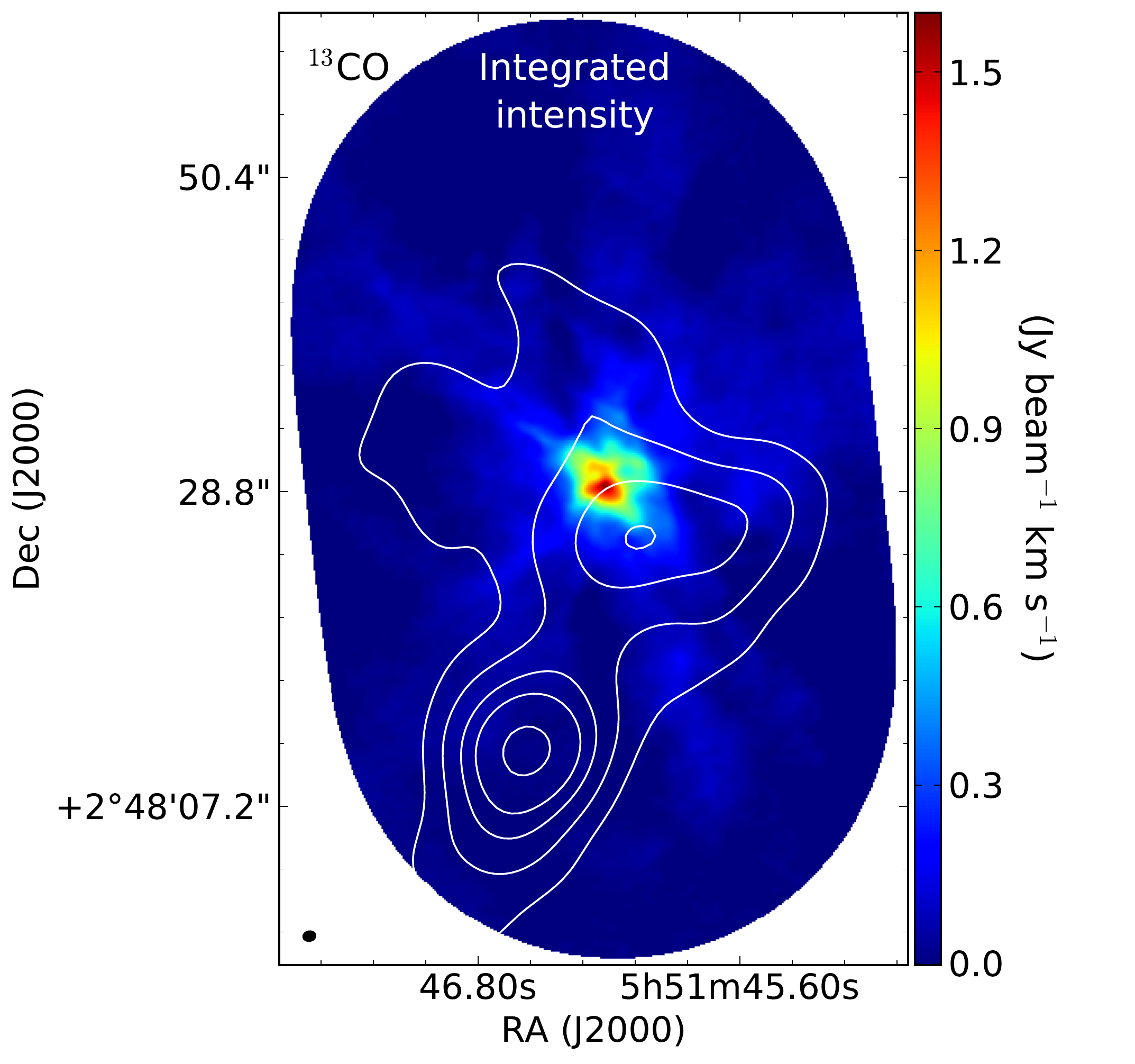}
\includegraphics[width=0.45\textwidth]{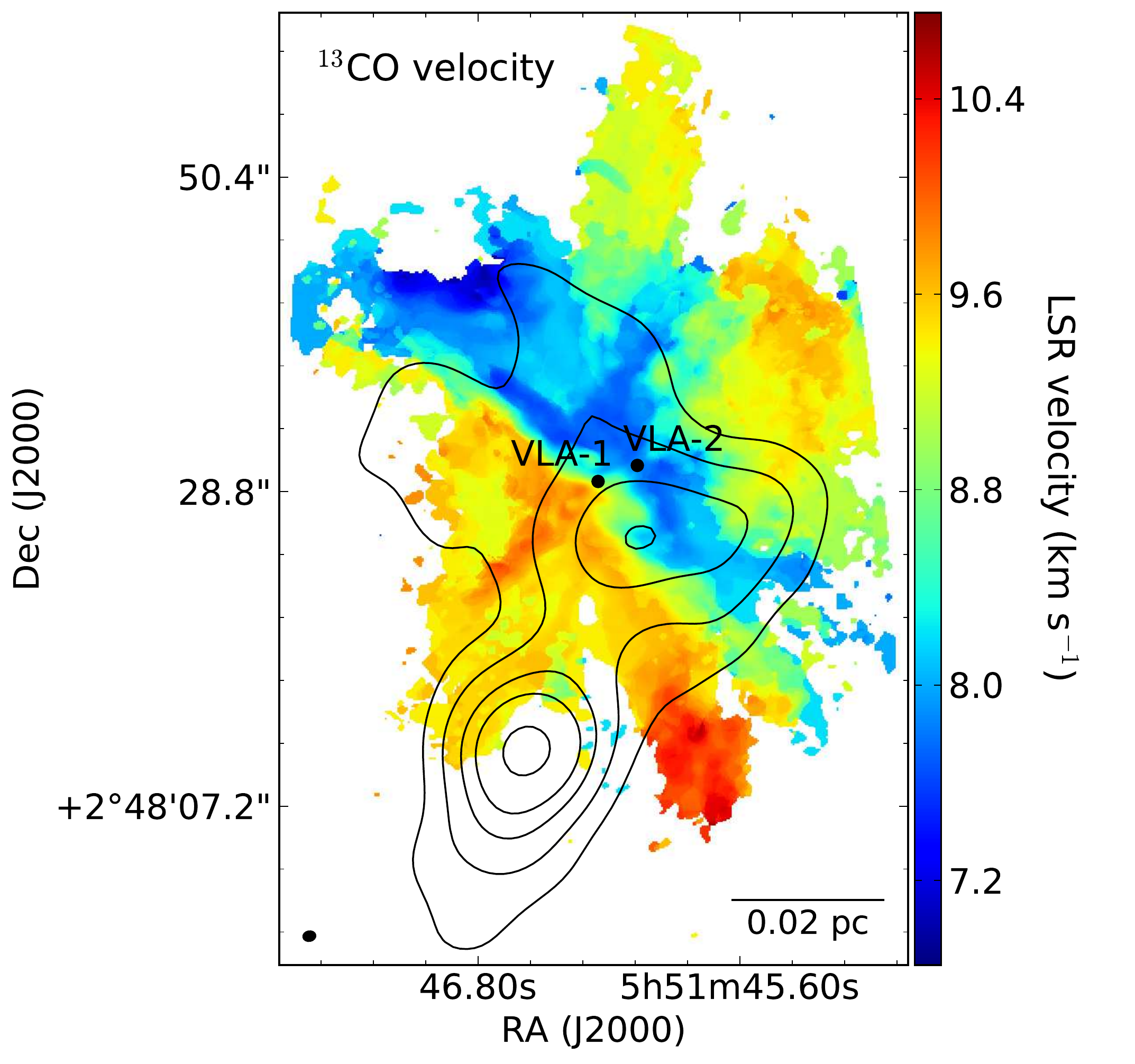}
\caption{The ALMA $^{13}$CO  integrated intensity (moment 0; {\it left}) and the LSR velocity (moment 1; {\it right}) maps. The $^{13}$CO emission in the velocity range from 7.0 to 11.0 km~s$^{-1}$ was used for moment 0 and moment 1 calculations.  Only pixels with a signal-to-noise ratio larger than 10 were included in the moment 1 calculations. The white contours represent the NH$_{3}$ (1,\,1) emission; the contour levels as in Fig.~\ref{f:mom0}. No $^{13}$CO emission was detected toward the center of NH$_{3}-$S. In both images, the ALMA beam is shown in the lower left corner. \label{f:13COmom}}
\end{figure*}

\begin{figure*}[ht!]
\centering
\includegraphics[width=0.45\textwidth]{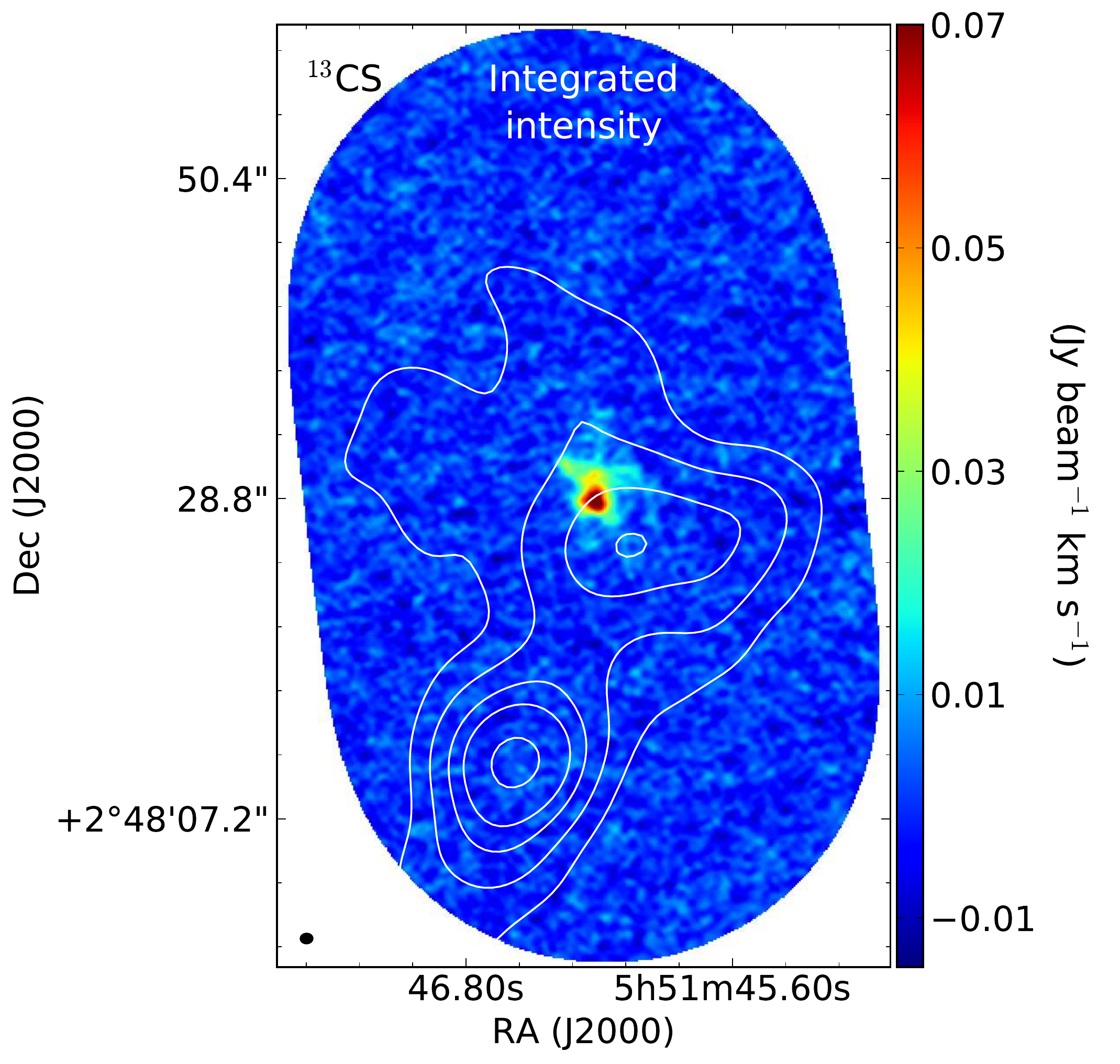}
\includegraphics[width=0.45\textwidth]{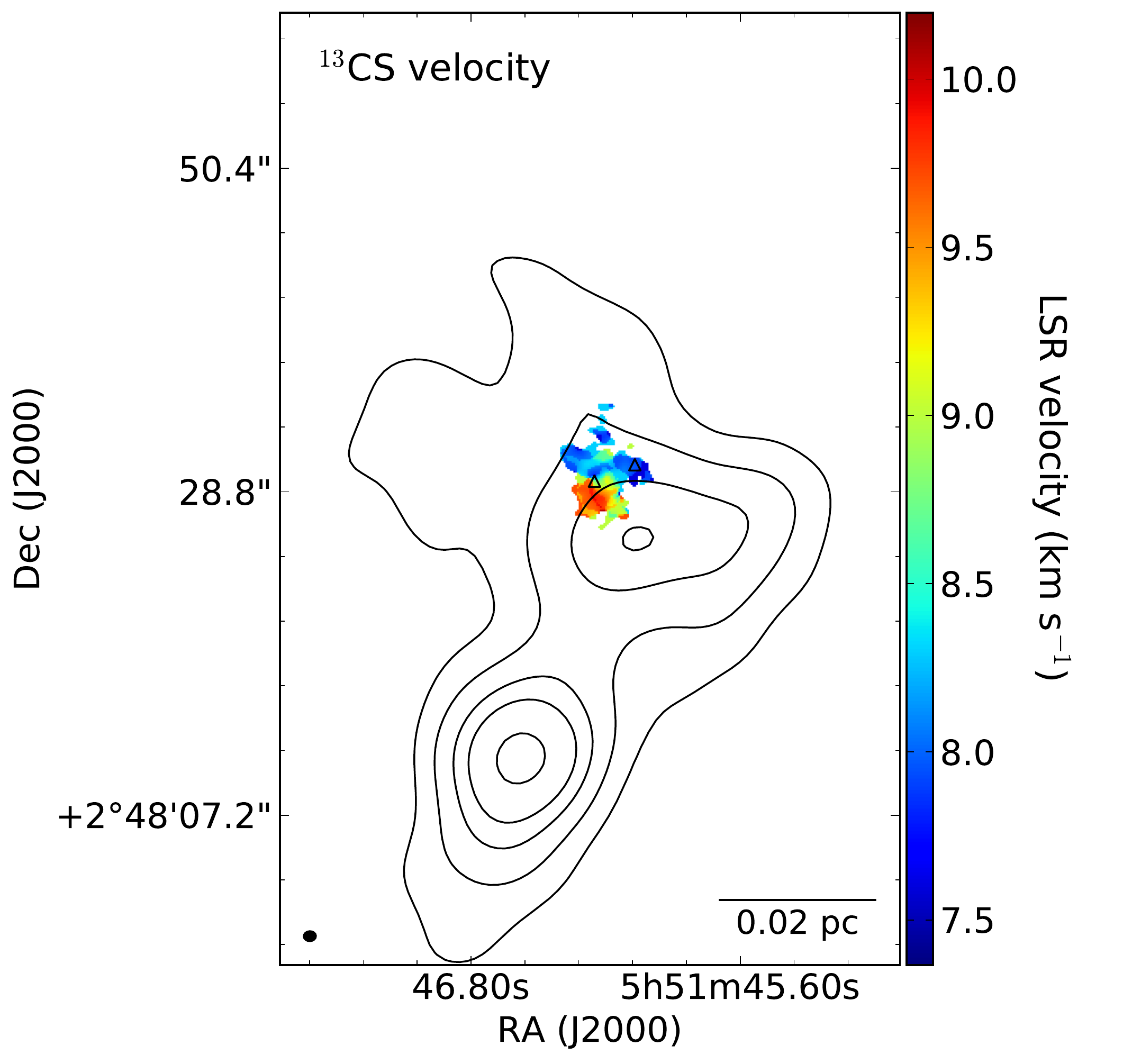}
\caption{The ALMA $^{13}$CS  integrated intensity (moment 0; {\it left}) and the LSR velocity (moment 1; {\it right}) maps. The velocity of the $^{13}$CS emission ranges from 5.5 to 11.1 km~s$^{-1}$; this velocity range was used for moment 0 and moment 1 calculations.  Only pixels with a signal-to-noise ratio larger than 5 were included in the moment 1 calculations. The white contours represent the NH$_{3}$ (1,\,1) emission; the contour levels as in Fig.~\ref{f:mom0}. The triangles indicate the positions of the VLA-1 and VLA-2 sources (see also Figs.~\ref{f:C18Omom} and \ref{f:13COmom}).  The $^{13}$CS emission is confined to the envelope of VLA-1; no $^{13}$CS was detected toward NH$_{3}-$S.  The ALMA beam is shown in the lower left corner.  \label{f:13CSmom}}
\end{figure*}

We investigate the distribution of the C$^{18}$O (2-1), $^{13}$CO (2-1), and $^{13}$CS (5-4) emission detected by ALMA to gain some insight into the nature of NH$_{3}-$S.  Figures ~\ref{f:C18Omom}, \ref{f:13COmom}, and \ref{f:13CSmom} show the integrated intensity images and the velocity distributions for C$^{18}$O, $^{13}$CO, and $^{13}$CS, respectively.  The $^{12}$CO (2-1)  integrated intensity contours are shown in Fig~\ref{f:12co}.  The ALMA C$^{18}$O line and 1.3 mm continuum data for HH\,111 is presented in \citet{lee2016} who studied the envelope and the disk of source VLA-1 in great detail.  

The carbon-bearing molecular emission traces the envelope and disk of the central source VLA-1 (C$^{18}$O and $^{13}$CS) and the molecular outflow (mainly $^{13}$CO and $^{12}$CO) in the Class I protostellar system HH\,111, as well as the molecular outflow associated with the HH\,121 jet. This region is associated mainly with the ammonia source NH$_{3}-$Main.   What is striking in Figs. ~\ref{f:12co}--\ref{f:13COmom} is the lack of the C$^{18}$O, $^{12}$CO, and $^{13}$CO emission in the center of NH$_{3}-$S.  However, the $^{13}$CO and C$^{18}$O emission wraps around the source roughly from east to west along its northern rim.  No $^{13}$CS was detected toward NH$_{3}-$S; the $^{13}$CS emission is confined to the envelope of VLA-1. 

The morphology of the $^{13}$CO and C$^{18}$O emission can be inspected in more details in Figs.~\ref{f:c18ochan}--\ref{f:3col13CO} in Appendix B.  Figure~\ref{f:c18ochan} shows the C$^{18}$O channel maps for the velocity range from 4.4 to 13.8 km~s$^{-1}$.  Figures~\ref{f:3colC18O} and \ref{f:3col13CO} show the three-color mosaics combining the C$^{18}$O and $^{13}$CO channel maps, respectively, with the {\it Spitzer} 4.5 $\mu$m image and the VLA NH$_{3}$ (1,\,1) channel maps; the C$^{18}$O/$^{13}$CO and NH$_3$ velocity range corresponds to the velocities of the NH$_3$ (1,\,1) main line emission toward NH$_{3}-$S (8.8--9.8 km~s$^{-1}$).  Several jet knots detected with \emph{Spitzer} at 4.5 $\mu$m allow us to relate the molecular line emission to the HH\,111 and HH\,121 jets. The $^{18}$CO and $^{13}$CO emission is filamentary south of VLA-1 toward and around NH$_3-$S.

\subsection{N$_{2}$D$^{+}$}
\label{s:n2dp}

\begin{figure*}[ht!]
\centering
\includegraphics[width=0.48\textwidth]{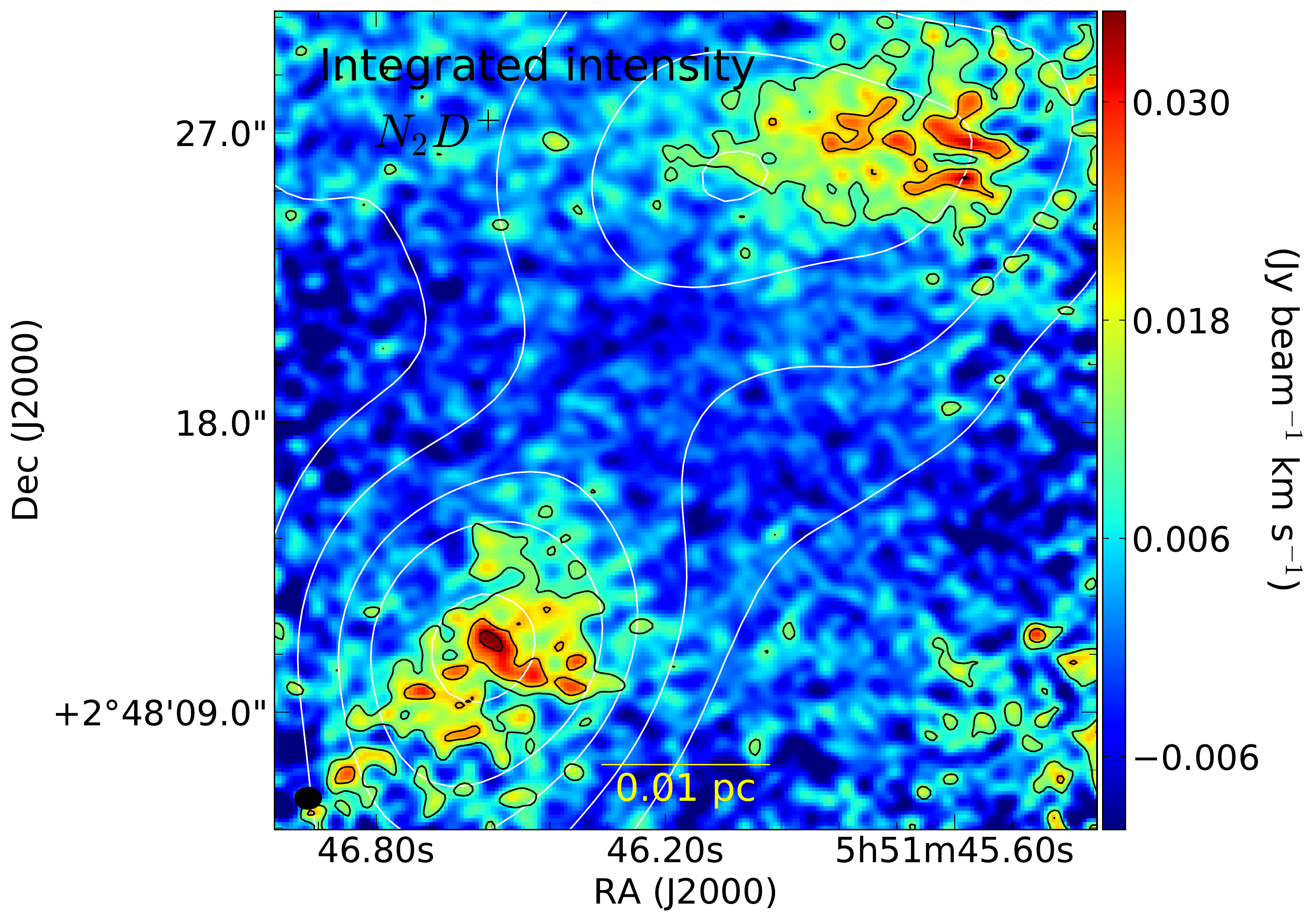} 
\includegraphics[width=0.45\textwidth]{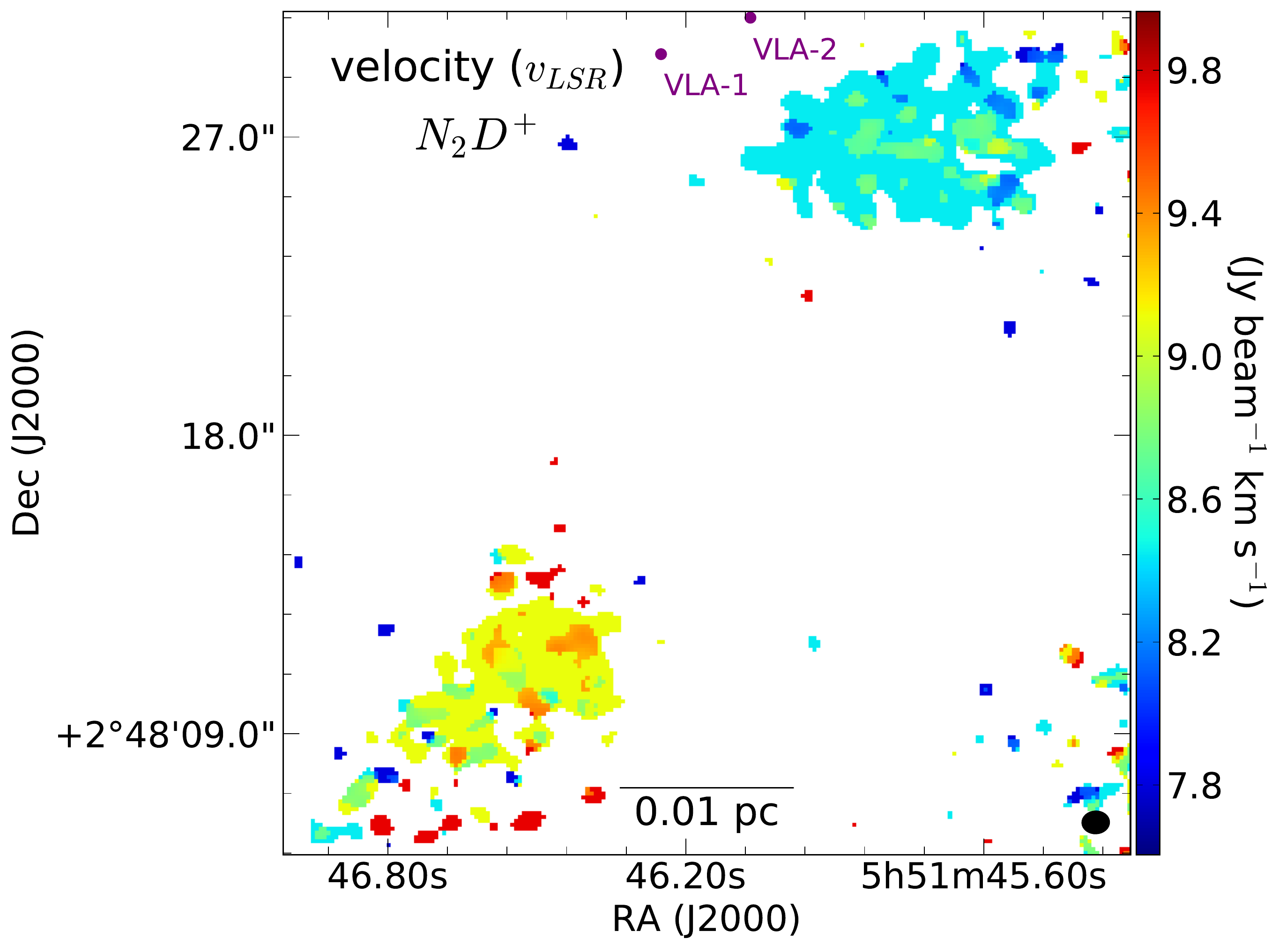} 
\caption{The N$_{2}$D$^{+}$ (3--2) integrated intensity (moment 0; {\it left}) and  $LSR$ velocity (moment 1; {\it right}) maps. The moment maps were made using the data cube corrected for the primary beam; only pixels with signal-to-noise ratio larger than 4 were included in the moment 1 calculations. The black contours in the left panel correspond to N$_{2}$D$^{+}$ integrated intensity with contour levels of (5, 10, 15) $\times$ 2.2 mJy beam$^{-1}$, the image rms noise.  The white contours show the distribution of the NH$_{3}$ (1,\,1) emission observed with the VLA; the contour levels as in Fig.~\ref{f:mom0}.  The size of the ALMA synthesized beam shown in the lower left/lower right ({\it left/right panel}) corner is 0$\rlap.{''}$81 $\times$ 0$\rlap.{''}$67, PA = -87$\rlap.^{\circ}$4.  The filled circles show the positions of protostars VLA-1 and VLA-2 (black circles in the moment 0 and purple circles in the moment 1 maps). \label{f:n2dpmom} }
\end{figure*}

The ALMA N$_{2}$D$^{+}$ (3-2) image shown in Fig.~\ref{f:n2dpmom} reveals two N$_{2}$D$^{+}$ condensations in the HH\,111/HH\,121 protostellar system. One of the condensations is associated with NH$_3-$S with the peak N$_{2}$D$^{+}$ emission coinciding with the peak of the NH$_3$ emission. The second N$_{2}$D$^{+}$ condensation is located in NH$_3-$Main; it is offset to the southwest from the location of the protostar VLA-1 and to the west from the peak of the NH$_3$ emission. The two N$_{2}$D$^{+}$ condensations show a velocity difference of $\sim$0.7 km~s$^{-1}$ (see Fig.~\ref{f:n2dpmom}), consistent with the NH$_3$ results. The N$_{2}$D$^{+}$ emission appears clumpy, possibly more extended emission has been filtered out by the interferometer.

As will be discussed in Section~\ref{s:nh3s},  observable abundances of N$_{2}$D$^{+}$ can only be achieved in the coldest and densest molecular cores where CO, the main destroyer of N$_{2}$D$^{+}$, is frozen-out onto dust grains (e.g., \citealt{caselli2002a}; \citealt{flower2006}).  N-bearing molecules such as N$_{2}$D$^{+}$ and NH$_3$ are used to study the cold material because they do not freeze out onto dust grains until a density of  $\sim$10$^6$ cm$^{-3}$ is reached (\citealt{bergin1997}; \citealt{flower2006}).  A non-detection of the CO emission toward the inner part of NH$_3-$S, shows that indeed CO is frozen-out where N$_{2}$D$^{+}$ is detected. Also, no CO emission has been detected toward the center of the N$_{2}$D$^{+}$ condensation in NH$_3-$Main in the velocity range corresponding to that of the N$_{2}$D$^{+}$ emission (see Fig.~\ref{f:C18Omom} and \ref{f:13COmom}). 

The eastern edge of the N$_{2}$D$^{+}$ condensation (as defined by the contour at the level of 20\% of the peak) in NH$_3-$Main and the maximum N$_{2}$D$^{+}$ emission pixel  are at a distance of $\sim$2$\rlap.{''}$7 ($\sim$0.005 pc or 1060 AU)  and $\sim$10$\rlap.{''}$8 ($\sim$0.02 pc or 4300 AU), respectively, from the protostar VLA-1.  These offsets can be explained by the destruction of the N$_{2}$D$^{+}$ molecules in regions of bright CO emission associated with the protostar.  

The NH$_3$ emission peak coincides with the eastern edge of the N$_{2}$D$^{+}$ condensation, but is offset by $\sim$7$''$ ($\sim$0.013 pc or 2800 AU) from the maximum N$_{2}$D$^{+}$ emission pixel.  The N$_{2}$D$^{+}$ emission is distributed along the jet rather than in the direction perpendicular to it. 

The relative offset between the distribution of the N$_{2}$D$^{+}$ and N$_{2}$H$^{+}$ emission has been detected in eight Class 0/I protostellar envelopes by \citet{tobin2013}. The observations show that N$_{2}$H$^{+}$ and NH$_{3}$ appear to trace  the same kinematics and physical conditions (e.g., \citealt{tobin2011}), thus we can expect the offset between the peak NH$_{3}$ and N$_{2}$D$^{+}$ emission.   The distribution of the N$_{2}$D$^{+}$ emission with respect to the N$_{2}$H$^{+}$ emission for several sources in the \citet{tobin2013} sample has similar morphology to that between N$_{2}$D$^{+}$ and NH$_{3}$ in HH\,111/HH\,121; for example,  L\,483 and L\,1165 have the N$_{2}$D$^{+}$ emission with the peak in the direction of the outflow with the  N$_{2}$H$^{+}$ and N$_{2}$D$^{+}$  offsets of $\sim$3200 AU and $\sim$2400 AU, respectively.  \citet{tobin2013} argue that the abundance peak offsets between N$_{2}$H$^{+}$ and N$_{2}$D$^{+}$  can be explained by an increased CO evaporation temperature due to ice mixtures and/or a gradient of the ${\it ortho}/{\it para-}$H$_2$ ratio  in the inner envelope.

\begin{figure*}[ht!]
\centering
\includegraphics[width=0.4\textwidth]{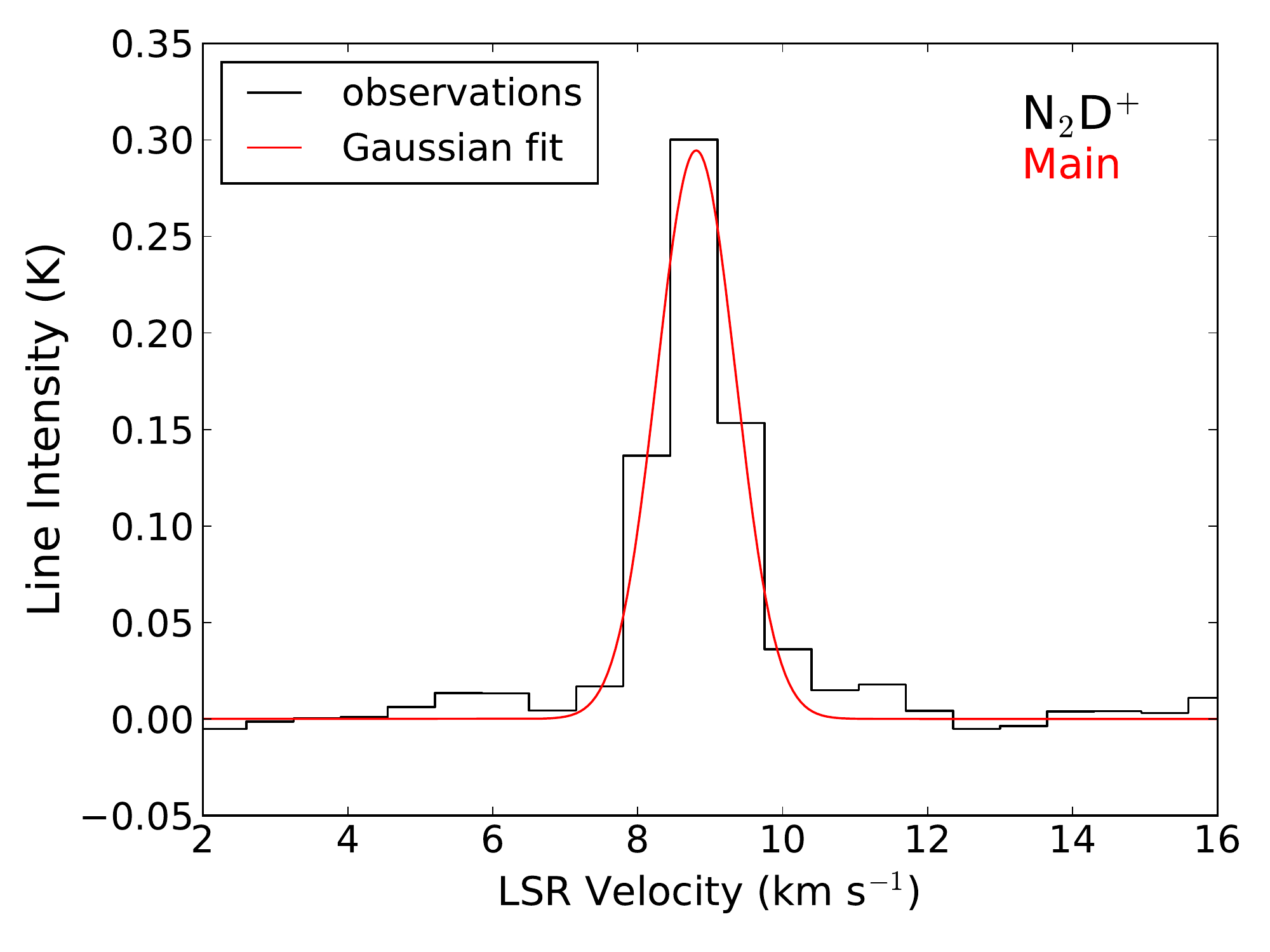} 
\includegraphics[width=0.4\textwidth]{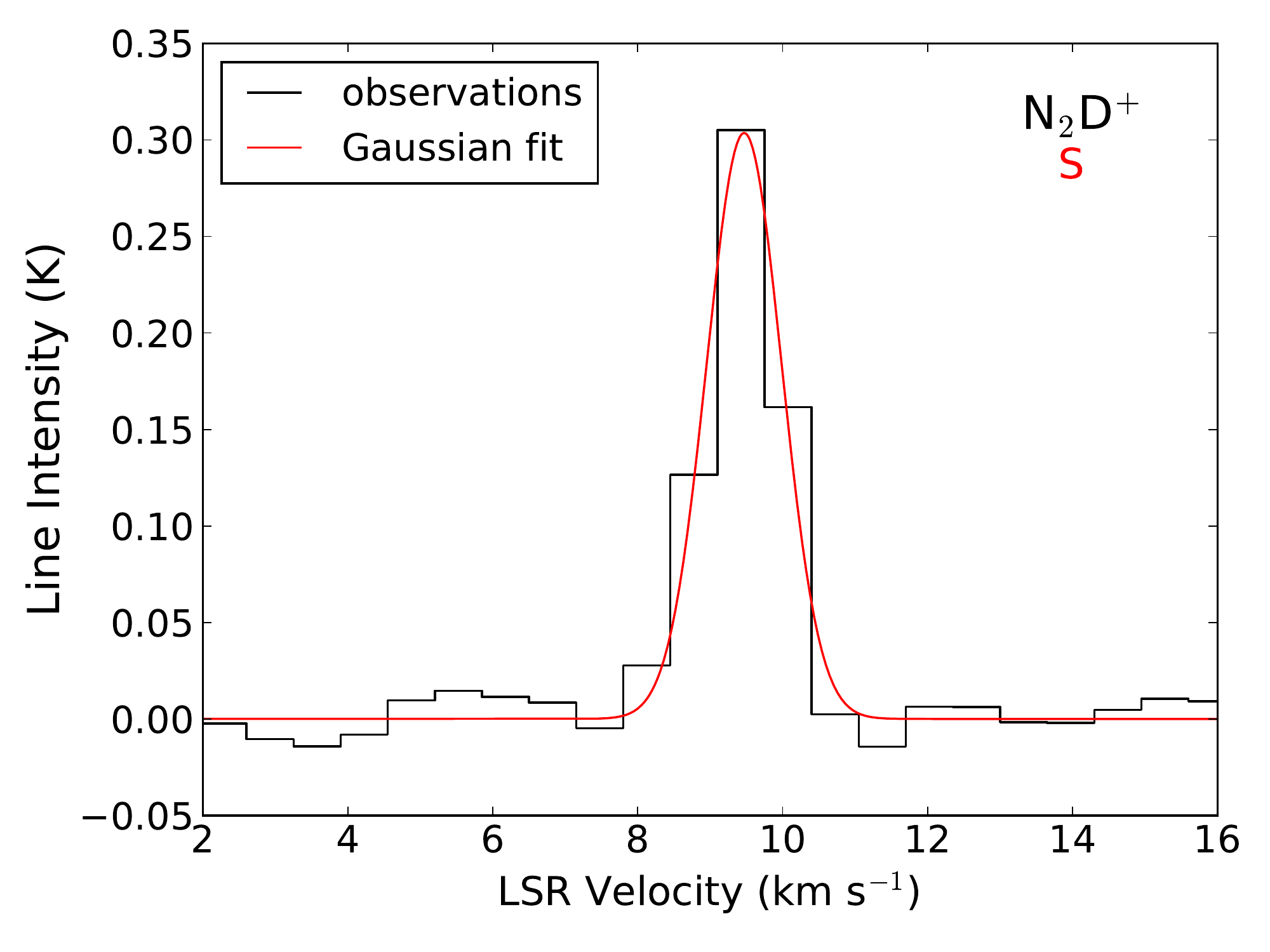} 
\caption{N$_{2}$D$^{+}$ (3--2)  line profiles for N$_{2}$D$^{+}$ clumps associated with NH$_3-$Main ({\it left}) and NH$_3-$S ({\it right}) averaged over the area enclosed by the 20\% of the peak contour for a corresponding source. The results of the Gaussian profile fitting are presented in Table~\ref{t:n2dpg}.  \label{f:n2dp}}
\end{figure*}

No velocity gradients are detected in N$_{2}$D$^{+}$ in HH\,111/HH\,121.  The N$_{2}$D$^{+}$ line profiles for both N$_{2}$D$^{+}$ concentrations are presented in Fig.~\ref{f:n2dp}. For each clump, the line profile was extracted as a mean over an elliptical region enclosing the contour with the value corresponding to the 20\% of the N$_{2}$D$^{+}$ emission peak. No hyperfine components have been resolved. The lines are broadened by the hyperfine component blending, non-thermal motions, and a relatively low velocity resolution. We thus fitted the N$_{2}$D$^{+}$ lines with the single Gaussian profiles; the results are listed in Table~\ref{t:n2dp}.  The N$_{2}$D$^{+}$ line parameters for both cores are very similar, but a higher velocity resolution observations are needed to estimate their physical parameters using the hyperfine emission line structure fitting.

\begin{deluxetable*}{ccccc}
\centering
\tablecaption{N$_{2}$D$^{+}$ (3--2) Line Parameters Obtained with a Gaussian Profile Fit\label{t:n2dp}}
\tablewidth{0pt}
\tablehead{
\colhead{Source} &
\colhead{$v_{LSR}$} &
\colhead{$\Delta$$v_{G}$\tablenotemark{a}} &
\colhead{$T_{peak}$} &
\colhead{$\int_{}^{} T dv$}  \\
\colhead{} &
\colhead{(km s$^{-1}$)} &
\colhead{(km s$^{-1}$)} &
\colhead{(K)} &
\colhead{(K km s$^{-1}$)} }
\startdata
Main & 8.81 $\pm$ 0.02 & 1.07 $\pm$ 0.04 & 0.29 $\pm$ 0.02  & 0.40 $\pm$ 0.01\\
S & 9.47 $\pm$ 0.02  & 1.00 $\pm$ 0.04 &  0.30 $\pm$ 0.02 & 0.39 $\pm$ 0.01
\enddata
\tablenotetext{a}{The line width (FWHM) corrected for instrumental broadening: $\Delta v_{G} = \sqrt{\Delta v_{G, obs}^2 - \Delta v_{instr}^2}$, where $\Delta v_{G, obs} $ is the observed line width and  $\Delta v_{instr}$ is the channel width of 0.65 km s$^{-1}$. \label{t:n2dpg}}
\end{deluxetable*}

\begin{figure}[ht!]
\centering
\includegraphics[width=0.5\textwidth]{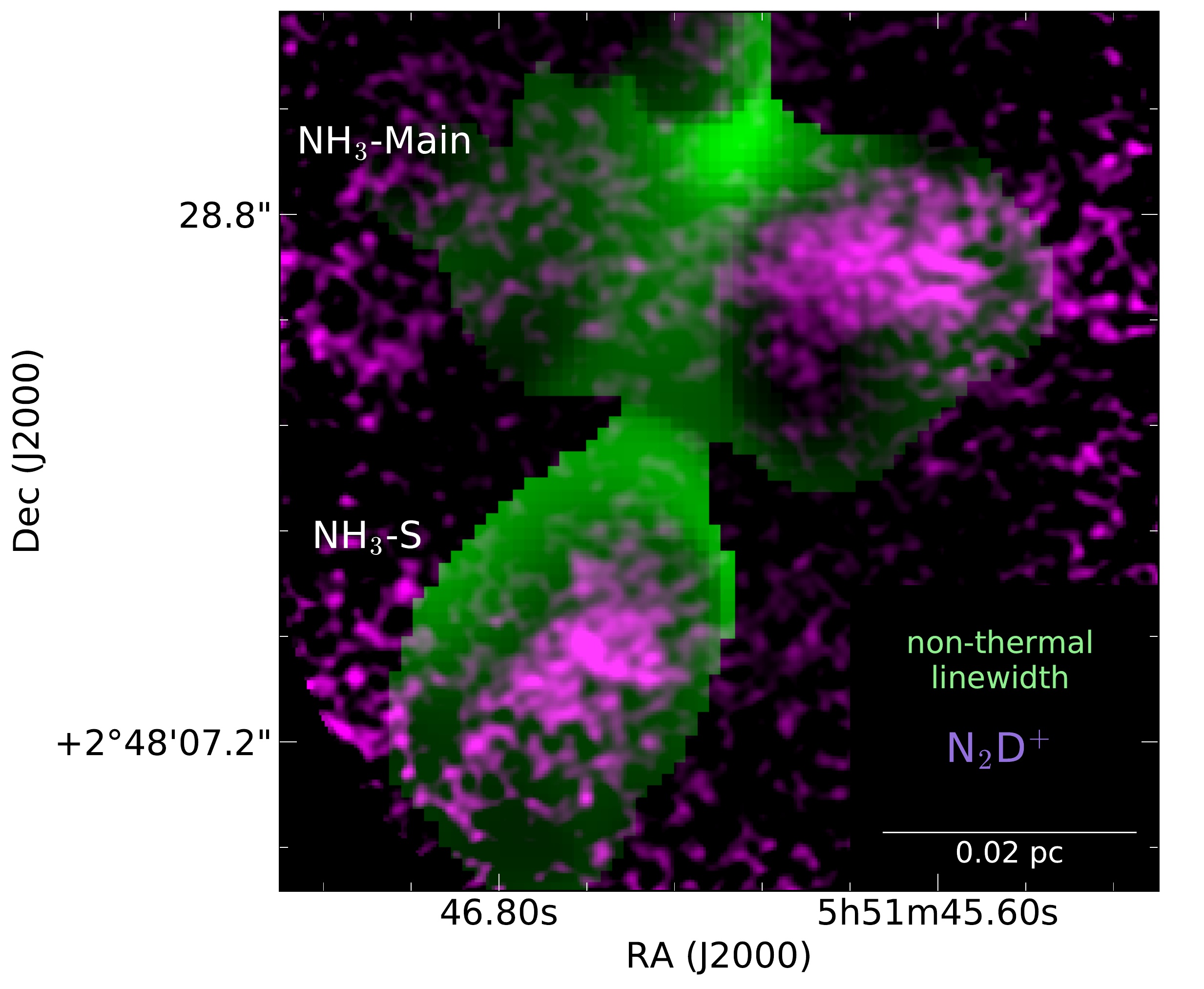} 
\caption{The image comparing the N$_2$D$^{+}$ integrated intensity map ({\it purple}) and non-thermal linewidths ({\it green}); it reveals the anti-correlation between these two quantities.  A comparison to Fig.~\ref{f:nonth} shows that the N$_2$D$^{+}$ emission is associated with regions characterized by the subsonic turbulent velocity dispersion. \label{f:n2dp_nth} }
\end{figure}


\section{Discussion}
\label{s:nh3s}

The properties of NH$_3-$S indicate that it is a starless core located in a turbulent medium with turbulence induced by the Herbig-Haro jets and associated outflows.  Dense cores are density enhancements of the cloud material with masses of 0.5--5 M$_{\odot}$, sizes of 0.03--0.2 pc, mean densities of 10$^{4}$--10$^{5}$ cm$^{-3}$, velocity extents of 0.1--0.3 km~s$^{-1}$,  and gas temperatures of 8--12 K (see a review by \citealt{bergin2007}).  The {\it Spitzer} 8.0 $\mu$m image shows that NH$_3-$S is located in the dark cloud and there is no indication of the presence of the central object in the available observations ranging from near-IR to mm wavelengths (see Section~\ref{s:nh3}).   The chemical structure of the NH$_3-$S core show evidence for ``selective'' freeze-out, an inherent property of dense cold cores.

NH$_3-$S has characteristics of a  ``coherent core'' (e.g., \citealt{goodman1998}; \citealt{caselli2002}).  The ``coherent core''  has subsonic internal motions (see Fig.~\ref{f:nonth}), indicating that turbulent motions contribute less to the gas pressure than the thermal component, thus representing a minor contribution to the core support (e.g., \citealt{myers1983}; \citealt{tafalla2004}).  In the ``coherent cores'', the observed linewidths remain approximately constant.  

As shown in Section~\ref{s:nonth}, the turbulent contribution to the linewidths increases toward the peripheries in the upper half of the NH$_3-$S core, reaching maximum values in regions exposed the most to the of Herbig-Haro jets and outflows that induce turbulence into the environment.  This pattern resembles the `transition to coherence' observed in other dense cores. For example, \citet{pineda2010} report a sharp transition between the coherent core and the more turbulent gas surrounding it in the B5 region in Perseus; in the transition region, the velocity dispersion changes by a factor of 2 over less than a beam width ($<$0.04~pc).  The transition between subsonic and supersonic turbulence has been observed in several other regions covering a range of environments and star formation activities (e.g., \citealt{pagani2010}; see also \citealt{andre2014}).  In HH\,111/HH\,121, we can study a dense core in a very violent star formation environment near the Herbig-Haro jets. 

The shape and the position of the HH\,121 jet with respect to NH$_{3}-$S poses the interesting question of whether its southern lobe  (emanating from VLA--1 at an angle of 15$^{\circ}$--20$^{\circ}$ east with respect to the northern lobe; \citealt{gredel1993})  was deflected off the dense material in the core, and as a result, changed its direction toward the west. During such a collision, the jet would have been strongly shocked. Indeed, the knots of the HH\,121 jet become strong beyond the area where such a collision would have taken place.  The HH\,110 jet located to the north from HH\,111 is an example of the jet deflected on a dense clump \citep{reipurth1991,reipurth1996}. The observational properties of HH\,110 such as the morphology and kinematics are in agreement with predictions of the analytic and numerical models of jet-cloud collisions (e.g., \citealt{raga1996}; \citealt{degouveia1999}).  The models show that  the emitting jet knots are still seen as coherent structures after the jet/cloud collision, as observed in HH\,121 (e.g., \citealt{raga1995}).   A collision between the HH\,121 jet with the core is a speculation at this point; a thorough analysis of the geometry and jet proper motions is needed to provide some insight into this possibility.  
  
 \subsection{Selective Freeze-out}
 
The chemical structure of the NH$_3-$S core shows evidence for ``selective'' freeze-out, an inherent characteristic of dense cold cores.  The abundance of carbon-bearing species such as CO and CS in the dense core centers can be a few orders of magnitude lower than at their edges, while the nitrogen-hydrogen bearing species such as N$_2$D$^{+}$ and NH$_3$ have a constant or slowly decreasing abundance (e.g., \citealt{caselli1999}; \citealt{caselli2002a}; \citealt{caselli2002b}; \citealt{bacmann2002}; \citealt{bergin2001,bergin2007}; \citealt{bergin2007}).  These abundance gradients are formed as a result of the gas-grain interactions that dominate the chemistry of the cores and lead to the freeze-out of important gaseous species (e.g., CO) and a subsequent formation of new species in the chemically altered environment. For example, the depletion of CO from the gas phase in the dense core centers leads to the production of species that are normally destroyed by CO, e.g. N$_2$H$^{+}$ and NH$_3$.

One of the consequences of the CO freeze-out is a great enhancement of the deuterium fractionation, i.e., the ratio of a deuterated species over its counterpart containing H (e.g., \citealt{roberts2000}; \citealt{busquet2010}; \citealt{bergin2007}). The deuterium fractionation toward HH\,111 was measured by \citet{hatchell2003} using the  [NH$_2$D]/[NH$_3$] ratio. 
\citet{hatchell2003} report spectroscopic observations centered on source VLA-1 in HH\,111 (and several other protostellar cores) of the NH$_3$ (1,\,1)--(4,\,4) lines with the Effelsberg 100 m telescope (HPBW$\sim$37$''$) and the NH$_2$D 1$_{11}$--1$_{01}$ line with the IRAM 30m telescope (HPBW$\sim$28$''$). They only detected the main hyperfine lines for NH$_3$ (1,\,1) ($v_{LSR}$ = 8.5 km~s$^{-1}$) and NH$_2$D toward HH\,111. Adopting the rotational temperature of 14.6 K derived for another source in Orion and correcting for different beam areas of the NH$_2$D and NH$_3$ observations, they derived the [NH$_2$D]/[NH$_3$] ratio for HH\,111 of 11\%. This value is much higher than 10$^{-5}$, the elemental value in the interstellar medium within the $\sim$10 K gas (e.g., \citealt{watson1974}; \citealt{oliveira2003}). The high value of deuterium fractionation determined based on the single dish spectroscopic observations indicate the presence of the dense cold material with depletion due to freeze-out in the HH\,111/HH\,121 protostellar system. The ALMA interferometric observations revealed the location of these regions. 

The chemistry of the dense cores changes with its evolution.  The effects of freeze-out are not important for less evolved starless cores when the density is less than a few times 10$^{4}$ cm$^{-3}$ (see e.g., \citealt{bergin2007}). At this time, the emission from molecules such as CO, C$^{18}$O, CS, and HCO$^{+}$ can be observed throughout the core. For more evolved sources, when the density is higher than a few times 10$^{4}$ cm$^{-3}$, the CO is frozen-out in the core center and nitrogen-bearing species (e.g., N$_2$H$^{+}$ and NH$_3$) become the best molecular tracers of the core gas.  With the CO frozen-out in its center,  NH$_3-$S seems to be at a later stage of the starless core evolution. 
 
\subsection{Alfv\'en waves}
 Two intriguing  findings from this study are that the $\rm  N_2D^+$ emission region actually lies at the center of the CO depletion region in NH$_3-$S, displaced from the  shell/rim where $\sigma_{nth}$$\ne$0, and that $I({\rm N_2D^+})\,\propto\,\sigma_{\rm nth}^{-1}$ (see Fig.~\ref{f:n2dp_nth}).   These observations are  consistent with theoretical predictions  for  chemistry in cold gas subject to the passage of magnetohydrodynamic (MHD) waves, presumably related to the existence of MHD turbulence \citep{charnley1998}.  The essential point is that the MHD waves in molecular clouds with the longest lifetimes are Alfv\'en waves \citep{arons1975}. In partially-ionized molecular clouds, the ion-electron plasma  experiences  MHD wave perturbations and moves relative to the neutral particles, undergoing collisional damping. The resultant relative ion-neutral streaming (i.e., ambipolar diffusion) imparts additional kinetic energy to collisions involving ions and neutral molecules and so this can nonthermally drive  endoergic chemical reactions   that would otherwise be inhibited at low temperatures ($\sim$10 K; e.g., \citealt{draine1980}).  
 
Low amplitude Alfv\'en waves can impart this additional kinetic energy to chemical reactions to drive them without significant gas heating.  In particular, the reaction underlying the gas-phase deuteration of interstellar molecules 
\begin{equation}
{\rm H_3^+ + HD ~~~\rightleftharpoons~~~ H_2D^+ + H_2 + 225\,K}
\end{equation}
is exothermic in the forward direction and proceeds rapidly at low temperatures. The rate of the reverse process depends on the quantum spin state of the H$_2$ molecules: when they are present in the LTE ${\it ortho}/{\it para}$ ratio (OPR) of 3:1 then the internal energy of  
${\it ortho}-$H$_2$ collisions can drive the reverse reaction at low temperatures (e.g., \citealt{pagani2011}). However, in cold molecular clouds it is expected that most of the H$_2$, formed and ejected from dust grains with an  OPR of 3:1, will be converted to  ${\it para}-$H$_2$ in ion-molecule spin-exchange reactions (\citealt{pagani2011}; \citealt{wirstrom2012}).
The most direct  effect of Alfv\'en waves connected to MHD turbulence would therefore be destruction of $\rm  H_2D^+$ in cold gas and a suppression of the $\rm  H_2D^+$/$\rm  H_3^+$ ratio. However, the most sensitive and easily  detectable effect, in the millimeter wavelength region, is connected to the $\rm  N_2D^+$/$\rm  N_2H^+$ ratio \citep{charnley1998} which is predicted to be suppressed over the ion-neutral collisional damping length, $L_{ni}$, given by: 
\begin{IEEEeqnarray}{r}
L_{ni} [{\rm cm}]  = 3.45 \times 10^{16}
\left( B \over 100 \mu \rm G \right)
\left( n_{\rm H} \over 10^4 \rm cm^{-3} \right)^{-3/2}  \\ 
\times \left( x_e \over 10^{-7} \right)^{-1}, ~~\rm \nonumber
\end{IEEEeqnarray}
where $ n_{\rm H}$ is the hydrogen nucleon density, $x_e$ is the fractional ionization, and $B$ is the magnitude of the magnetic field \citep{markwick2000}. 

The VLA--1/VLA--2 region or the HH\,111 and HH\,121 outflows could be the sources of these putative  Alfv\'en waves. Numerical simulations by \citet{decolle2005} show that ejection of high-density clumps can generate Alfv\'en waves  in the ambient material perpendicular to the direction of the jet motion.   Assuming constant physical conditions between the possible wave sources and NH$_3-$S, we can estimate $L_{ni}$ to see if this is plausible. Taking  $x_e  = 3\times10^{-8}$ and $ n_{\rm H}= 2\times 10^4$ cm$^{-3}$, typical of dark clouds, and  B =160 $\mu \rm G$ at this density\citep{crutcher2012}, we find $L_{ni} = 5.86 \times 10^{16}$ cm.  The distance between VLA-1 and the 50\% of the NH$_3$ (1,\,1) emission peak contour in HH\,111/HH\,121 is 10$''$ which corresponds to $5.98 \times 10^{16}$ cm and so, given the approximations made, VLA--1 could possibly be a source of waves.  Alternatively, Fig.\ref{f:12co} shows that any waves originating in the HH\,121 and HH\,111 outflows, and  emanating perpendicular to the outflow direction, are closer to 50\% of the NH$_3$ emission peak contour and so would  indeed impact the regions of  NH$_3-$S that show no $\rm  N_2D^+$ emission.  It would be  interesting to produce a complementary map of the  $\rm  N_2H^+$ emission to obtain the  spatial $\rm  N_2D^+$/$\rm  N_2H^+$ to see if it is also inversely proportional to $\sigma_{\rm NT}$ and test the Alfv\'en wave scenario further. For sufficiently large amplitudes such MHD waves will steepen into weak C-shocks and these can also have observable effects \citep{pon2012}.

\subsection{External Illumination}
The high intensity of the NH$_{3}$ emission from NH$_{3}-$S suggests that the source may also be affected by the strong UV radiation from HH\,111 and HH\,121 jets.  Observational surveys have detected compact regions of enhanced emission (compared to quiescent dark clouds) in several molecules, among them NH$_{3}$ and HCO$^{+}$, just ahead of Herbig-Haro objects (e.g., \citealt{torrelles1992}; \citealt{girart1994}; \citealt{girart1998}), as well as along the jets (e.g., \citealt{christie2011}). These ``externally illuminated clumps''  are quiescent, cool ($\sim$10--20 K), have sizes of 10$''$--20$''$ (0.019--0.038 pc at 400 pc), and have similar chemical properties.  The high molecular abundances from some species found in these clumps (NH$_{3}$ and HCO$^{+}$, but also  CH$_{3}$OH, H$_{2}$CO, SO$_{2}$, and others) suggest a chemical alteration of the high-density quiescent clumps in molecular clouds induced by the radiation generated in the Herbig-Haro shocks (e.g.,  \citealt{girart1994}; \citealt{viti1999}).  In this theoretical picture, UV radiation can photodesorb molecules from icy grain mantles  containing H$_{2}$O, NH$_{3}$, CH$_{4}$, CH$_{3}$OH, H$_{2}$CO, etc. \citep{boogert2015} and drive an active gas-phase photochemistry.  This interpretation is supported by both observations and theoretical models (both static and dynamic where the radiation source is moving; \citealt{taylor1996}; \citealt{christie2011}).  The classical examples of externally illuminated clumps are those found near HH\,1/2 (e.g., \citealt{girart2002}),  HH \,7-11 (e.g.,  \citealt{dent1993}),  HH \,34  (e.g., \citealt{rudolph1992}),  and HH\,80N (e.g., \citealt{girart1994,girart1998,girart2001}).  \citet{girart2001} found star formation signatures in the HH\,80N dense clump. Unlike NH$_{3}-$S, these clumps have been detected in CO and its isotopologues. 

\begin{figure*}[ht!]
\centering
\includegraphics[width=0.32\textwidth]{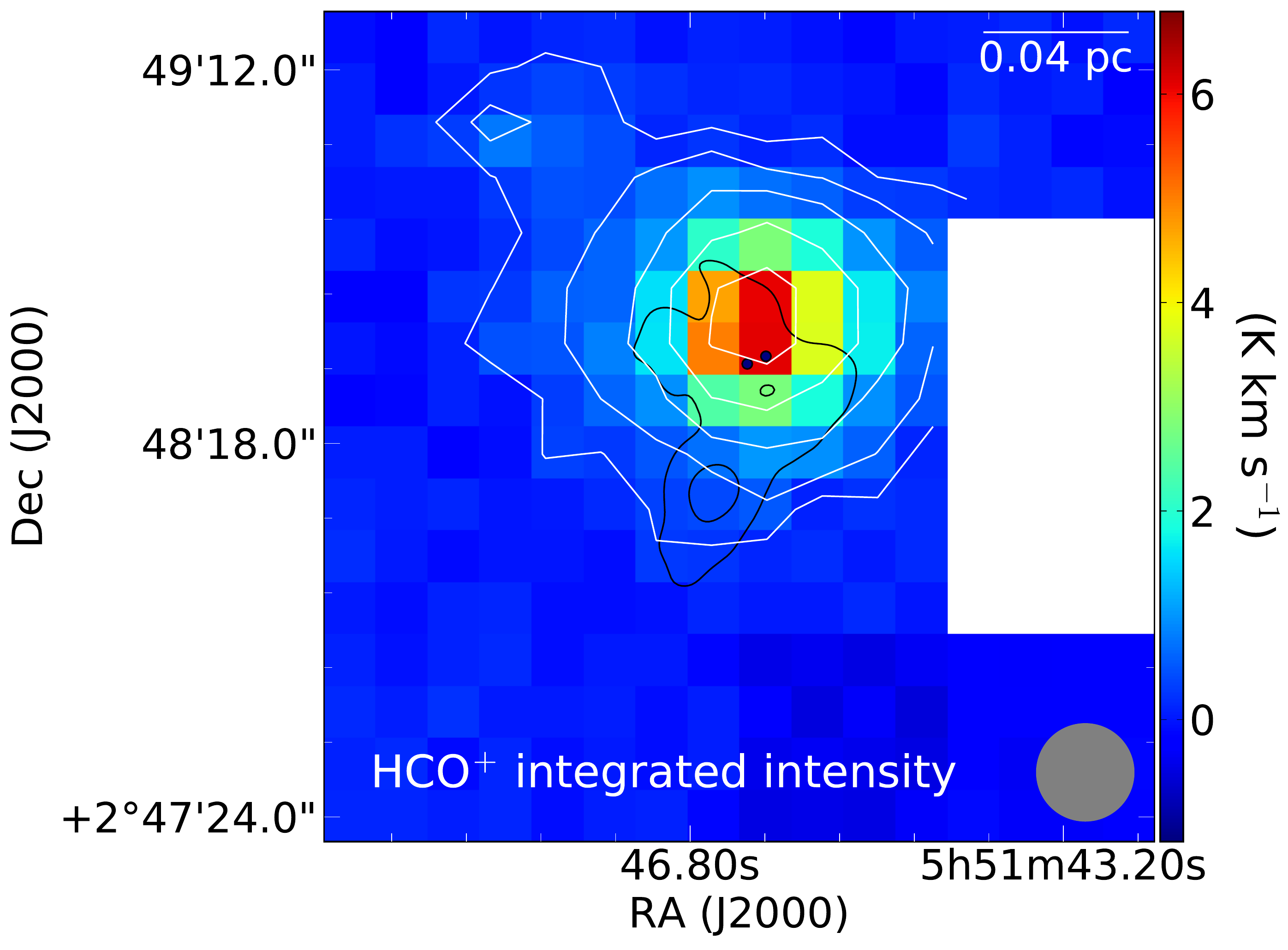}
\includegraphics[width=0.33\textwidth]{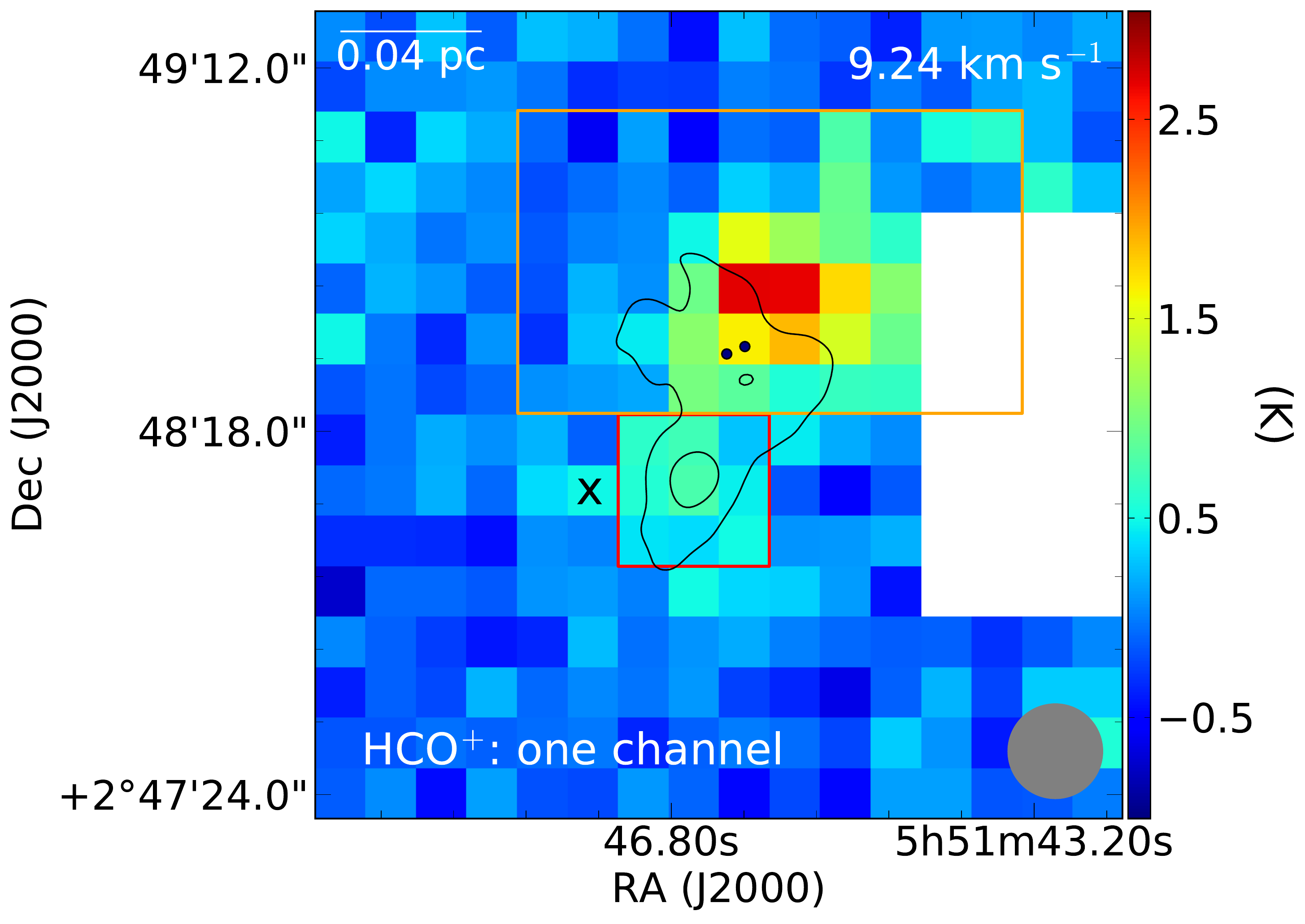}
\includegraphics[width=0.33\textwidth]{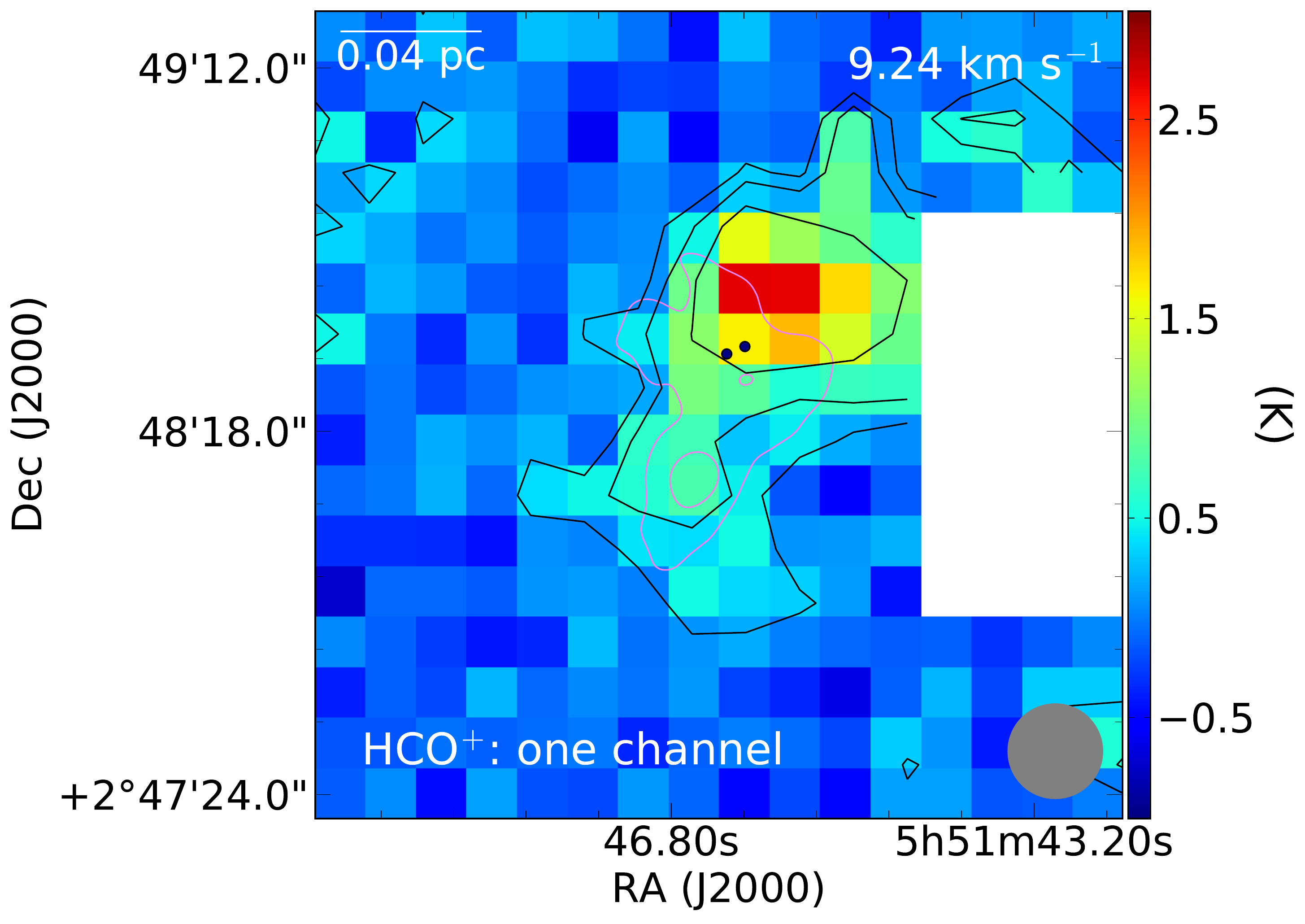}
\caption{{\it Left}: The HCO$^{+}$ (4--3) integrated intensity image. The HCO$^{+}$ contours are shown in white; the contour levels are (5, 10, 20, 40, 80)\% $\times$ 6.08 K~km~s$^{-1}$, the maximum value of the integrated intensity. The NH$_3$ (1,\,1) contours levels shown in black correspond to the 20\% and 75\% of the peak NH$_3$ emission (see Fig.~\ref{f:mom0}).  {\it Center} and {\it Right}: The HCO$^{+}$ channel map corresponding to the velocity of 9.24 km~s$^{-1}$.  The NH$_3$ contours as in the {\it left} panel.  The HCO$^{+}$ contours in the {\it right} panel correspond to (10, 20, 40, 60, 90)\% $\times$ 2.68 K, the peak HCO$^{+}$ emission. The red square in the {\it middle} panel indicates pixels with spectra shown in Fig.~\ref{f:spechcop}, and the orange rectangle those shown in Fig.~\ref{f:spechcopmain}.  The spectrum extracted from the pixel indicated with a black `$\times$' sign is shown in Fig.~\ref{f:hcopblue}. The positions of protostars VLA-1 and VLA-2 are indicated with black filled circles.  Two out of 16 HARP receptors (the 4$\times$4 array; see Section~\ref{s:jcmtdata}) were not working during the observations, resulting in a white space seen in the images. The size of the JCMT beam is shown in the lower right in each image (HPBW$\sim$14$''$).  \label{f:hcop}}
\end{figure*}

\begin{figure}[ht!]
\centering
\includegraphics[width=0.45\textwidth]{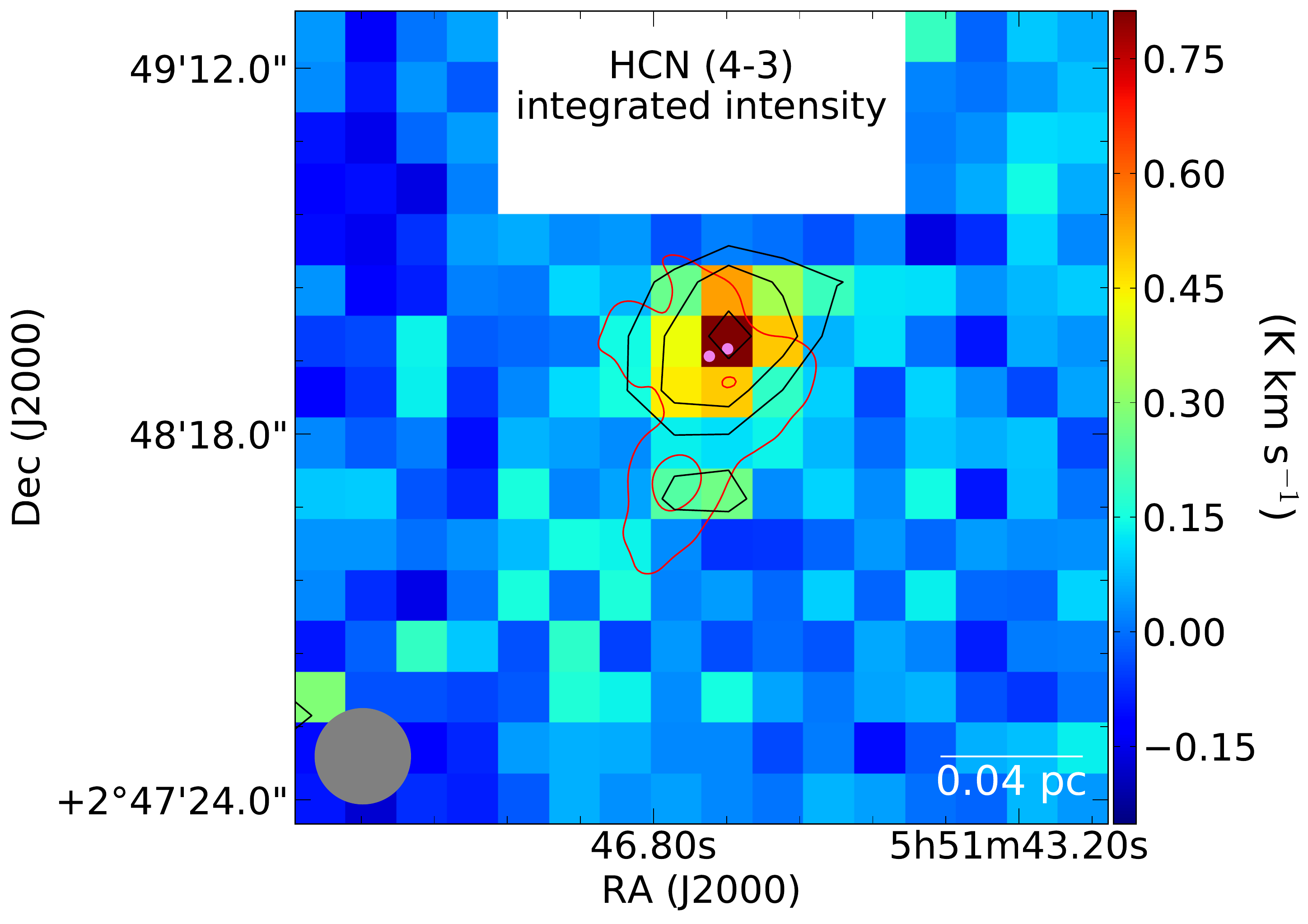}
\caption{The HCN (4--3) integrated intensity image. The black contours represent the HCN emission; the contour levels are (20, 40, 80)\% $\times$ 0.94 K~km~s$^{-1}$, the peak HCN emission. The NH$_3$ contours shown in red as in Fig.~\ref{f:hcop}. The VLA-1 and VLA-2 positions are indicated with violet filled circles. The size of the JCMT beam is shown in the lower left  (HPBW$\sim$14$''$).  \label{f:hcn}}
\end{figure}

\begin{figure*}[ht!]
\centering
\includegraphics[width=0.3\textwidth]{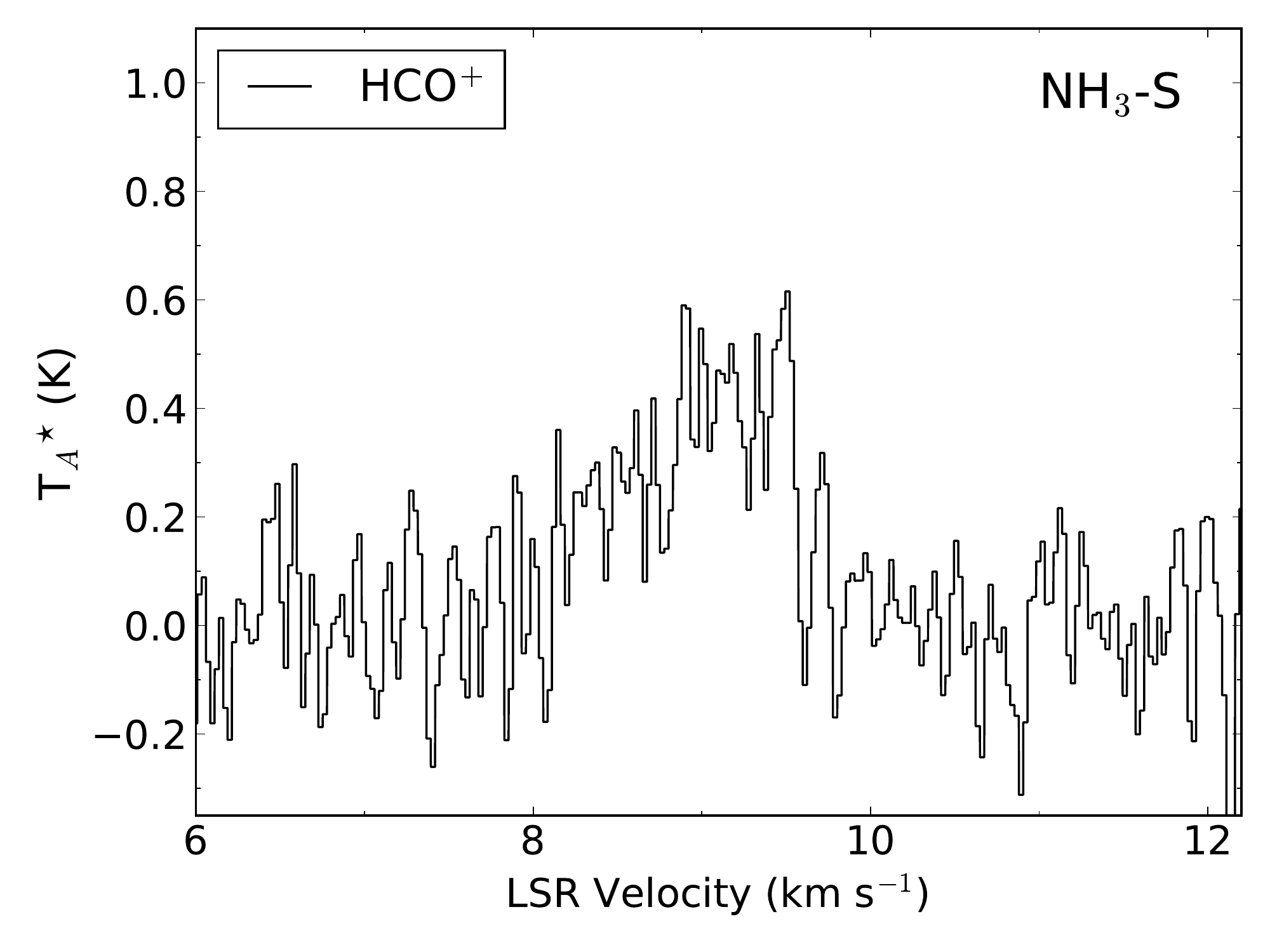}
\includegraphics[width=0.3\textwidth]{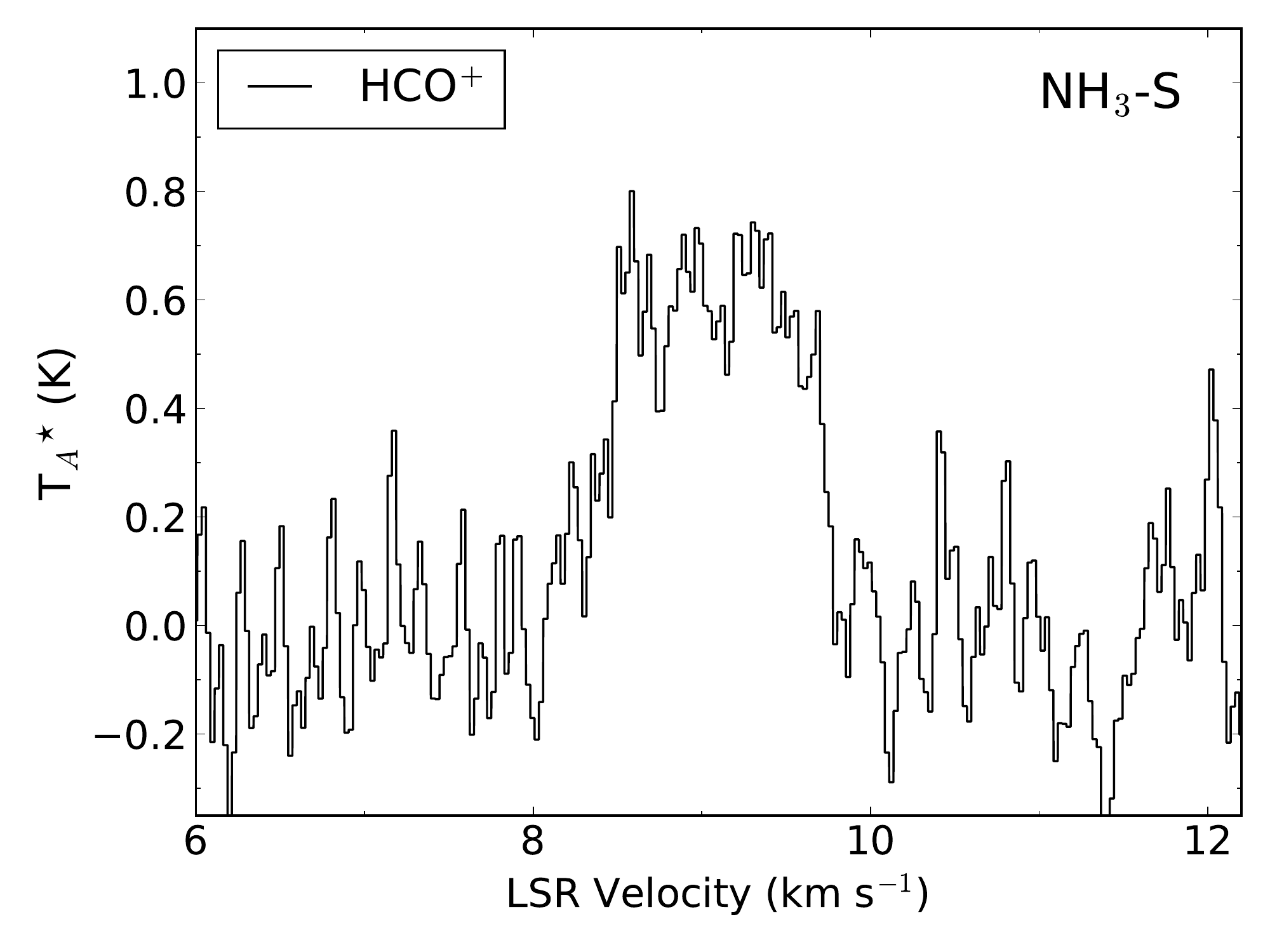}
\includegraphics[width=0.3\textwidth]{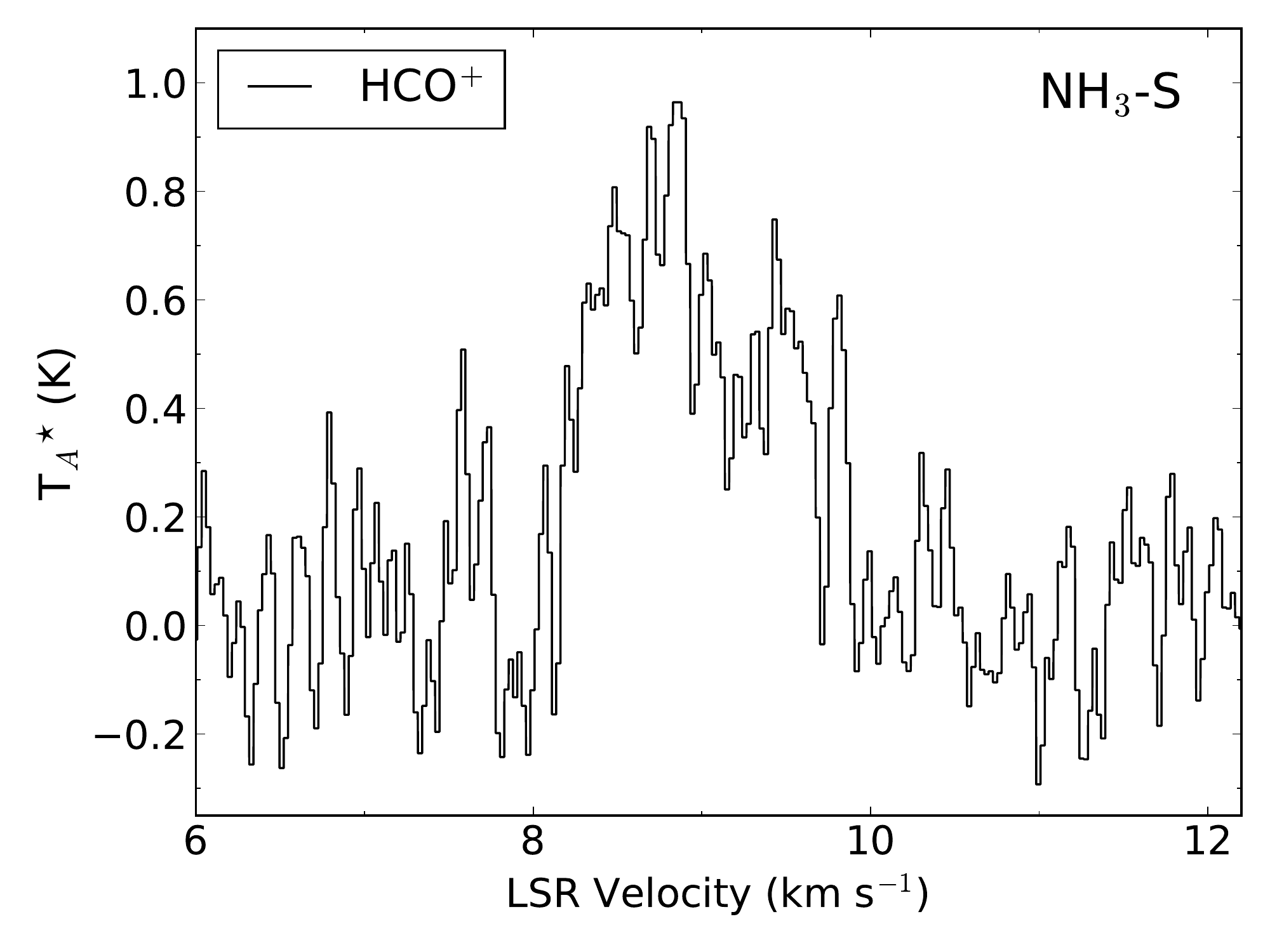}
\includegraphics[width=0.3\textwidth]{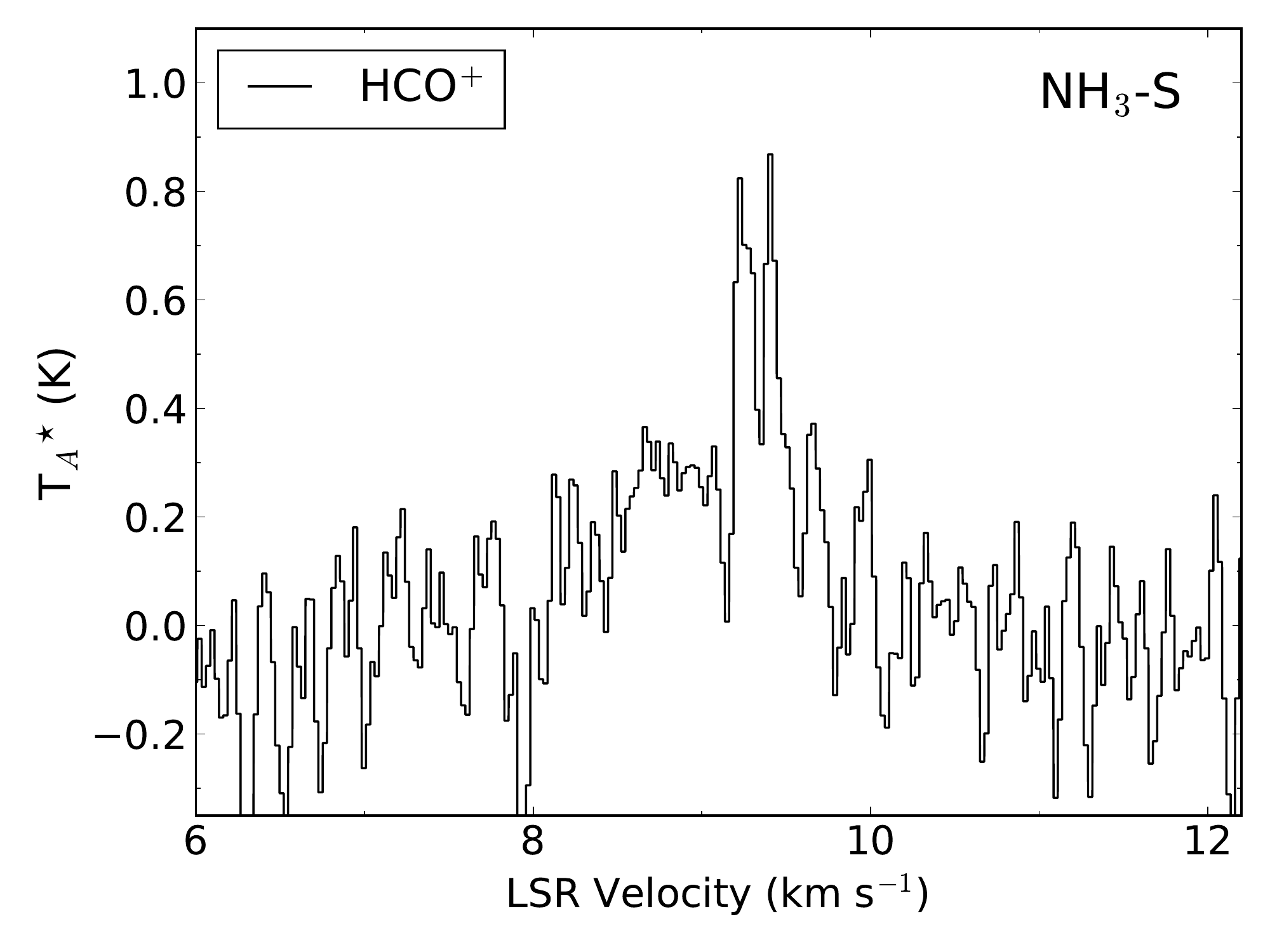}
\includegraphics[width=0.3\textwidth]{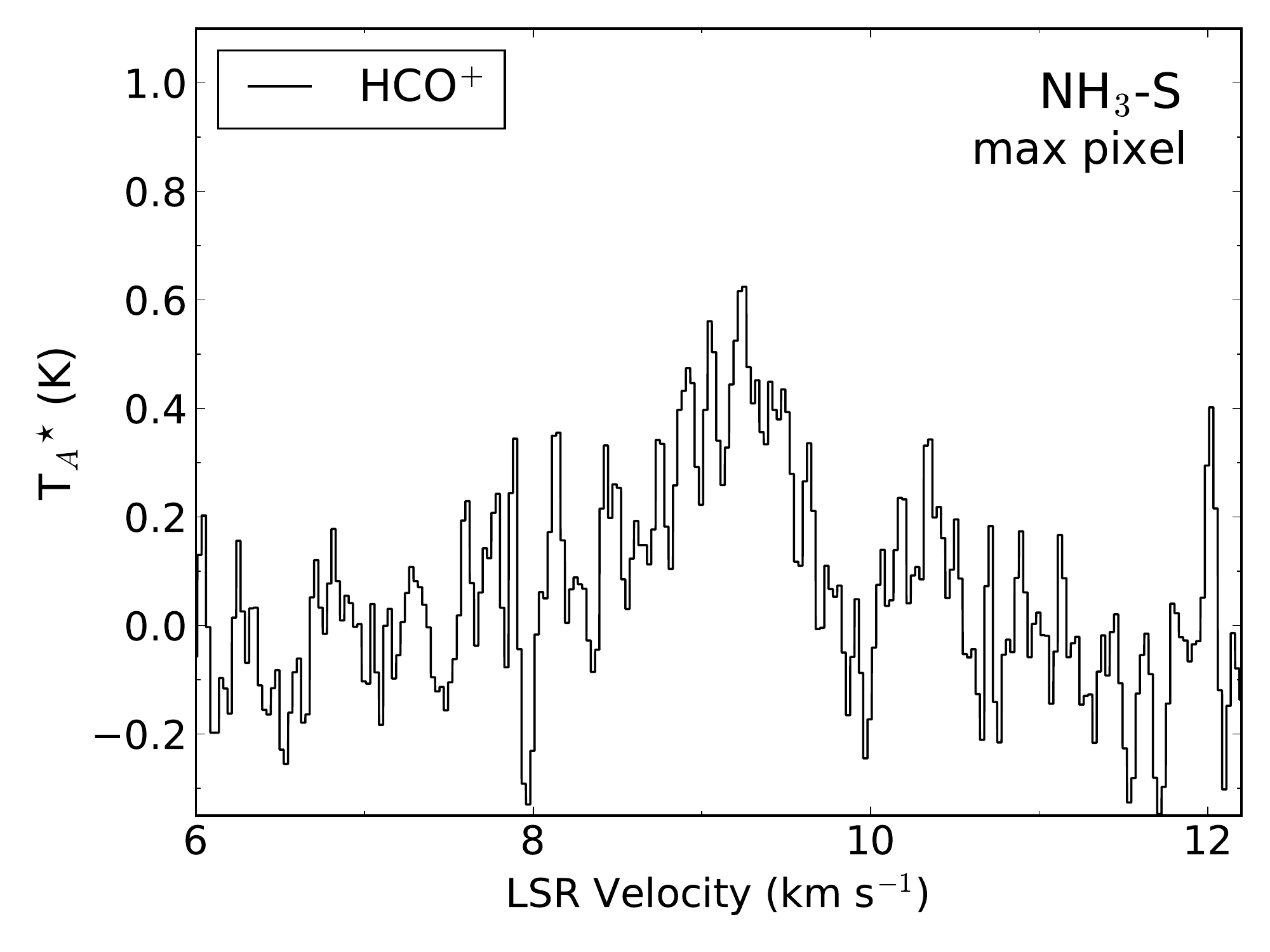}
\includegraphics[width=0.3\textwidth]{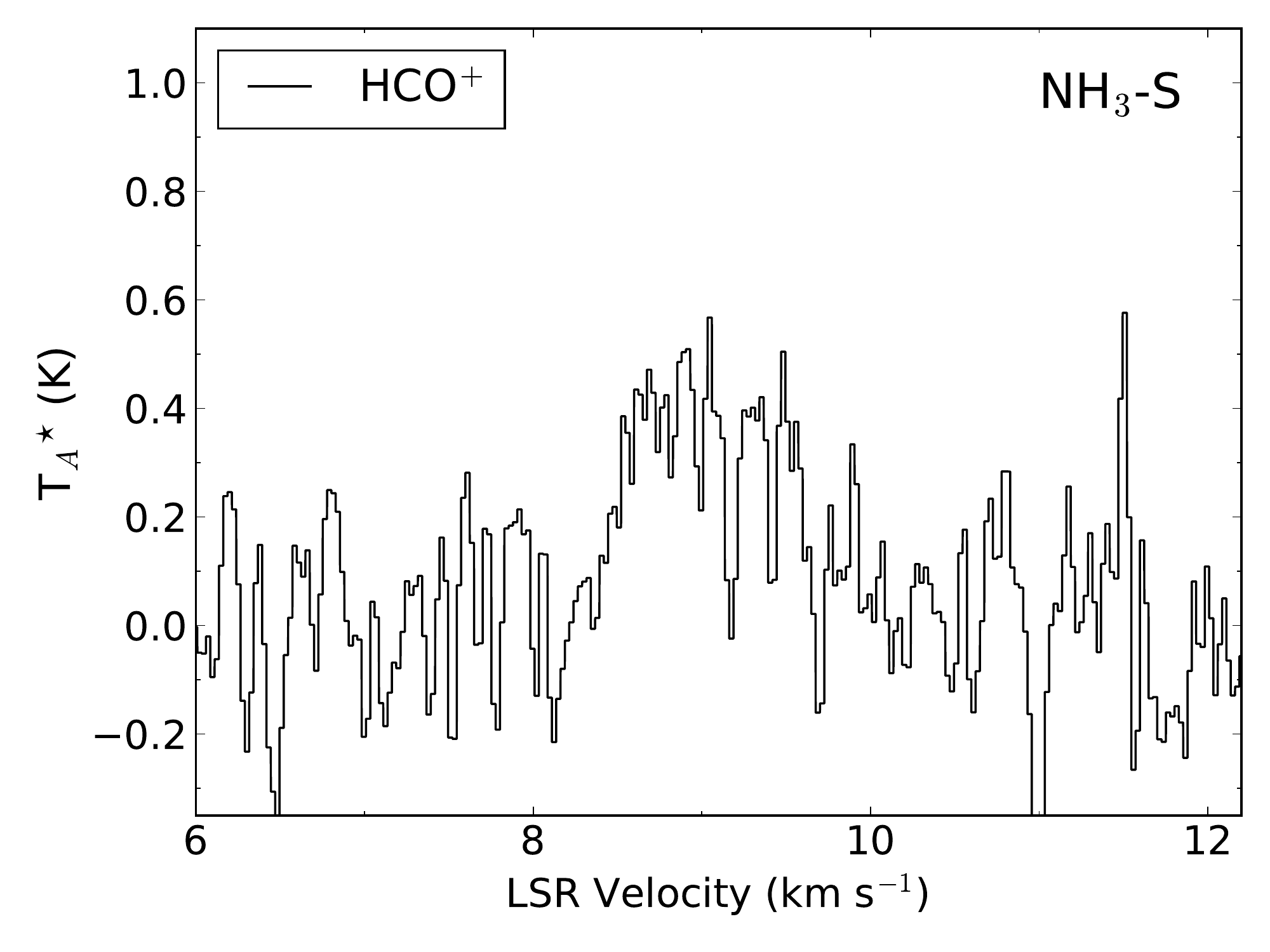}
\includegraphics[width=0.3\textwidth]{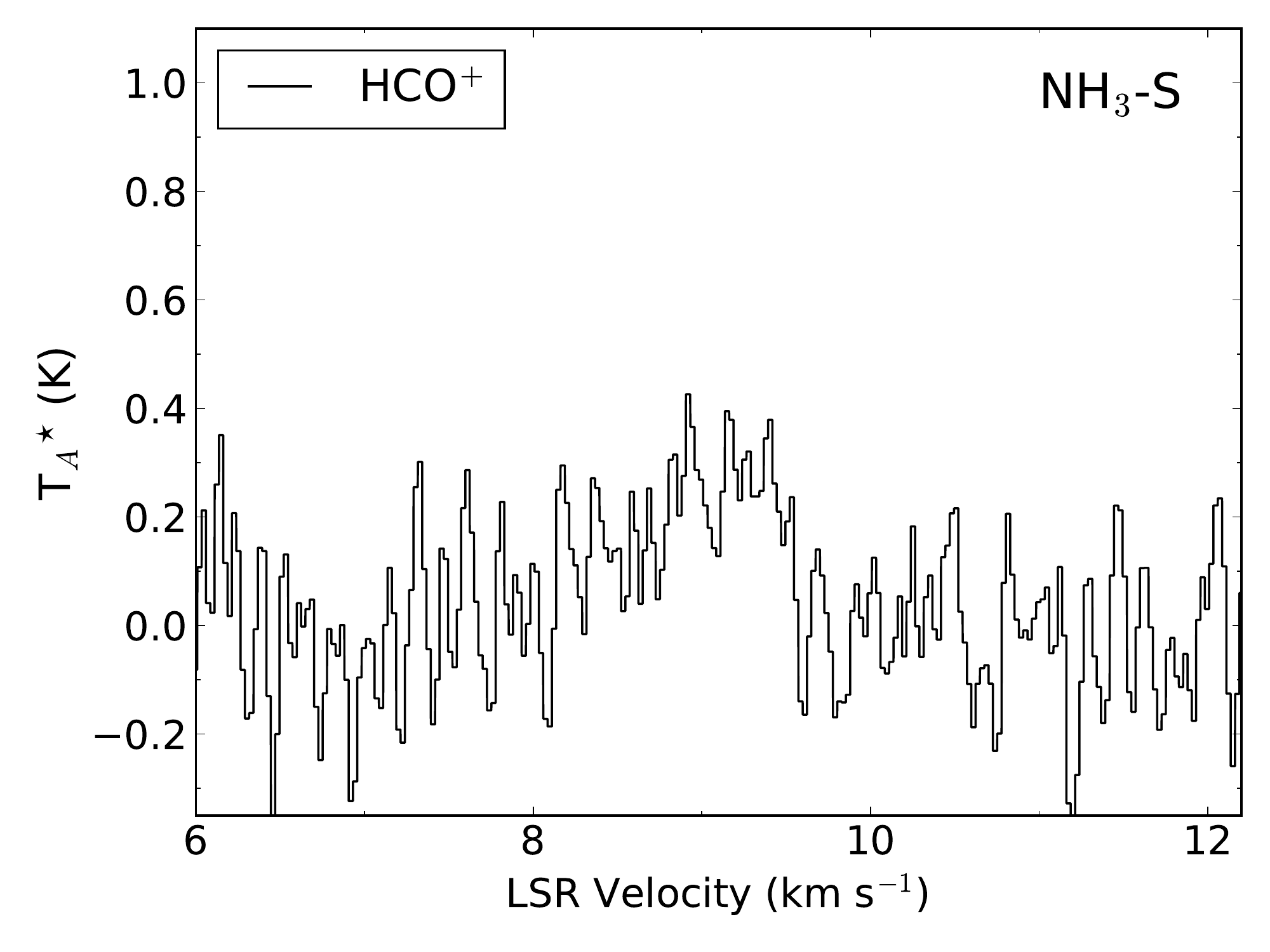}
\includegraphics[width=0.3\textwidth]{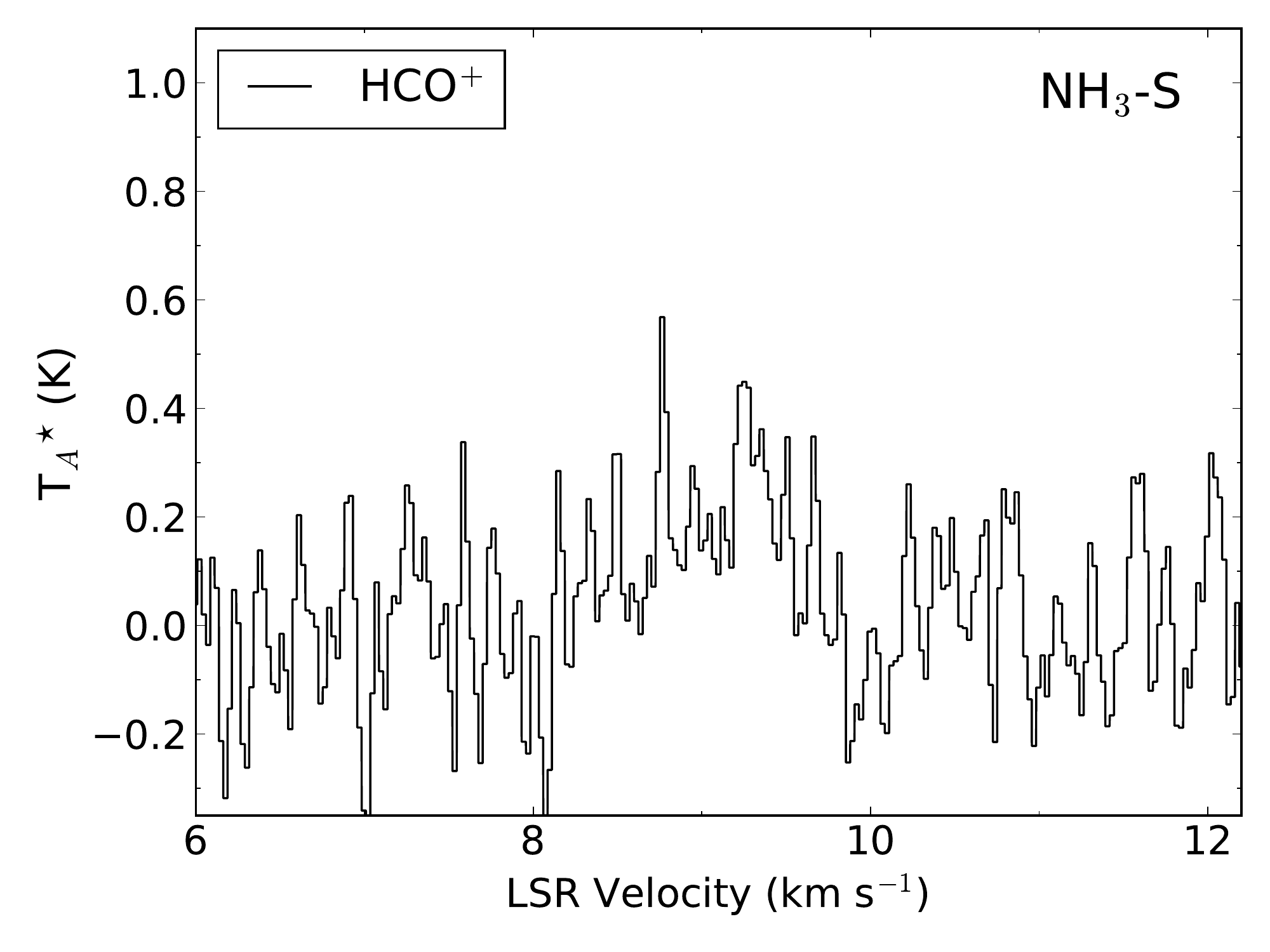}
\includegraphics[width=0.3\textwidth]{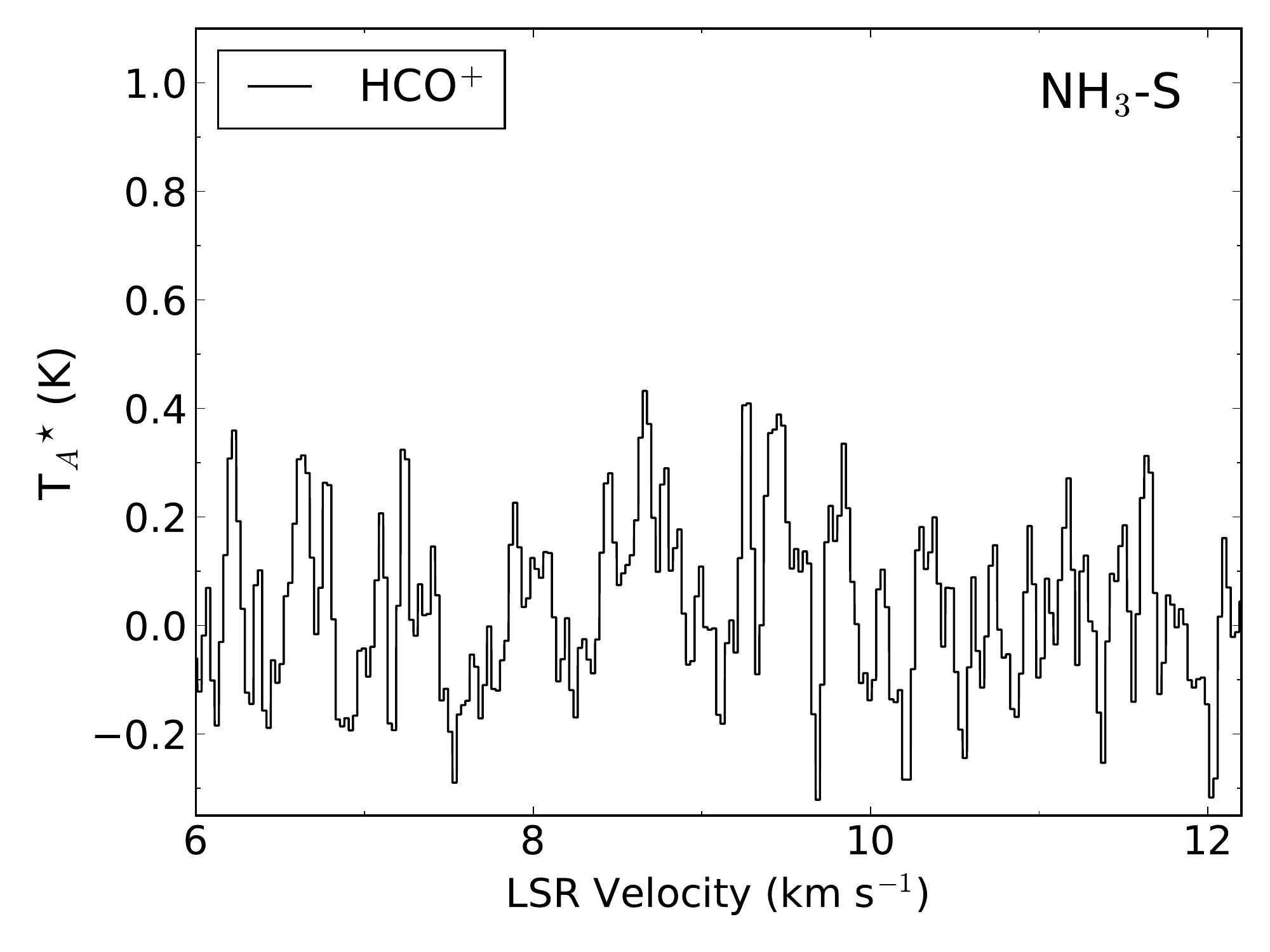}
\caption{The HCO$^{+}$ (4-3) spectra for individual pixels enclosed in the red square in Fig.~\ref{f:hcop}, centered on the pixel associated with the NH$_3-$S HCO$^{+}$ peak emission in the $\sim$9.24 km s$^{-1}$ channel.  The spectra were Hanning smoothed with the smoothing kernel width of three. \label{f:spechcop}}
\end{figure*}

\subsection{HCO$^{+}$ and HCN}
\label{s:hcop}

We conducted the JCMT HCO$^{+}$ (4--3) and HCN (4--3) observations to test the idea that NH$_3-$S is externally illuminated by UV radiation from the Herbig-Haro objects.  Observations of externally illuminated clumps show that the HCO$^{+}$ emission is enhanced in these regions with line intensities much stronger than expected in quiescent dark clouds, which was successfully explained by theoretical models (see above; e.g., \citealt{taylor1996}; \citealt{viti1999}; \citealt{girart2002}; \citealt{viti2003}; \citealt{christie2011}).   HCN is expected to be underabundant (e.g., \citealt{girart2002}).  The HCO$^{+}$ data can also be used to confirm the CO freeze-out in the center of NH$_3-$S. If CO is completely frozen-out, no HCO$^{+}$ should be detected. 

HCO$^{+}$ is an optically thick molecule that is sensitive to the infall and outflow motions due to the presence of self-absorption. Blue asymmetries in the line profiles are a good indicator of gas infall motions, while red asymmetries indicate an outflow (e.g., \citealt{myers1996}; \citealt{evans1999}; \citealt{chira2014}).  \citet{tobin2013} found that HCO$^{+}$ is a good tracer of the warm, inner envelope ($<$1000 AU) in protostars, but sufficiently high resolution interferometric observations are required to confirm that the HCO$^{+}$ emission originates in the envelope rather than the outflow.     

Figures~\ref{f:hcop}  and \ref{f:hcn}  show the HCO$^{+}$ and HCN integrated intensity images, respectively.  Both images show only source NH$_3-$Main.  An inspection of the data cubes confirmed that an HCN emission is not detected toward NH$_3-$S. However, NH$_3-$S can be clearly separated from NH$_3-$Main in the HCO$^{+}$ channel corresponding to the velocity of 9.24 km~s$^{-1}$ (see Fig.~\ref{f:hcop}).  The peak HCO$^{+}$ emission for NH$_3-$S at this velocity is at the $\sim$4$\sigma$ level and its emission peak coincides with the peak of the ammonia emission.   A detection of faint HCO$^{+}$ emission toward NH$_{3}-$S is not consistent with the predictions of theoretical models of external illumination by the strong ultraviolet radiation. The peak of the HCO$^{+}$ emission in NH$_3-$Main is offset from the position of the protostar VLA-1 in the direction of the HH\,111 jet, indicating that the HCO$^{+}$ emission at this velocity traces the outflow.

\begin{figure}[h!]
\centering
\includegraphics[width=0.4\textwidth]{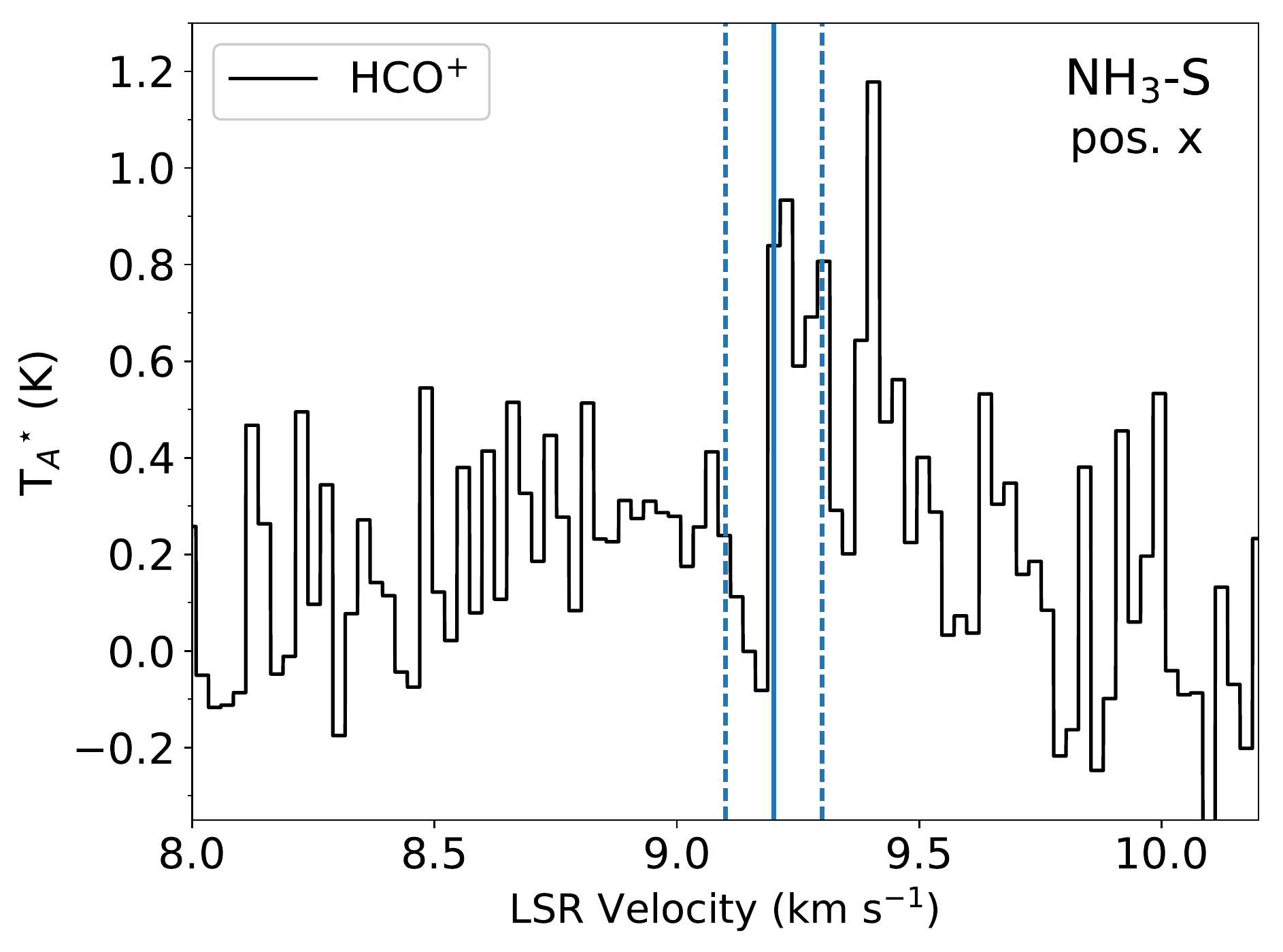}
\caption{The HCO$^{+}$ line profile extracted from a single pixel marked with an `$\times$' sign in Fig.~\ref{f:hcop}b; this is an unsmoothed version of the spectrum shown in the middle of the left panel in Fig.~\ref{f:spechcop}.  The blue solid line indicates the velocity of NH$_3-$S measured from the ammonia data; the dashed lines indicate the 1$\sigma$ uncertainty in velocity (see Table~\ref{t:physpar}). \label{f:hcopblue}}
\end{figure} 
 
Figure~\ref{f:spechcop} shows the HCO$^{+}$ line profiles for a grid of 3$\times$3 pixels centered on the peak HCO$^{+}$ emission in NH$_3-$S (see Fig.~\ref{f:hcop}).  The line profiles show multiple velocity components and the lines get brighter toward the position of NH$_3-$Main, indicating that the bright emission from the envelope and outflow associated with protostar VLA-1 contributes to the line profiles. The JCMT observations have relatively low spatial resolution (HPBW$\sim$14$''$) and disentangling the contribution to the line profiles from two neighboring sources with a high HCO$^{+}$ intensity contrast is difficult.  The high spatial and spectral resolution observations with high sensitivity are needed to search for the infall and outflow signatures in NH$_3-$S using HCO$^{+}$. 

Figure~\ref{f:spechcop} shows that the brightest HCO$^{+}$ emission that can be kinematically associated with NH$_3-$S originates at the eastern peripheries of the source, outside the area with CO depletion (see e.g., Fig.~\ref{f:C18Omom}).  The HCO$^{+}$ line profile extracted from a single 7$\rlap.{''}$5$\times$7$\rlap.{''}$5 pixel shows signatures of infall - it is self-absorbed, double peaked with the blue peak brighter than the red peak  (see Fig.~\ref{f:hcopblue} and the middle plot in the left panel in Fig.~\ref{f:spechcop}).  However, these velocity peaks may be two separate velocity components.  The origin of the higher velocity component at $\sim$9.4 km s$^{-1}$ is uncertain. There is not enough evidence to draw firm conclusions on the presence of infall. 

For a comparison, in Fig.~\ref{f:spechcopmain} in Appendix~C, we show the HCO$^{+}$ line profiles for NH$_3-$Main for a grid of 10$\times$6 pixels; this area covers the VLA-1 envelope and the base of the outflows (see Fig.~\ref{f:hcop}).  The HCO$^{+}$ line profiles show the signatures of infall and outflow.   A detailed analysis of the HCO$^{+}$ and HCN data for NH$_3-$Main is out of scope of this paper.

\subsection{HCO${^+}$/CO ratio}
\label{s:hcopco}

An interesting fact is that whilst CO and its isotopologues are clearly depleted in ${\rm NH_3-S} $  and elsewhere, HCO$^+$ emission is nevertheless detectable. Although this may appear counterintuitive, it can be shown to be  a natural consequence of ion-molecule chemistry by a simple analysis (\citealt{charnley1997}; see Appendix D).   In this case the $\rm HCO{^+}/CO$ number density ratio is given by 
  \begin{IEEEeqnarray}{r}
 $${ 
 {n({\rm HCO^+})\over n({\rm CO}) } =      \left[  {\alpha \over k_i } +   {n({\rm CO}) \over  n_e }  \right]^{-1}
  }$$ 
\end{IEEEeqnarray}
 where $n_e $ is the electron number density and $\alpha$ and $k_i $ are generic rate coefficients for electron dissociate recombination and proton transfer, respectively.

If  we adopt typical 10 K rate coefficients of $k_i \approx\rm10^{-9}~cm^3~s^{-1}$ and  $\alpha\approx\rm10^{-7}~cm^3~s^{-1}$, and taking $n_e\sim10^{-8} $, we  can identify two limiting cases depending on the $n({\rm CO})/n_e $ ratio.  Assuming   CO is present at its typical undepleted abundance,  where $n({\rm CO})\sim10^{-4}\,n({\rm H_2})$, then  $ {n({\rm HCO^+})/ n({\rm CO}) }\sim10^{-4}$, as is typically found. In the case where significant CO depletion has occurred, $n({\rm CO})\sim n_e$,  $ {n({\rm HCO^+})/ n({\rm CO}) }\sim$ $k_i / \alpha\sim10^{-2}$. Thus, even when CO is  depleted to abundances of $\sim10^{-8}-10^{-7} $, the ${\rm HCO^+}$ abundance could be in the range $\sim10^{-10}-10^{-9} $ and remain detectable;  ${\rm H^{13}CO^+}$ may therefore also be detectable in NH$_3-$S.

\subsection{The Virial Mass}

To investigate whether NH$_3-$S is unstable to gravitational collapse, we calculate the virial parameter ($\alpha$) defined as $\alpha = M_{{\rm vir}} / M$, where $M_{\rm{vir}}$ is the virial mass:
\begin{equation}
\label{e:vir}
M_{{\rm vir}} = \frac{5 \sigma_v^2 R}{G}
\end{equation}
(e.g., \citealt{maclaren1988}; \citealt{mckee1992}; \citealt{enoch2008}; \citealt{kauffmann2013}), and $M$ is the observed source mass. In Eq.~\ref{e:vir}, $\sigma_v$ is the total linewidth of the molecular gas, $R$ is the radius of the core, and $G$ is the gravitational constant. The total linewidth is the combination of non-thermal gas motions ($\sigma_{nth}$ calculated from NH$_{3}$) and the thermal motions of the particle of mean mass (or a thermal sound speed, $c_s$): $\sigma_v = \sqrt{c_s^2 + \sigma_{nth}^2}$.  For $R$, we use the ``effective radius'' ($FWHM_{eff}$/2, see above). To account for the central condensation of the core, $M_{{\rm vir}}$ calculated using Eq.~\ref{e:vir} can be divided by a parameter `a' given by $a = \frac{1-p/3}{1-2p/5}$ for a power-law density profile $\rho (r) \propto r^{-p}$. We adopt $p$ = 1.5, giving $a = 1.25$.   We derive $M_{{\rm vir}}$ of 0.75 M$_{\odot}$ for NH$_3-$S using Eq.~\ref{e:vir} or 0.6 M$_{\odot}$ after correcting for non-uniform density profile.

Since the virial parameter is related to the ratio of the kinetic to potential energy, it can be used to assess the stability of the core (see e.g., \citealt{kauffmann2013} and references therein).  For $\alpha \gg 1$, the kinetic energy dominates and clumps/cores will expand and disperse, while those with  $\alpha \ll 1$ are often unstable and will likely collapse. The homogenous and spherical clumps/cores with $\alpha = 1$ are considered gravitationally bound and virialized. In the above estimate of the virial parameter, only the gravity and velocity dispersion are considered. The external pressure that may confine the clumps or magnetic fields that can support the cores against self-gravity are neglected. 

For NH$_3-$S, we derive the virial parameter $\alpha$ of 2.5 using $M_{{\rm vir}}$ of 0.75 M$_{\odot}$, and the observed mass $M$ of 0.3 M$_{\odot}$ derived from the NH$_3$ data. Theoretical models show that non-magnetized clumps/cores with $\alpha  \lesssim 2$ are gravitationally bound (e.g., \citealt{bertoldi1992}; \citealt{kauffmann2013}).  Due to large uncertainties of the virial parameter estimation, it is not clear whether NH$_3-$S is a marginally gravitationally bound or unbound (pressure-confined) starless core. The observations (e.g., \citealt{tachihara2002}; \citealt{morata2005}) and theoretical results (e.g., \citealt{klessen2005} for the turbulent fragmentation model) indicate that starless cores have virial masses larger than their actual masses or are near equipartition (e.g., \citealt{caselli2002}).  The models show that gravitationally unbound cores may still collapse if they are compressed by turbulence  (e.g., \citealt{gomez2007}).

\section{Summary and Conclusions}
\label{s:summary}

We present the results of  VLA NH$_3$ (1,\,1) and (2,\,2) observations of the HH\,111/HH\,121, combined with the analysis of  JCMT HCO$^{+}$ and HCN observations, and archival ALMA $^{13}$CO, $^{12}$CO, C$^{18}$O, N$_2$D$^{+}$, and $^{13}$CS data.  We detected two ammonia sources in HH\,111/HH\,121. One of the ammonia sources (NH$_3-$Main) is associated with  HH\,111 and traces the envelope of the protostar that is the source of the Herbig-Haro jet. The second ammonia source (NH$_3-$S) located $\sim$15$''$ ($\sim$0.03 pc) toward south-east is a new detection.  The HH\,111/HH\,121 protostellar system and its surroundings have been thoroughly covered by the observations from optical to cm wavelengths, yet NH$_{3}-$S remained undetected until our NH$_{3}$ observations. 

We use the NH$_3$ data to derive the kinematic and physical properties of NH$_3-$Main and NH$_3-$S, including the velocity and velocity dispersion, the kinetic temperature, excitation temperature, NH$_3$ column density, and mass. NH$_3-$Main and NH$_3-$S are two distinct velocity components with the NH$_3$ line-center velocities separated by $\sim$1 km~s$^{-1}$.  The reason for multiple velocity components in a single region (also observed toward other protostars) could be related to the initial conditions in the clouds in agreement with the the theory of gravoturbulent star formation.  

The carbon-bearing molecular emission traces the envelope and disk of the central source VLA-1 and molecular outflows associated with the HH\,111 and HH\,121 jets, i.e.  the region coinciding with NH$_{3}-$Main.  No C$^{18}$O, $^{12}$CO, and $^{13}$CO emission is detected in the center of NH$_{3}-$S.  However, the $^{13}$CO and C$^{18}$O emission wraps around the source roughly from east to west along its northern rim.  The $^{13}$CS emission is confined to the envelope of VLA-1. 

There are two N$_{2}$D$^{+}$ condensations in HH\,111/HH\,121.  One of the condensations is located in NH$_{3}-$Main. The second N$_{2}$D$^{+}$ condensation is associated with NH$_3-$S with the peak N$_{2}$D$^{+}$ emission coinciding with the peak of the NH$_3$ emission.  The observable abundances of N$_{2}$D$^{+}$ can only be achieved in the coldest and densest molecular cores where CO is frozen-out onto dust grains. A non-detection of CO in the center of NH$_3-$S provides evidence for this ``selective'' freeze-out, which is an inherent property of dense cold cores.

Based on the ammonia data, we determined the turbulent velocity dispersions in the region.  In NH$_{3}-$Main, the turbulent velocity dispersion is supersonic close to the protostar VLA-1 which is the source of the jet, indicating that the jet is interacting with the envelope material.  NH$_3-$S has subsonic internal motions and roughly constant observed linewidths, consistent with it being a ``coherent core''.  

Two interesting results of this study are that the N$_2$D$^+$ emission region lies at the center of the CO depletion region in NH$_3-$S, displaced from the rim where the non-thermal velocity dispersion is enhanced and that the intensity of the N$_2$D$^+$ emission is inversely proportional to the non-thermal velocity dispersion.  These observations are  consistent with theoretical predictions  for  chemistry in cold gas subject to the passage of MHD waves, presumably related to the existence of MHD turbulence. The MHD waves in molecular clouds with the longest lifetimes are Alfv\'en waves, which can be generated by the HH\,111 and HH\,121 outflows in the direction perpendicular to the jet motion and impact the regions of NH$_3-$S that show no N$_2$D$^{+}$ emission. 

Another interesting fact is that whilst CO and its isotopologues are clearly depleted in NH$_3-$S and elsewhere, HCO$^+$ emission is nevertheless detectable.  We show that it is a natural consequence of ion-molecule chemistry.   We also investigated a possibility that NH$_3-$S is an externally illuminated clump.  A detection of faint HCO$^{+}$ emission toward NH$_{3}-$S in our JCMT observations is not consistent with the predictions of theoretical models of external illumination by the strong ultraviolet radiation. 

The physical and chemical properties of NH$_3-$S, the fact that it is located in the dark cloud and there is no indication of the presence of the central object suggest that NH$_3-$S is a starless core.  The environment of the core is turbulent with turbulence induced by two Herbig-Haro jets and associated outflows, and may be an important factor in the core's formation and evolution.  Based on the currently available data, we cannot fully explain the nature of NH$_3-$S.  Further molecular line observations and very high sensitivity submm/mm continuum observations are essential to gain more insight into the nature of NH$_3-$S and on the interaction between Herbig-Haro jets and nearby dense cores. The starless core studies provide a great opportunity to determine the initial conditions of  star formation.

\acknowledgements 
We thank the anonymous referee for insightful comments and suggestions which helped us improve the paper.  The work of M.S. was supported by an appointment to the NASA Postdoctoral Program at the Goddard Space Flight Center, administered by Universities Space Research Association under contract with NASA. S. C. acknowledges the support from the NASA's Emerging Worlds Program. J.E.P. acknowledges the financial support of the European Research Council (ERC; project PALs 320620).  S.-L.Q. is supported by NSFC under grant No 11373026, and by Top Talents Program of Yunnan Province (2015HA030). We would like to thank Art Duke, Nabil Afram, and the Hubble Space Telescope SAMS team for expert assistance with the VLA data retrieval.   This research made use of APLpy, an open-source plotting package for Python \citep{robitaille2012}. The National Radio Astronomy Observatory is a facility of the National Science Foundation operated under cooperative agreement by Associated Universities, Inc. This paper makes use of the following ALMA data: ADS/JAO.ALMA\#2012.1.00013.S. ALMA is a partnership of ESO (representing its member states), NSF (USA) and NINS (Japan), together with NRC (Canada), NSC and ASIAA (Taiwan), and KASI (Republic of Korea), in cooperation with the Republic of Chile. The Joint ALMA Observatory is operated by ESO, AUI/NRAO and NAOJ. The James Clerk Maxwell Telescope is operated by the East Asian Observatory on behalf of The National Astronomical Observatory of Japan, Academia Sinica Institute of Astronomy and Astrophysics, the Korea Astronomy and Space Science Institute, the National Astronomical Observatories of China and the Chinese Academy of Sciences (Grant No. XDB09000000), with additional funding support from the Science and Technology Facilities Council of the United Kingdom and participating universities in the United Kingdom and Canada. The JCMT Program ID is M16BP057.  The archival JCMT SCUBA-2 450 $\mu$m image shown in Fig.~\ref{f:fir2mm} was obtained as part of the project with Program ID M97BU88. 


\bibliographystyle{apj}
\bibliography{refs.bib}

\appendix
\section{ALMA Science Data Model UIDs}
The complete list of ALMA Science Data Model (ASDM) UIDs used:  \\
uid://A002/X7f18fb/X72, uid://A002/X800eb6/X52, uid://A002/X7fc9da/X27a1, uid://A002/X7ebc8f/X666, uid://A002/X75ab74/X11f4, uid://A002/X75ab74/Xeb1, uid://A002/X75bfbf/X1130, uid://A002/X75bfbf/Xab8, uid://A002/X75bfbf/Xd9e, uid://A002/X78774a/X842, uid://A002/X788be1/X438, uid://A002/X78e6fe/X3c7, uid://A002/X78e6fe/X6f8, uid://A002/X7fc9da/X2d75, uid://A002/X7fc9da/X5642, uid://A002/X7fc9da/X5b35, uid://A002/X966cea/X156e, uid://A002/X966cea/X183d, uid://A002/X969646/X1ef6, uid://A002/X96bfab/X12a7, uid://A002/X96e770/X128

\section{The ALMA C$^{18}$O and $^{13}$CO Channel Maps}
\setcounter{figure}{0}
\makeatletter 
\renewcommand{\thefigure}{B.\arabic{figure}}

\begin{figure*}[ht!]
\centering
\includegraphics[width=0.9\textwidth]{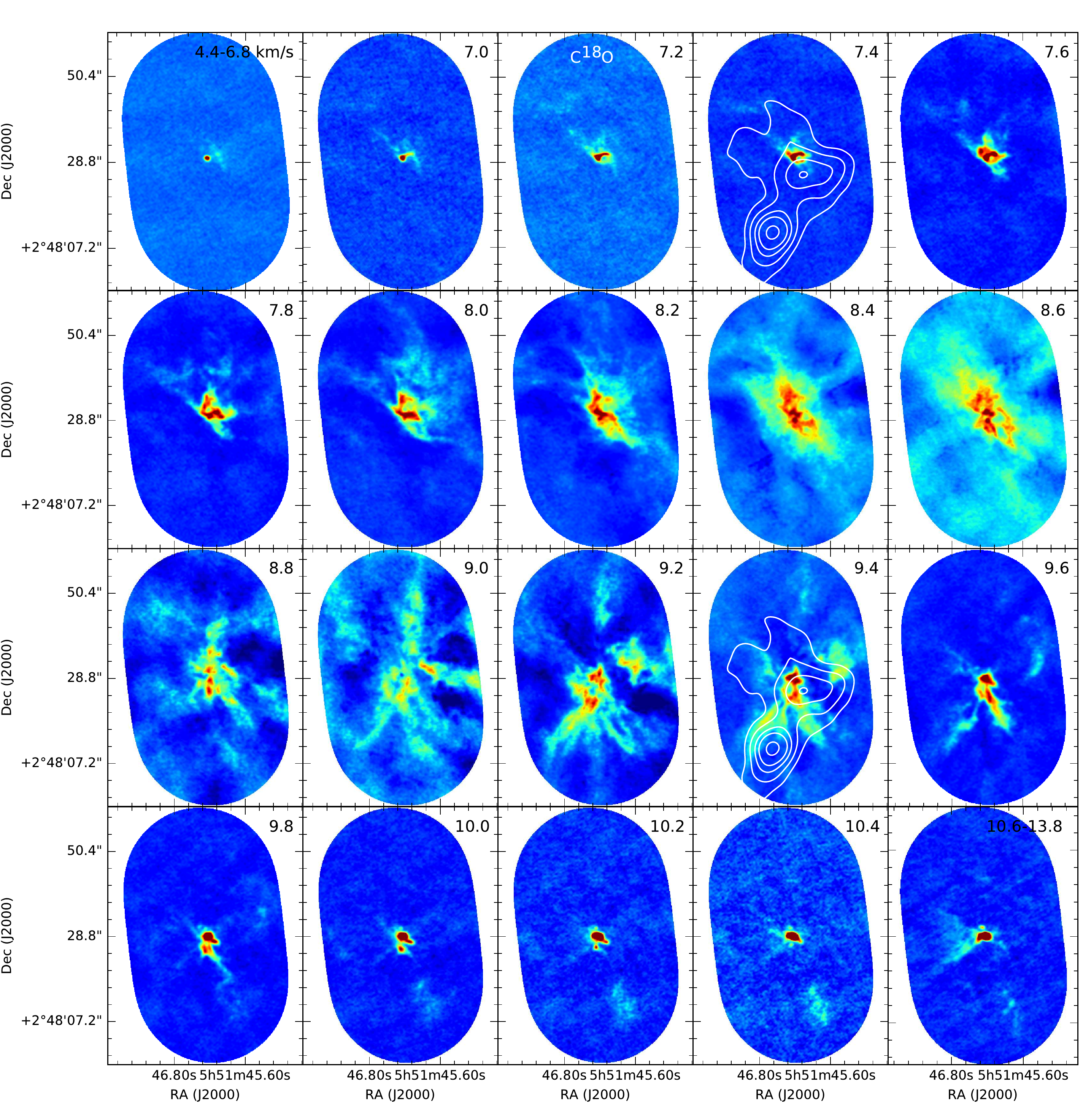} 
\caption{The ALMA C$^{18}$O channel maps of the HH\,111/HH\,121 protostellar system.  The NH$_{3}$ (1,\,1) contours are shown in white in selected images; the contour levels as in Fig.~\ref{f:mom0}.  The LSR velocities  (or velocity ranges in case of the first and the last image) in km~s$^{-1}$ corresponding to each channel (or image showing the integrated emission) are indicated in the upper right corners of the images.  \citet{lee2016} used the C$^{18}$O data to study the envelope and the disk of the HH\,111 protostellar system (or NH$_{3}-$Main). \label{f:c18ochan} }
\end{figure*}

\begin{figure*}[ht!]
\centering
\includegraphics[width=0.33\textwidth]{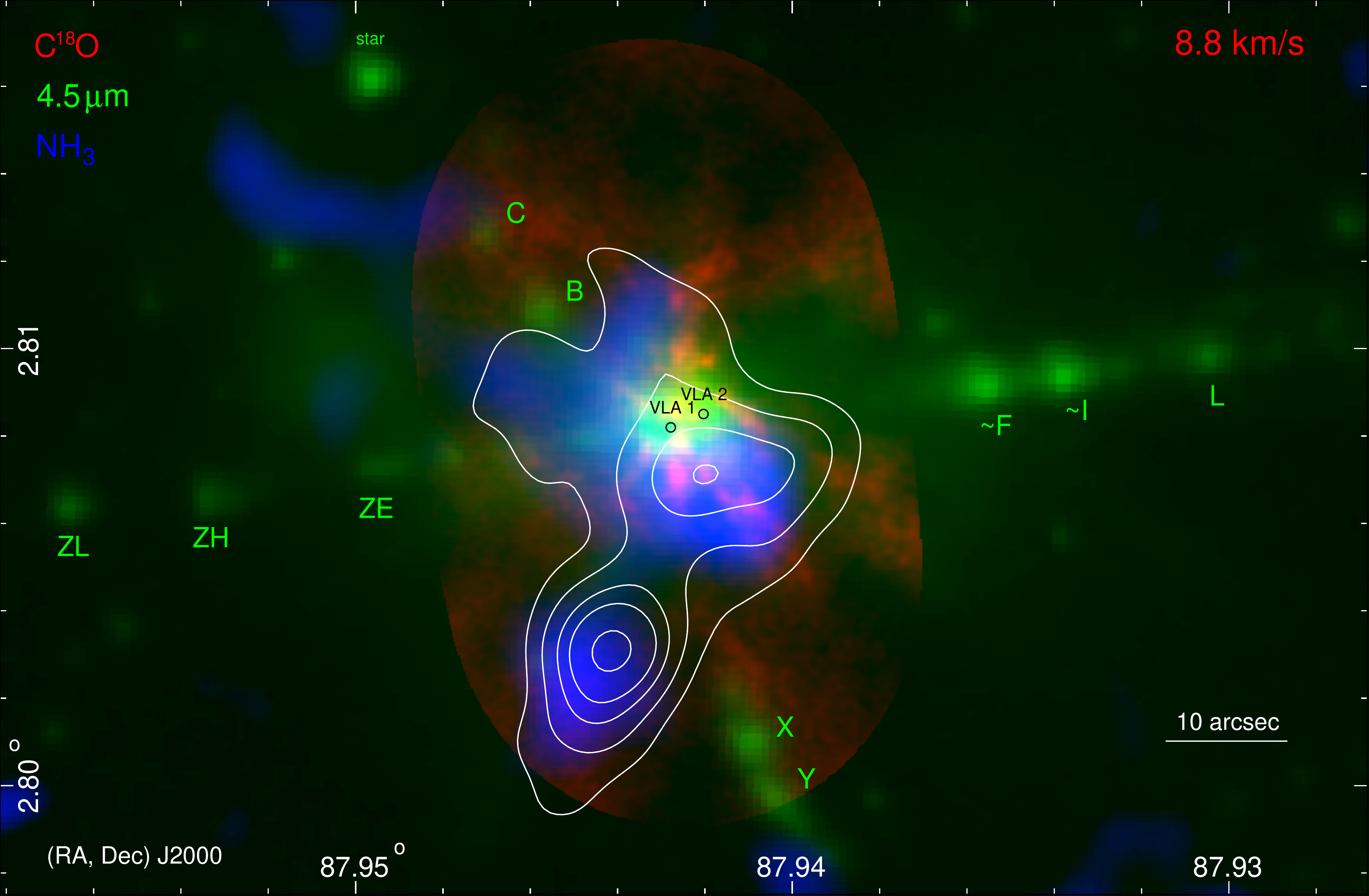} 
\includegraphics[width=0.33\textwidth]{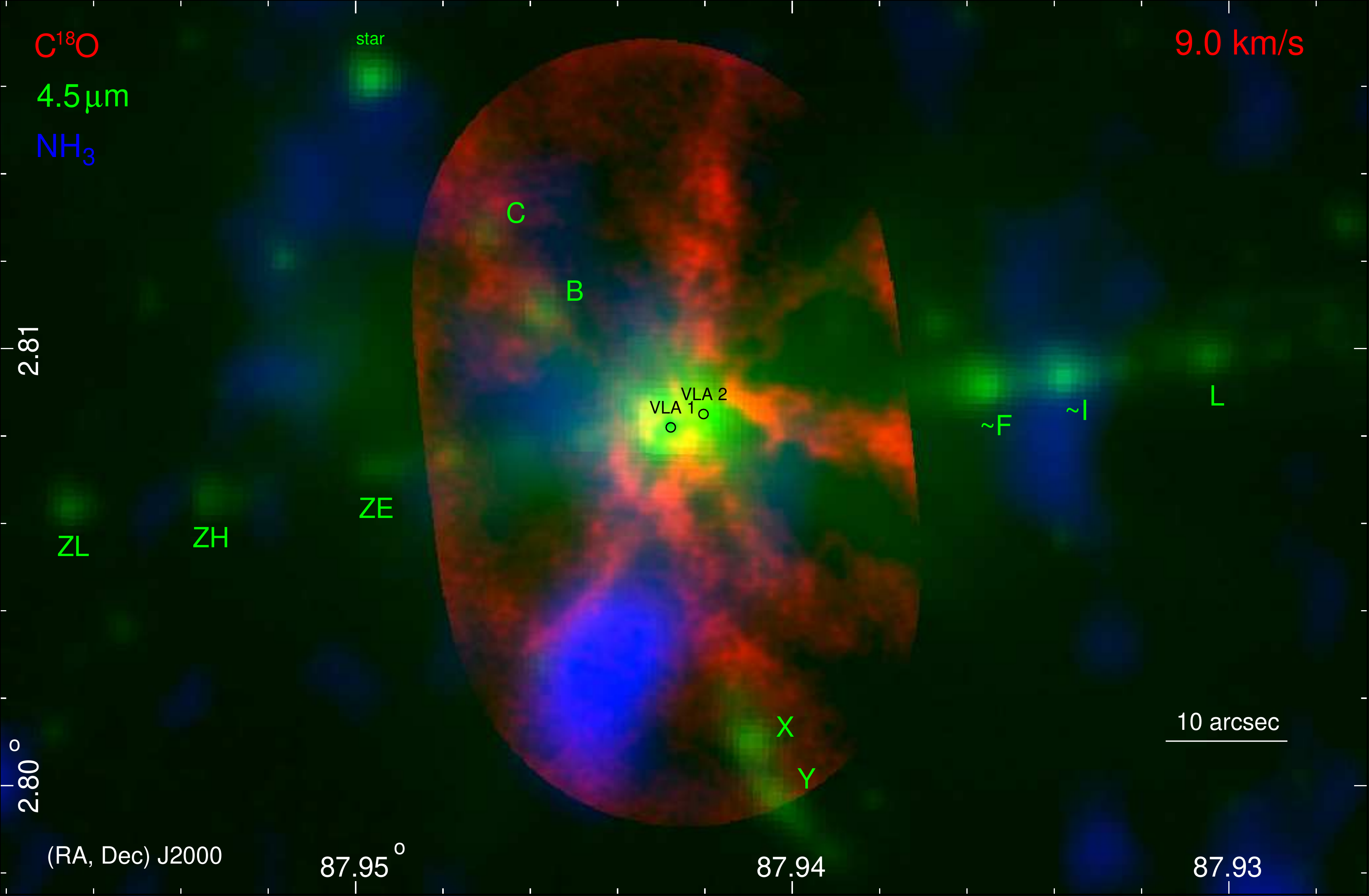} 
\includegraphics[width=0.33\textwidth]{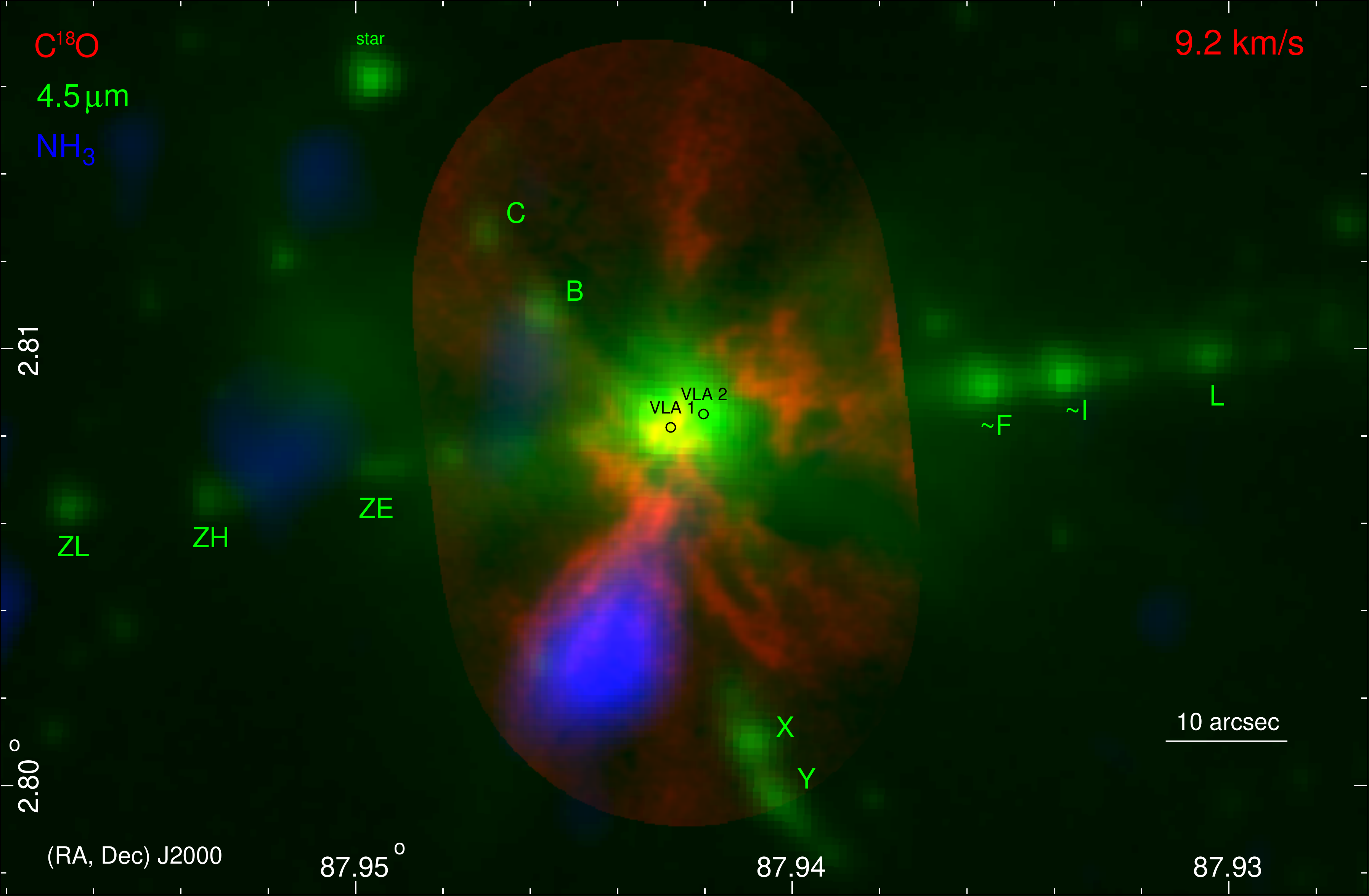} 
\includegraphics[width=0.33\textwidth]{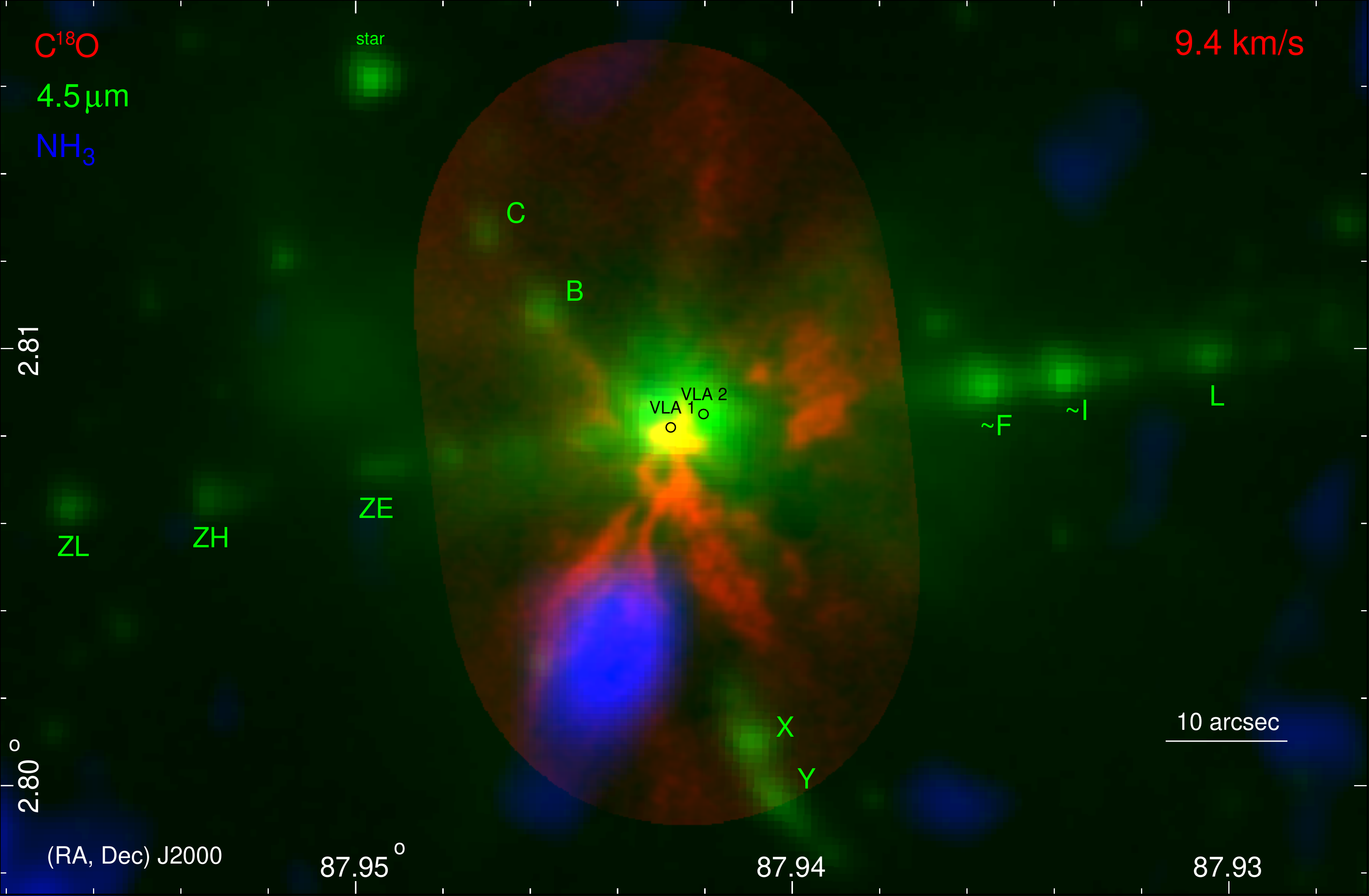} 
\includegraphics[width=0.33\textwidth]{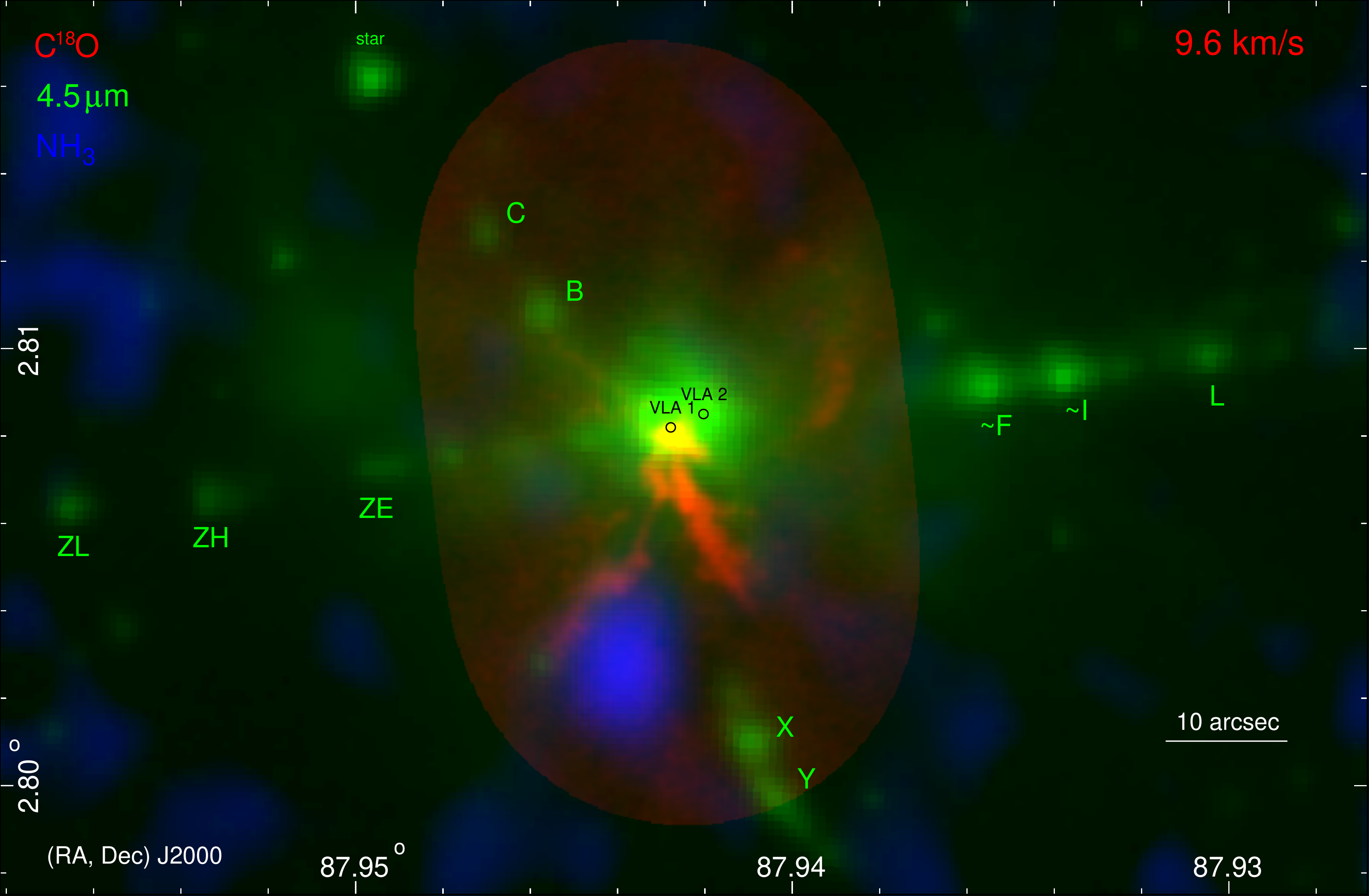} 
\includegraphics[width=0.33\textwidth]{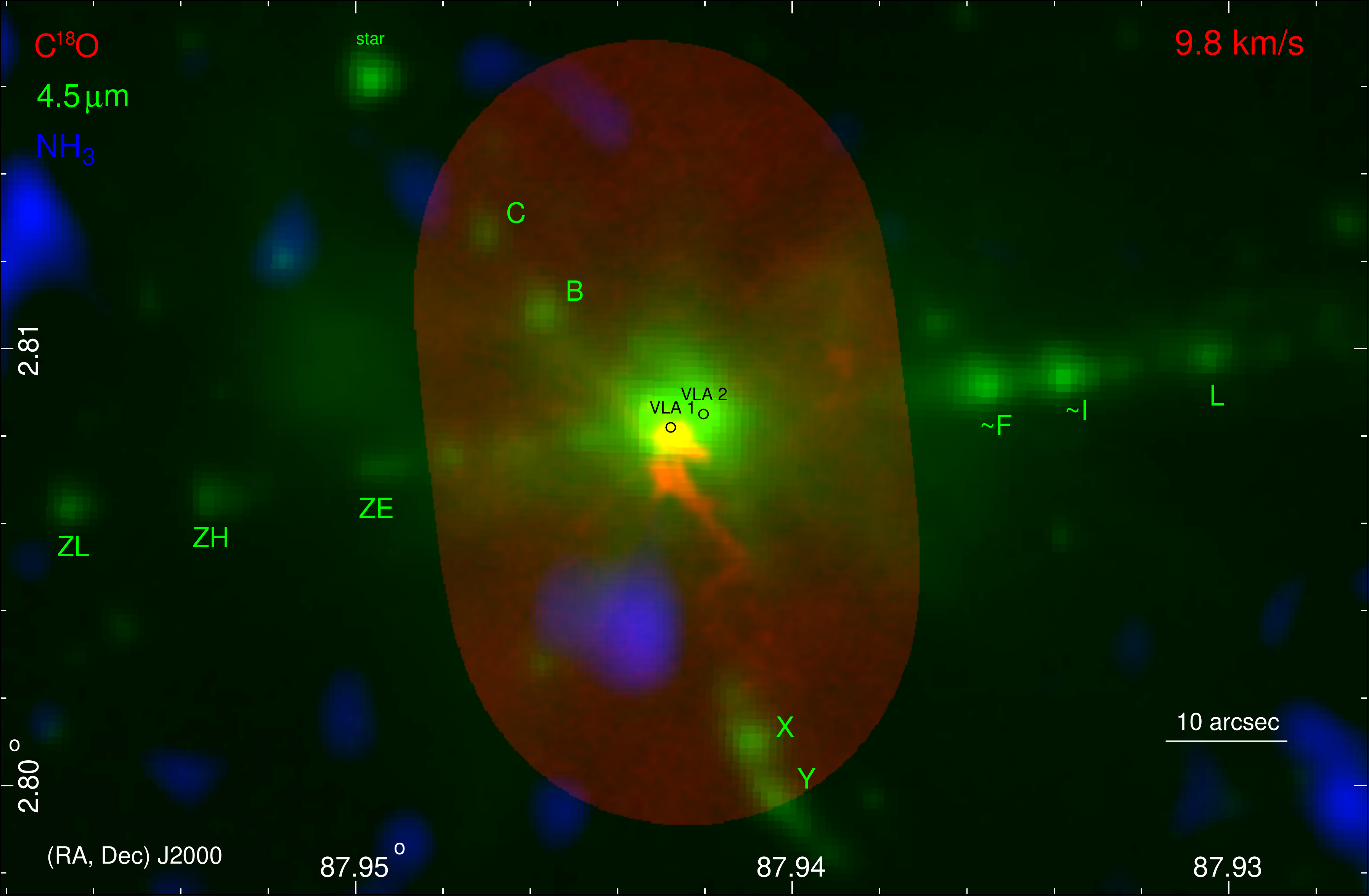} 
\caption{The three-color composite images combining the ALMA C$^{18}$O channel maps ({\it red}; the LSR velocities are indicated in the upper right corners), {\it Spitzer} IRAC 4.5 $\mu$m ({\it green}), and VLA NH$_{3}$ channel maps  ({\it blue}; the LSR velocities are within 0.004--0.05 km~s$^{-1}$ the velocities of C$^{18}$O) images.  The shown velocity range corresponds to the NH$_{3}$ (1,\,1) main line emission toward NH$_{3}-$S. Selected HH\,111 and HH\,121 jet knots visible in the IRAC 4.5 $\mu$m image are labelled (e.g., \citealt{reipurth1997}; \citealt{gredel1994}). \label{f:3colC18O}}
\end{figure*}

\begin{figure*}[ht!]
\centering
\includegraphics[width=0.33\textwidth]{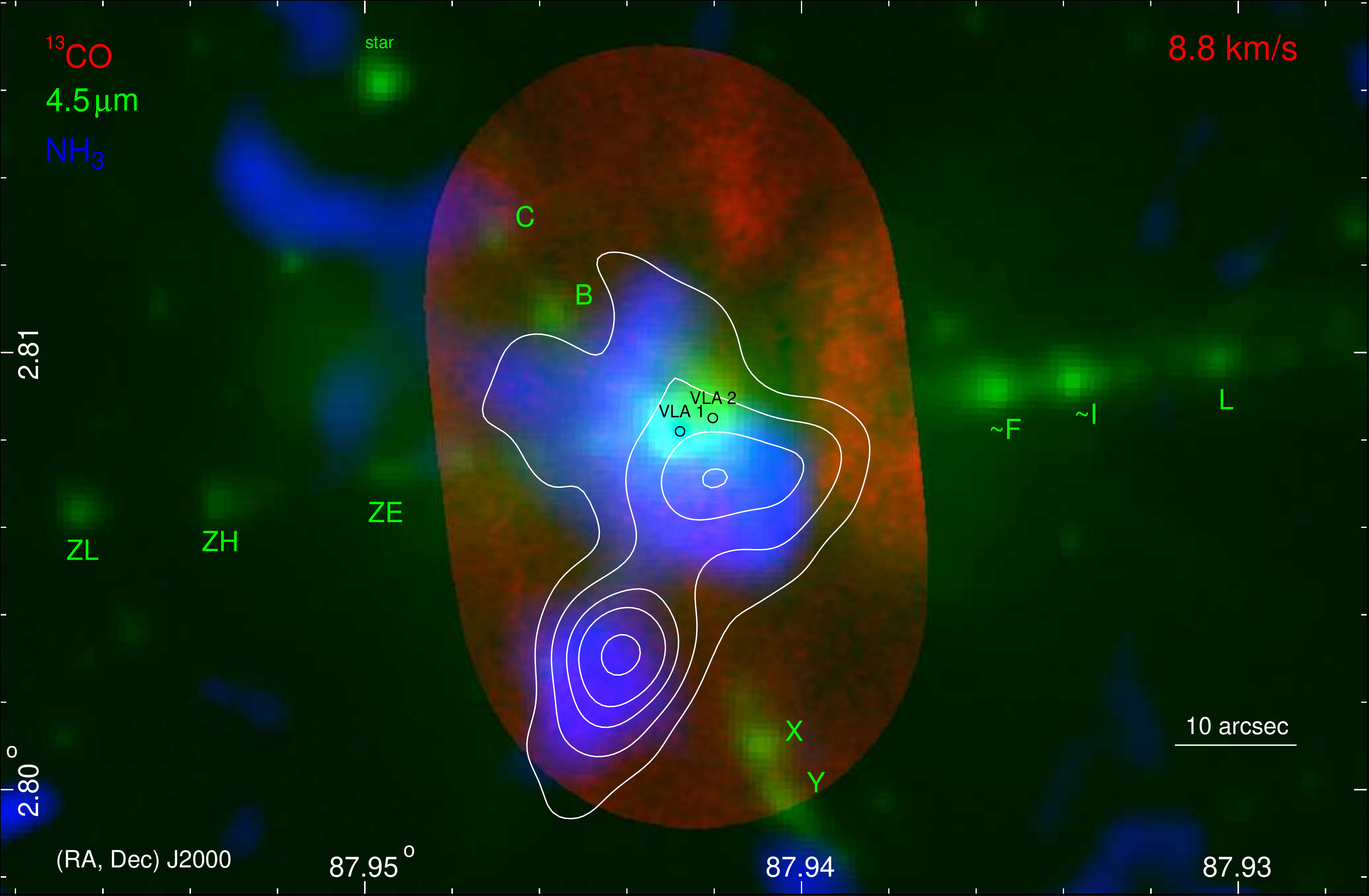} 
\includegraphics[width=0.33\textwidth]{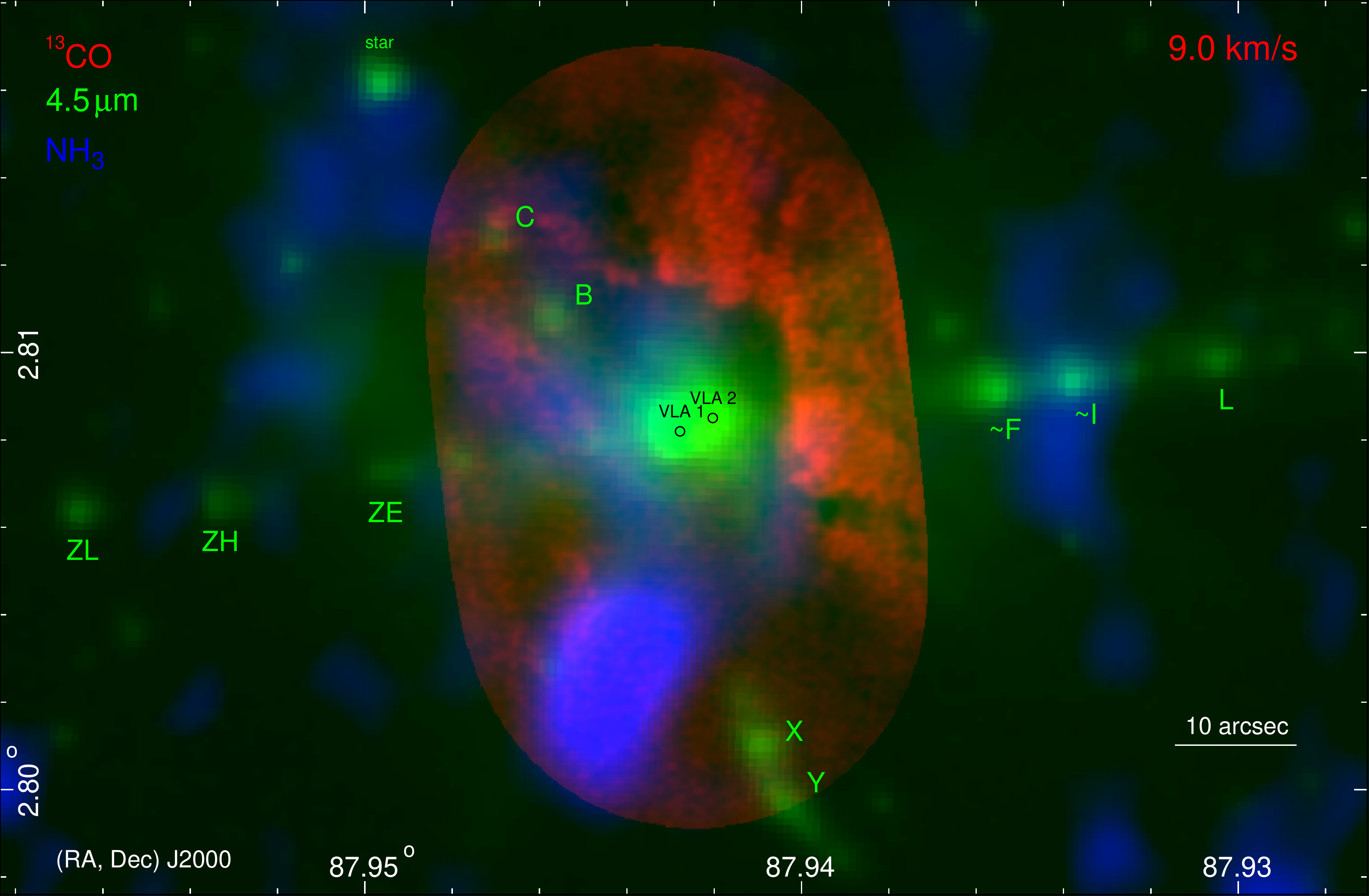} 
\includegraphics[width=0.33\textwidth]{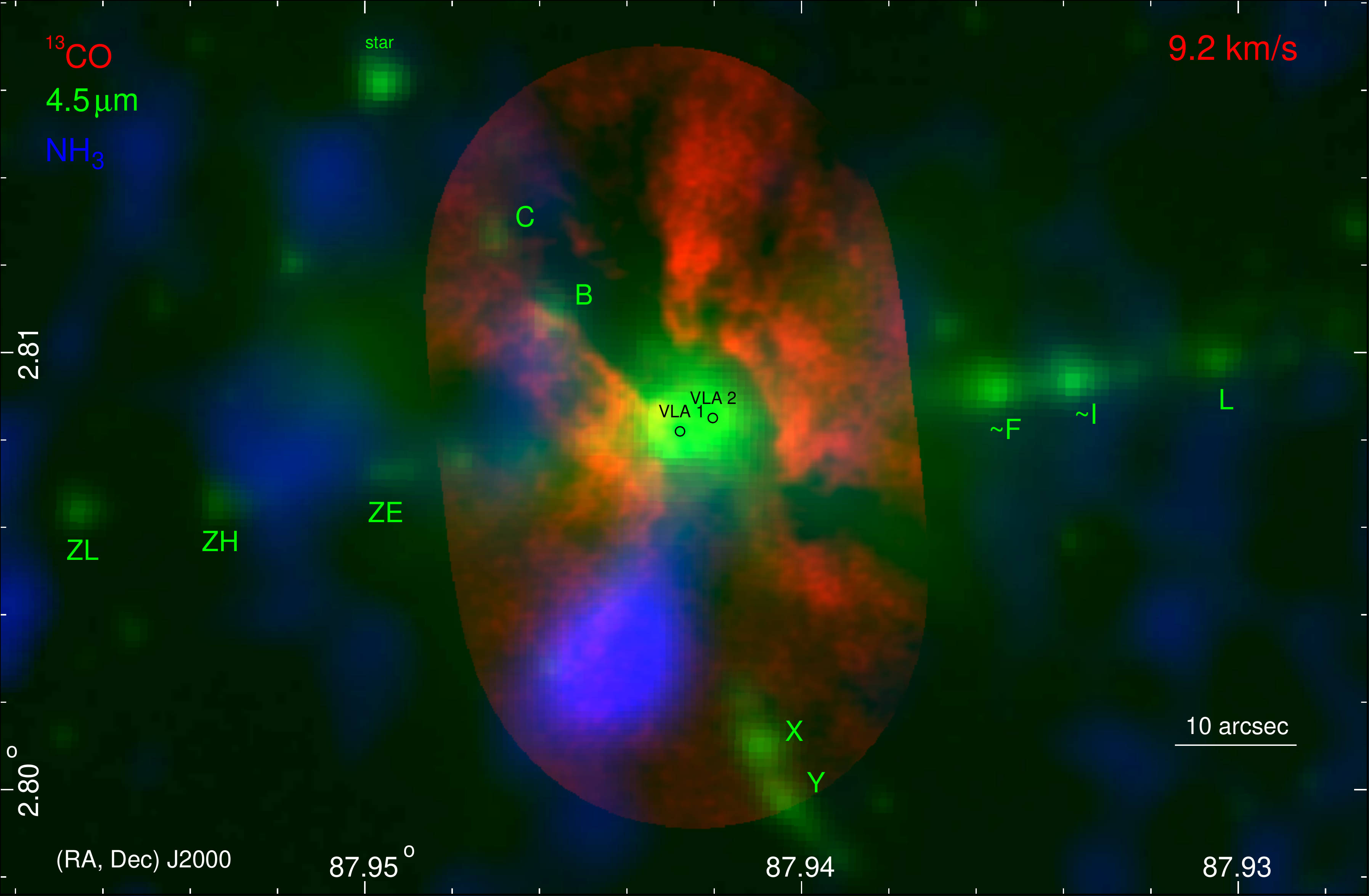} 
\includegraphics[width=0.33\textwidth]{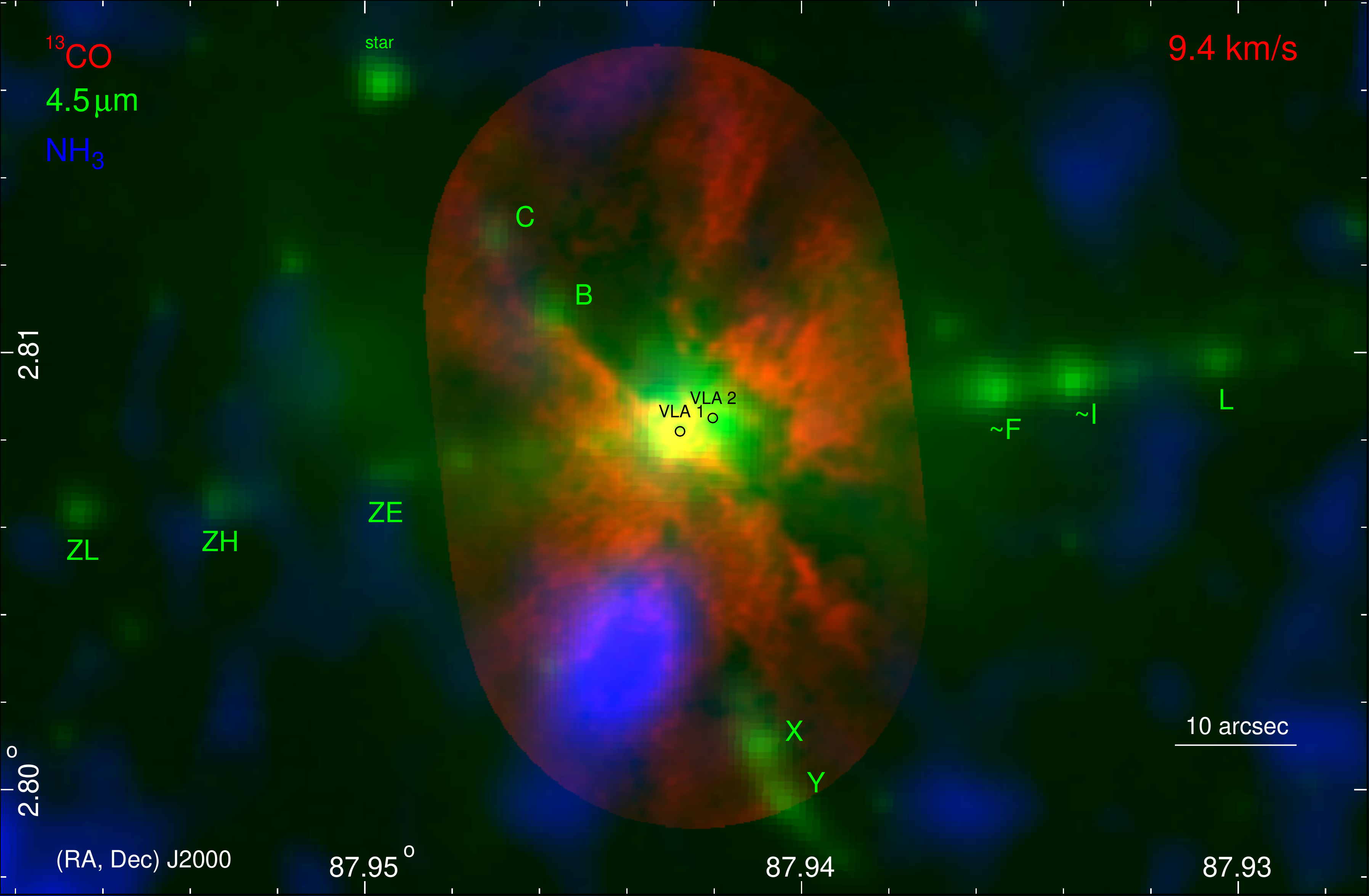} 
\includegraphics[width=0.33\textwidth]{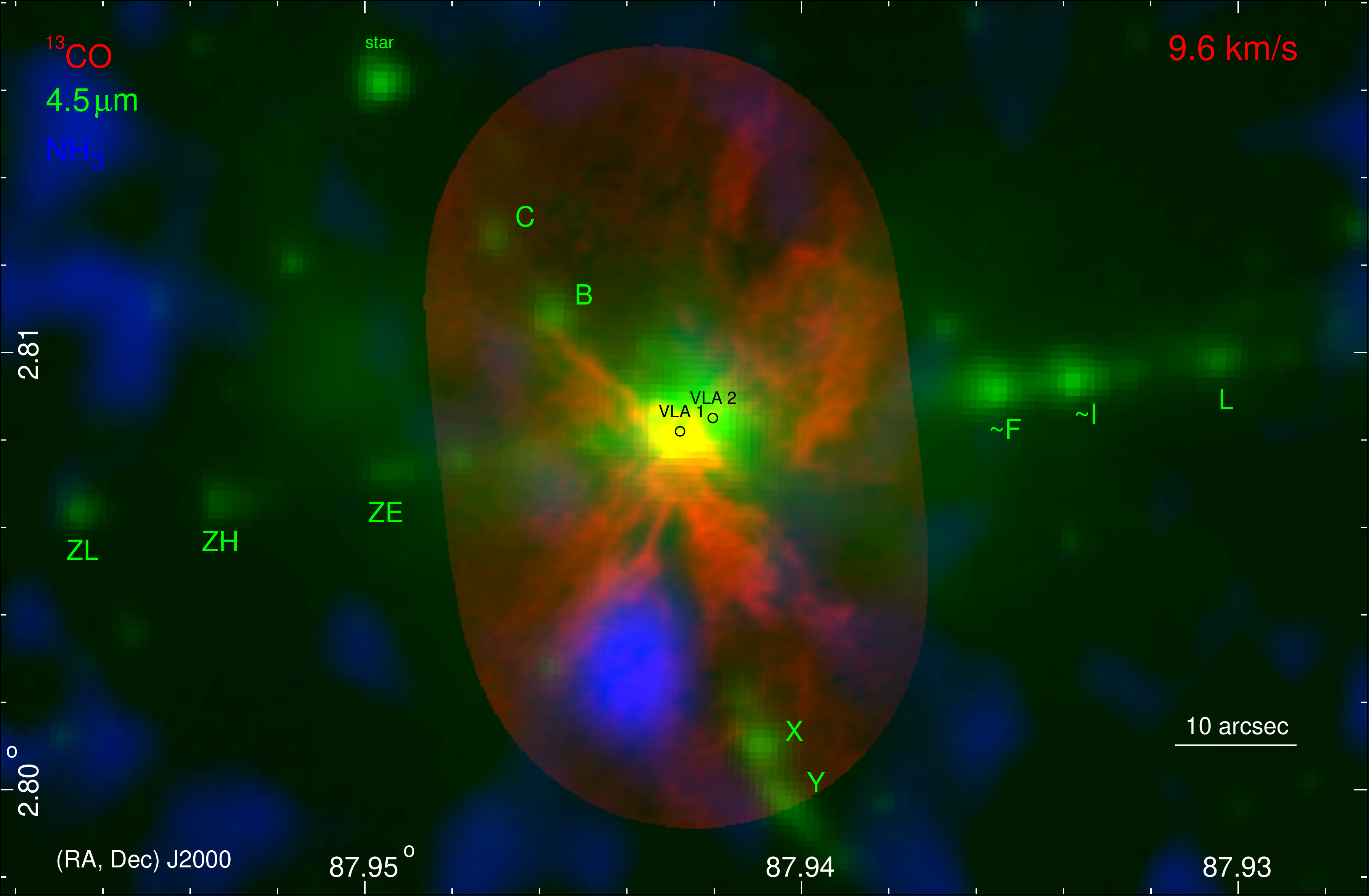} 
\includegraphics[width=0.33\textwidth]{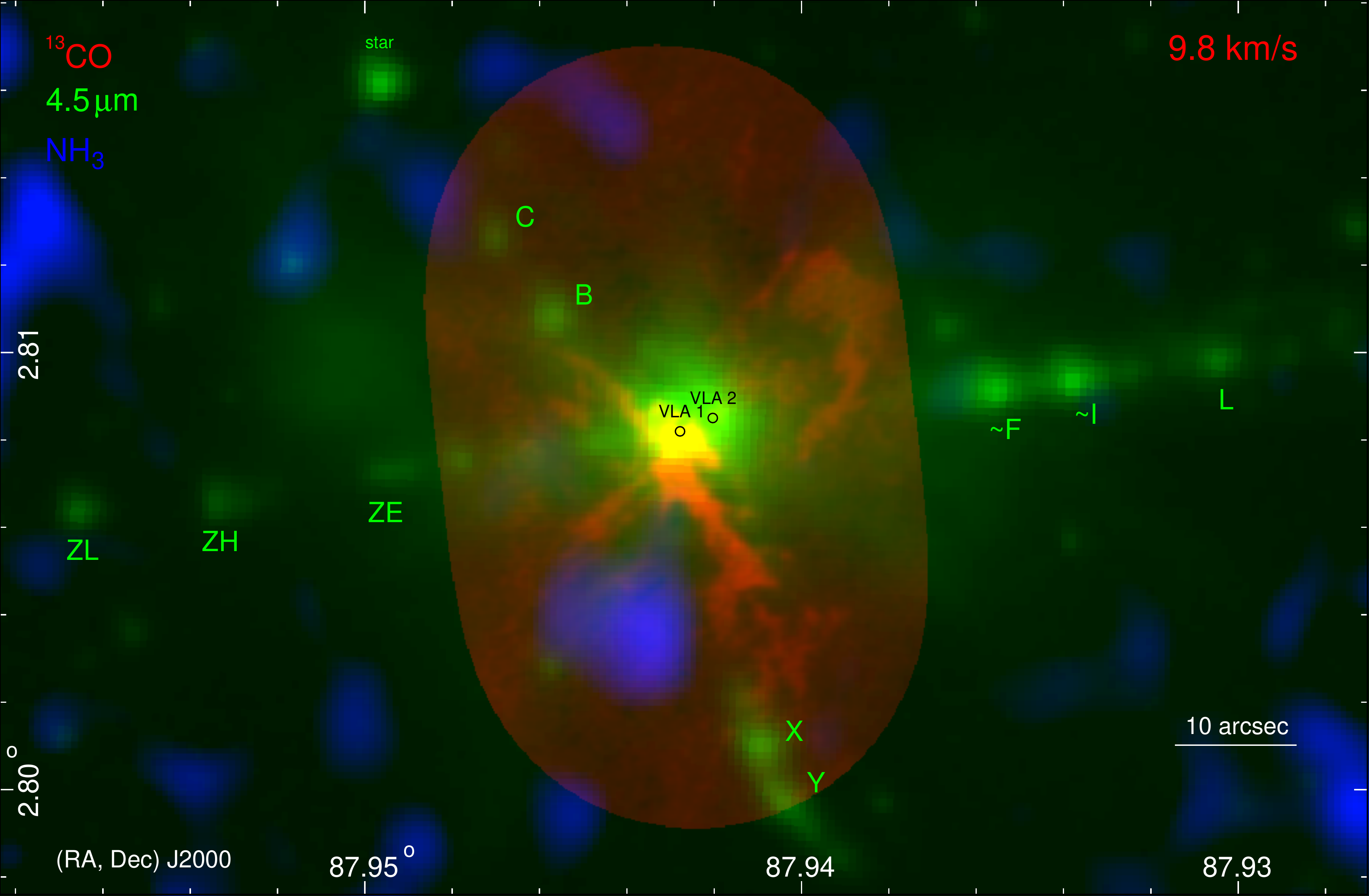} 
\caption{The three-color composite images combining the ALMA $^{13}$CO channel maps ({\it red}; the LSR velocities are indicated in the upper right corners), {\it Spitzer} IRAC 4.5 $\mu$m ({\it green}), and VLA NH$_{3}$  ({\it blue}; the LSR velocities are within 0.004--0.05 km~s$^{-1}$ the velocities of $^{13}$CO) images.  The shown velocity range corresponds to the NH$_{3}$ (1,\,1) main line emission toward NH$_{3}-$S. Selected HH\,111 and HH\,121 jet knots visible in the IRAC 4.5 $\mu$m image are labelled (e.g., \citealt{reipurth1997}; \citealt{gredel1994}). \label{f:3col13CO}}
\end{figure*}

We present the figures showing the ALMA C$^{18}$O (2-1) channel maps (Fig.~\ref{f:c18ochan}) and the C$^{18}$O (2-1) /$^{13}$CO (2-1) channel maps combined with the NH$_{3}$ (1,\,1) channel maps for corresponding LSR velocities and the {\it Spitzer} 4.5 $\mu$m image (Figs.~\ref{f:3colC18O}/\ref{f:3col13CO}). The figures are described in details in Section~\ref{s:almaco}. 

\section{The HCO$^{+}$ (4--3) Line Profiles for HH\,111 / NH$_3-$Main}
\setcounter{figure}{0}
\makeatletter 
\renewcommand{\thefigure}{C.\arabic{figure}}

\begin{figure*}[ht!]
\centering
\includegraphics[width=\textwidth]{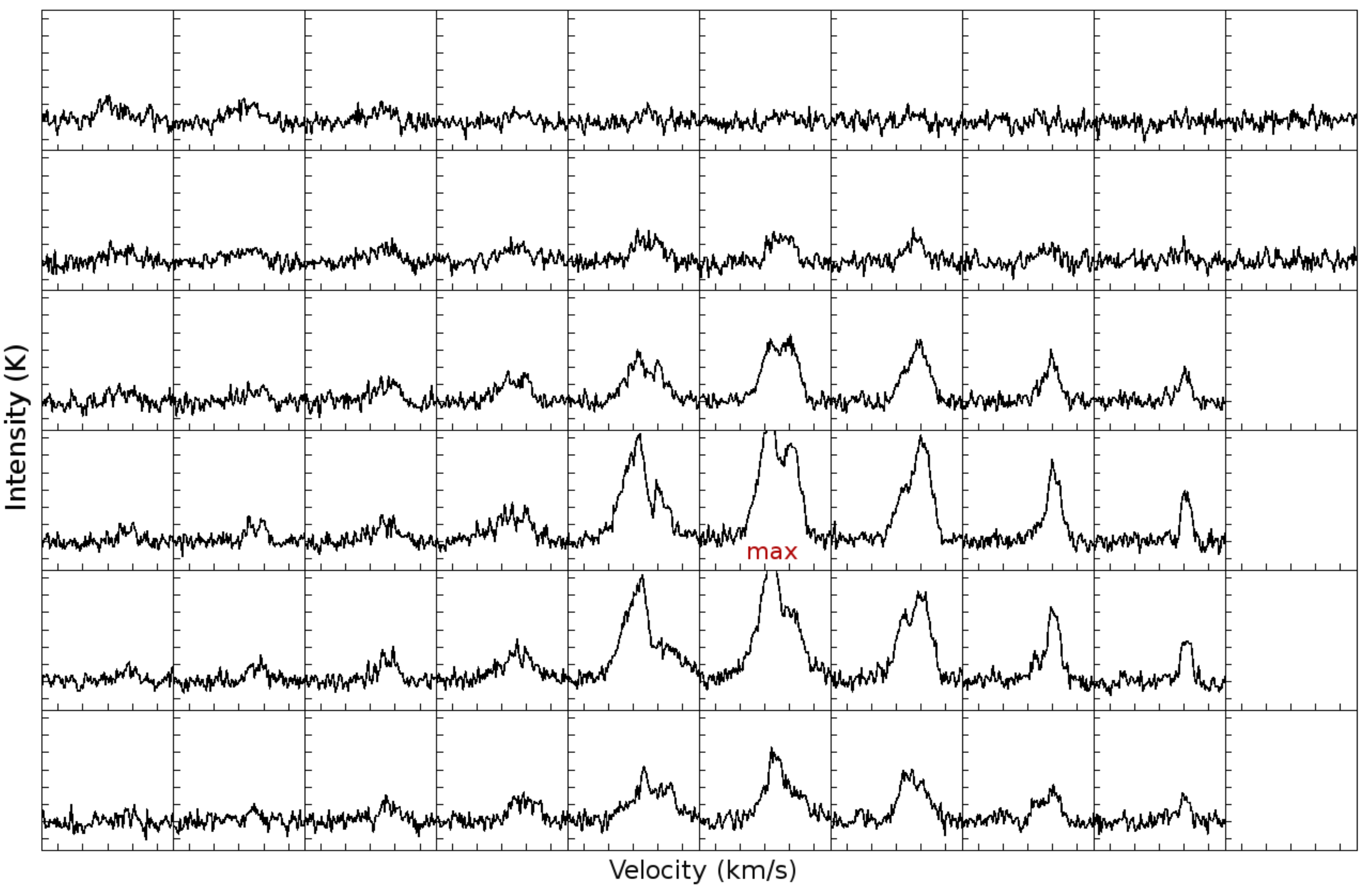}
\caption{The HCO$^{+}$ (4--3) spectra for individual pixels enclosed in the orange rectangle in Fig.~\ref{f:hcop}; the region covers the source of the HH\,111 jet and the base of the molecular outflow.  The spectra were Hanning smoothed with the smoothing kernel width of three.  \label{f:spechcopmain}}
\end{figure*}

In Figure~\ref{f:spechcopmain}, we present the JCMT HCO$^{+}$ (4--3) line profiles for HH\,111 that corresponds to the ammonia source NH$_3-$Main.  Figure~\ref{f:spechcopmain} shows the HCO$^{+}$ spectra for individual pixels enclosed in the orange rectangle in Fig.~\ref{f:hcop}.  The figure is briefly discussed in Section~\ref{s:hcop}.  

\section{The HCO$^{+}$/CO Number Density Ratio}
\setcounter{equation}{0}

We follow the analytic treatment of \citet{charnley1997}.  Cosmic ray ionization of $\rm H_2$ produces $ {\rm H_3^+}$, at a rate $\zeta$ ionizations s$^{-1}$ which then forms ${\rm HCO^+}$ and ${\rm N_2H^+}$ by  proton transfers to CO and N$_2$;  ${\rm N_2H^+}$ is also destroyed by proton transfer to CO.  We  assume a generic rate $k_i $ for these processes. The three molecular ions are then assumed to undergo electron dissociative recombination (neglecting recombination on negatively charged dust grains)  at rates $\alpha_3$,  for $ {\rm H_3^+}$,  and $\alpha$, for CO and N$_2$.  The number density of ${\rm HCO^+}$ in steady-state is then               
  \begin{IEEEeqnarray}{r}
  { n({\rm HCO^+})~=~ { { k_i  n({\rm H_3^+}) + k_i  n({\rm N_2H^+}) } \over { \alpha n_e     } }  n({\rm CO}) }
  \label{e:appd1}
  \end{IEEEeqnarray}
  
  If all atomic oxygen is frozen out on grains as water ice, then 
   \begin{IEEEeqnarray}{r}
 { n({\rm H_3^+})= { { \zeta  n({\rm H_2}) } \over {k_i n({\rm CO})+k_i n({\rm N_2})+ \alpha_3  n_e  } } }
   \end{IEEEeqnarray}
   
  \begin{IEEEeqnarray}{r} 
  { n({\rm N_2H^+})~=~ { {k_i  n({\rm H_3^+})n({\rm N_2}) } \over 
{  k_i n({\rm CO})+\alpha n_e   } }} 
   \end{IEEEeqnarray}
   
  Substituting for $n({\rm H_3^+})$ and $n({\rm N_2H^+})$ in Eq.~\eqref{e:appd1},  we obtain
  \begin{IEEEeqnarray}{r} 
 { n({\rm HCO^+})= { { k_i  \zeta n({\rm H_2}) } \over {\alpha_3 
 n_e +  k_i n({\rm CO})+k_i n({\rm N_2})    } }  
 \left[ 1+    { k_i n({\rm N_2})   \over   \alpha  n_e  +  k_i n({\rm CO})  } \right]{n({\rm CO})\over  \alpha  n_e }}
   \end{IEEEeqnarray}

Now  assuming $\alpha_3 = \alpha$  and making use of the result that the total electron number density $n_e$ is given by 
 \begin{equation} 
n_e ^2  = { \zeta n({\rm H_2})  \over \alpha} 
\end{equation}
(e.g., \citealt{oppenheimer1974}), then the $\rm HCO{^+}/CO$ number density ratio is given by 
 \begin{IEEEeqnarray}{r} 
 {{n({\rm HCO^+})\over n({\rm CO}) } = 
  { k_i  n_e     \over  \alpha n_e +  k_i n({\rm CO}) }  =    { 1  \over  {\alpha \over k_i } +   {n({\rm CO}) \over  n_e }  }}
 \end{IEEEeqnarray}

\end{document}